\documentclass[prx,twocolumn,english,superscriptaddress,nolongbibliography]{revtex4-2}
\usepackage[T1]{fontenc}
\usepackage[latin9]{inputenc}
\usepackage{graphicx}
\usepackage{amsmath}  
\usepackage{xfrac}
\usepackage{txfonts}
\usepackage{float}
\usepackage[mathscr]{euscript}

\makeatletter
\usepackage{babel}
\usepackage{color}
\usepackage[colorlinks = true,
            linkcolor = blue,
            urlcolor  = blue,
            citecolor = blue,
            anchorcolor = blue]{hyperref}
\usepackage{tikz}

\makeatother

\newcommand{\yto}{Yb\textsubscript{2}Ti\textsubscript{2}O\textsubscript{7}}

\begin{document}

\title{Intrinsic quantum disorder in \yto{} and the quantum $S=1/2$ pyrochlore phase diagram} 

\author{Shang-Shun Zhang}
\email{szhang57@utk.edu}
\affiliation{Department of Physics and Astronomy, The University of Tennessee, Knoxville, Tennessee 37996, USA}

\author{Anish Bhardwaj}
\affiliation{Department of Physics, St. Bonaventure University, New York 14778, USA}

\author{S.M. Koohpayeh}
\address{Institute for Quantum Matter and Department of Physics and Astronomy, Johns Hopkins University, Baltimore, MD 21218} 

\author{D.M. Pajerowski}
\affiliation{Neutron Scattering Division, Oak Ridge National Laboratory, Oak Ridge, TN 37831, USA}

\author{Jeffrey G. Rau}
\affiliation{Department of Physics, University of Windsor, 401 Sunset Avenue, Windsor, Ontario, N9B 3P4, Canada}

\author{Hitesh J. Changlani}
\email{hchanglani@fsu.edu}
\affiliation{National High Magnetic Field Laboratory, Tallahassee, Florida 32310, USA}
\affiliation{Department of Physics, Florida State University, Tallahassee, Florida 32306, USA}

\author{Allen Scheie}
\email{scheie@lanl.gov}
\affiliation{MPA-Q, Los Alamos National Laboratory, Los Alamos, NM 87545, USA}

	\date{\today}

\begin{abstract}
We present an experimental and theoretical study of the anisotropic pyrochlore phase diagram. Inelastic field-dependent neutron scattering on Yb$_2$Ti$_2$O$_7$ shows intrinsic broadening and a flat low-energy magnon mode which is partially captured by interacting magnon models. Exact diagonalization reveals the existence of an emergent quantum phase between ferromagnetism and antiferromagnetism, in which Yb$_2$Ti$_2$O$_7$  Hamiltonian potentially resides. 
This behavior matches the phenomenology of quantum criticality in heavy fermion systems, and shows  Yb$_2$Ti$_2$O$_7$ is a clean system which can be field-tuned from well-defined magnons to a nontrivial quantum ground state. This suggests that quantum criticality is a generic feature of the dipolar phase diagram.

\end{abstract}

\maketitle

\section{Introduction}

Quasiparticles are one of the foundational concepts
of condensed matter physics. They successfully explain 
many different kinds of materials behaviors, from bulk properties to spectral responses \cite{altland2010condensed}. 
A frontier in condensed matter then is where quasiparticle 
picture breaks down. 
Trivially, this occurs in disordered systems which can no longer be described as a smoothly varying quantum field. 
However, a more interesting scenario is when quasiparticle breakdown occurs in a system without extrinsic disorder \cite{pitaevskii1959properties,Nichitiu_2024_He4}.  
It includes the fractionalization of conventional quasiparticles into qualitatively different ones \cite{Laughlin_Nobel,broholm2020quantum,stone2006quasiparticle}, characteristic features of novel phases such as quantum spin liquid, and also the cases where quasiparticles disappear entirely, as in quantum critical metals \cite{hu2024quantum,abanov2003quantum,abanov2020interplay,eberlein2016hyperscaling,hartnoll2011quantum}. 

\begin{figure}
	\centering
\begin{tikzpicture}
    \node[inner sep=0] (image) at (0,0) {\includegraphics[width=0.48\textwidth]{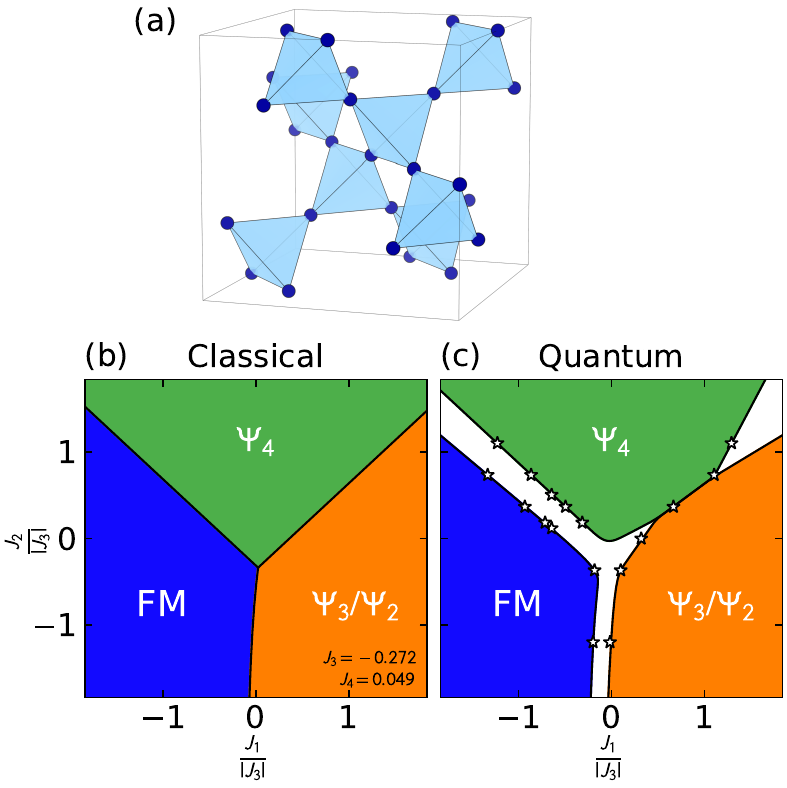}};
    \draw (2.8, 2.5) node {${\bf J} = \left( \begin{matrix} J_2 & J_4 & J_4 \\ -J_4 & J_1 & J_3\\ -J_4 & J_3 & J_1\end{matrix} \right)$};
 \end{tikzpicture}
	\caption{(a) Pyrochlore unit cell of corner sharing tetrahedra, with the nearest neighbor exchange matrix \cite{Ross_Hamiltonian}. (b) Classical phase diagram \cite{Yan2017}. (c) Quantum phase diagram from this study, where the white stars indicate phase boundaries from exact diagonalization. White regions are the emergent quantum phases surrounding classical degeneracies.} 
	\label{fig:pryochlore}
\end{figure}

In this context, we revisit a well-studied material \yto{}. This compound has effective $J=1/2$ magnetic Yb$^{3+}$ ions arranged in a highly frustrated pyrochlore lattice of corner sharing tetrahedra \cite{Blote1969,HodgesCEF,Onoda_2011} (see Fig. \ref{fig:pryochlore}). 
The cleanest crystals show ferromagnetic order at $T_c=270$~mK \cite{FM_order2003,Chang_2012_Higgs,Shafieizadeh2018,Yaouanc_order,Scheie2017,Saubert_2020} (although $\sim 2\%$ crystalline disorder suppresses the magnetic order \cite{SampleDependence_HC,SampleDependence_Ross,Ross_stuffing,GaudetRoss_CEF,Mostaed2017,DOrtenzio_noGsOrder,Bowman2019,SeyedPaper}). 
And yet, even in clean samples with robust magnetic order, the zero-field neutron spectrum (which measures the dynamic spin correlations) shows a diffuse continuum \cite{Ross2009,Ross_Hamiltonian,Thompson_2017,Scheie2020} instead of sharp magnon modes typically seen in well-ordered magnets \cite{fishman2018spin}. Furthermore, above the magnetic ordering transition there is anomalous thermal conductivity \cite{MonopolesConductivity,hirschberger2019enhanced} and quantum-critical-like scaling in the dynamic structure factor \cite{Scheie_2022_Dynamical}, indicating non-trivial physics at play.

For a time, the strange behavior of \yto{} was interpreted as evidence for a quantum spin ice state \cite{Ross_Hamiltonian,Hayre2013,Hamachi_2016,Armitage_monopoles_2016,ApplegateSpinIce,SpinIce_review}, but refinements of its spin exchange Hamiltonian \cite{Robert2015,Thompson_2017,Scheie2020} and clear evidence of ferromagnetism \cite{Yaouanc_order,Scheie2017,Saubert_2020} have refuted this idea. The refined spin exchange Hamiltonian is extremely close to a phase boundary between ferromagnetism (FM) and antiferromagnetism (AFM) \cite{Robert2015,Thompson_2017,Yan2017,rau2019magnon,Scheie2020,Scheie_2022_Dynamical}, suggesting that its exotic behavior is caused (in a manner not completely clear) by ``multiple phase competition'' \cite{Jaubert2015,Yan2017}. 
Accordingly, it was previously suggested that the diffuse spectral features were from a coexistence of FM and AFM in \yto{}, perhaps within antiferromagnetic domain walls \cite{Scheie2020}. 

To test these ideas, we study high-purity single crystal \yto{} with inelastic neutron spectroscopy in a magnetic field to suppress the magnetic domains. 
We also compare the inelastic data to models of interacting magnons. 
We finally use exact diagonalization calculations to map the quantum pyrochlore phase diagram and put \yto{} in context,  
to guide research into other dipolar pyrochlore systems.  
We find that at low fields, the broadened magnon modes are intrinsic to \yto{}, and not the result of domains or phase coexistence. We also observe 
flattened magnon bands at low energy driven by quantum effects. 
Our phase diagram calculations show an extended region of emergent quantum phase, which explains the breakdown of magnon physics in \yto{}, and bears and strong resemblance to quantum criticality.

\section{Methods}

{We now briefly summarize the experimental and theoretical methods that were used in the paper. More details about these procedures can be found in the Supplemental Information \cite{SuppMat}.}

\subsection{Neutron Experiments}
\begin{figure*}
	\centering\includegraphics[width=\textwidth]{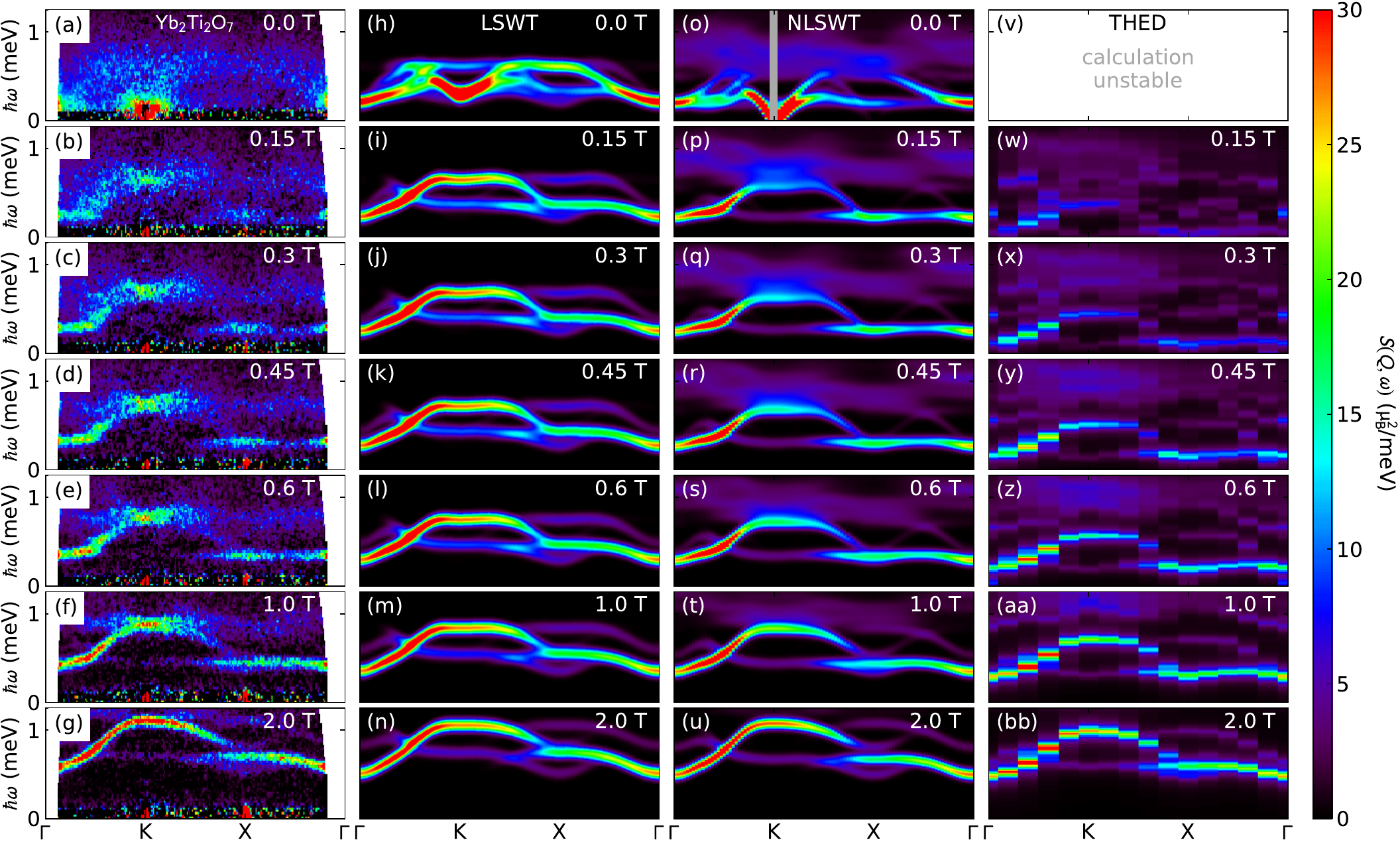}
	
	\caption{\yto{} inelastic neutron scattering along high symmetry directions $\Gamma (000) \rightarrow K (220) \rightarrow X (200) \rightarrow \Gamma$ at $T=0.15$~K at various fields between 0~T and 2~T along [001]. Panels (a)-(g) show experimental data. At 0~T the excitations are very broad and are most intense around $K$, but the smallest field completely suppresses the $K$-point low-energy scattering. As field increases, the modes gradually become sharper and less diffuse.
    Panels (h)-(n) show linear spin wave theory (LSWT) calculations of the inelastic spectrum using the Hamiltonian in Ref. \cite{Thompson_2017}.
    Panels (o)-(u) show nonlinear spin wave theory (NLSWT) calculations using the same Hamiltonian (the zero field calculations around the $K$-point are unstable, and are shown by a grey region).
    Panels (v)-(bb) show the THED calculations with the same Hamiltonian. At zero field the THED calculations are unstable, but at finite fields they show a significant amount of broadening.
    }
	\label{flo:measuredSpectra}
\end{figure*}

We measured the single crystal neutron spectrum of \yto{} using the CNCS spectrometer \cite{CNCS} at Oak Ridge National Laboratory's Spallation Neutron Source. Two crystals (the same as those used in Refs. \cite{Scheie2017,Scheie2020,Scheie_2022_Dynamical} grown with the traveling solvent floating zone technique \cite{SeyedPaper}) were coaligned in the $(hk0)$ scattering plane and mounted in a dilution refrigerator in a vertical [001] magnetic field. (The [001] magnetic field was measured previously, but in a sample with some ``stuffing'' disorder \cite{Thompson_2017} which introduces some extrinsic broadening from defects. Here we study a cleaner sample  to reveal the intrinsic behavior \cite{SeyedPaper,Shafieizadeh2018}.) Further details on experimental methods are in Appendix \ref{app:experiment}. The data are shown in Fig. \ref{flo:measuredSpectra}.

\subsection{Effective low-energy $S=1/2$ dipolar Hamiltonian and parameters}
Following previous literature~~\cite{Curnoe_2007,Onoda_2011,Ross_Hamiltonian} (and references therein), the spin-$1/2$ low-energy effective Hamiltonian on the pyrochlore lattice for dipolar spins, with nearest neighbor interactions and Zeeman coupling to an external field $h=(h_x,h_y,h_z)$ is given by
\begin{equation}
H = \frac{1}{2} \sum_{ij} J^{\mu\nu}_{ij} S^{\mu}_{i} S^{\nu}_{j} - \mu_{B} h^{\mu} \sum_{i} g^{\mu \nu}_{i} S^{\nu}_{i},
\label{eq:Ham_supp}
\end{equation}
where $i,j$ are nearest neighbors and $\mu,\nu$ refer to $x,y,z$, $S^{\mu}_i$ refer to the spin-1/2 components at site $i$, and
 $\mathbf{J}_{ij}$ and $\mathbf{g}_i$ are bond and site dependent spin-exchange interactions and coupling matrices respectively
(whose components have been explicitly written out in the SM). For example, 
for [1,1,0] bonds we have,
\begin{equation}
    {\bf J} = \left( \begin{matrix} J_1 & J_3 & J_4 \\ J_3 & J_1 & J_4\\ -J_4 & -J_4 & J_2\end{matrix} \right),
    \label{eq:exchangematrix}
\end{equation}
where $J_1,J_2,J_3,J_4$ are used to parameterize the coupling matrices. Ref.~\cite{Thompson_2017} determined $J_1 = -0.028$~meV, $J_2 = -0.326$~meV, $J_3 = -0.272$~meV, $J_4 = 0.049$~meV and $g$-tensor $g_{xy} = 4.17$,  $g_{z} = 2.14$.  Although other 
parameter sets have been proposed \cite{Ross_Hamiltonian,Robert2015,Scheie2020}, this set 
was fitted to the largest amount of data and at fields high enough to avoid nonlinear magnon effects which occur below 2~T \cite{rau2019magnon}. 
While we have associated this specific parameter set with Yb$_2$Ti$_2$O$_7$, we have also carried out an extensive scan of other Hamiltonian parameters with multiple theoretical techniques (discussed below). 

\subsection{ Interacting Magnons: Nonlinear spin wave theory and truncated Hilbert space exact diagonalization
}

We simulated the \yto{} inelastic neutron spectrum using  non-interacting linear spin wave theory (LSWT), perturbatively interacting nonlinear spin wave theory (NLSWT) and a non-perturbative  ``truncated Hilbert space exact diagonalization''  (THED) technique where magnon-magnon interactions are treated up to infinite order. The results of these calculations are shown in Fig. \ref{flo:measuredSpectra}. 

The LSWT calculations, which neglect magnon-magnon interactions, were performed using \texttt{SpinW} \cite{SpinW} on top of a canted ferromagnetic ground state.  
The NLSWT calculations include magnon-magnon interactions perturbatively to next to leading order in $1/S$. Details of these calculations for Yb$_2$Ti$_2$O$_7$ can be found in Ref.~\cite{rau2019magnon}. These magnon-interaction corrections include the static ``tadpole'' and ``Hartree'' diagrams as well as the dynamical ``bubble'' that can describe spontaneous magnon decay~\cite{zhito2013}. For additional details, see the Supplemental Information \cite{SuppMat} (in which is cited \cite{mourigal2013}).

To capture strong interaction effects beyond perturbation theory, we employ a non-perturbative semiclassical approach that projects the full spin Hamiltonian onto a truncated Hilbert space containing at most two magnons, followed by exact diagonalization of this reduced Hamiltonian. We refer to this method as truncated Hilbert-space exact diagonalization (THED) \cite{Legros_2021,zhang2025nonperturbative}. This approach is well justified when the single-magnon excitation possesses a finite energy gap much larger than the magnon-magnon interaction energy, ensuring that states with three or more magnons make negligible contributions to the low-energy sector. In Yb$_2$Ti$_2$O$_7$, this condition holds under sufficiently strong external fields ($\gtrsim 1$~T) but breaks down near zero field, where the single-magnon gap becomes small.

To make contact with NLSWT, the ``single-magnon states'' in the above construction correspond to those defined within LSWT, while the ``two-magnon states'' are their 
 product states in Fock space. 
The one- and two-magnon sectors are coupled via the cubic interaction vertex, responsible for the energy renormalization and finite lifetime of single-magnon excitations in NLSWT. Meanwhile, interactions among two-magnon states are mediated by the quartic vertex, which in perturbative treatments leads to Hartree-Fock corrections to the single-magnon dispersion. Within THED, however, the quartic vertex is treated non-perturbatively, allowing for the emergence of bound or resonant states and redistribution of spectral weight within the energy window of an excitation continuum \cite{Legros_2021,zhang2025nonperturbative}.

\subsection{Exact Diagonalization Phase Diagram Calculations}
In both the NLSWT and THED procedures discussed above, the starting point is a classical spin-ordered state on top of which additional quantum fluctuations are incorporated. While this works well for many ordered magnets, it is inadequate for capturing 
 exotic physics where either long-range order is absent or the notion of well-defined quasiparticles completely breaks down. 

With this motivation, we have carried out Lanczos exact diagonalization (ED) calculations on the Hamiltonian in Eq.~\eqref{eq:Ham_supp} 
 to avoid the above restrictions. 
The major limitation is the accessible system sizes, in this work we have carried out calculations on $N=16$ and $N=32$ sites, with a majority of our assertions based on the latter.
(More details about the geometry and the use of translational symmetry have been discussed in the Supplemental Information \cite{SuppMat} and in previous work~\cite{Changlani2017quantum}.)
We computed the equal-time spin correlations $S({\bf Q})$ across the nearest-neighbor exchange phase diagram, varying the elements of the nearest-neighbor exchange matrix in Eq. \eqref{eq:exchangematrix}. 
Specifically, we varied $J_1$ and $J_2$ while keeping $J_3$ and $J_4$ fixed to $J_3=-0.272$~meV and $J_4=0.049$~meV, corresponding to the Yb$_2$Ti$_2$O$_7$ Hamiltonian reported in Ref. \cite{Thompson_2017}. (In the Supplemental Information \cite{SuppMat}, we show the calculations repeated for $J_3=-0.3$~meV and $J_4=0.0$.)

\begin{figure*}
	\centering
	\includegraphics[width=\textwidth]{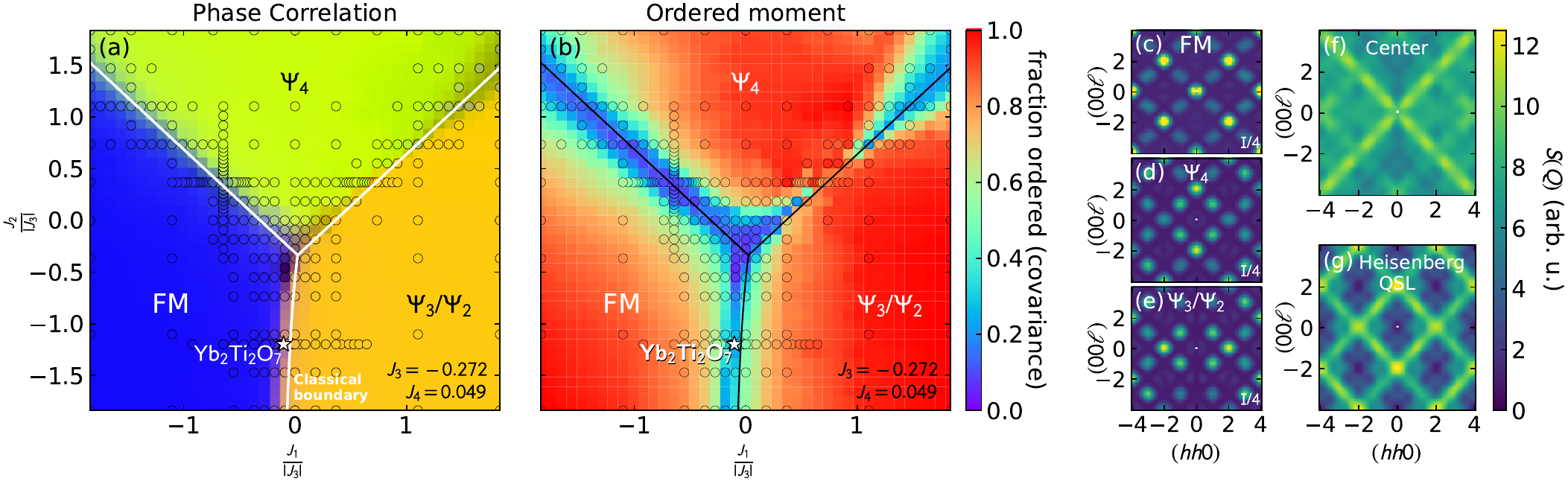}
	\caption{Calculated pyrochlore phase diagram assuming $J_3=-0.272$~meV, $J_4=0.049$~meV (the \yto{} parameters in Ref. \cite{Thompson_2017}). Panel (a) shows the phases determined by Pearson R correlations. Panel (b) shows the size of the ordered moment, calculated as the covariance with the reference $S({\bf Q})$. The black circles in panels (a) and (b) indicate parameters which were explicitly calculated; regions in between were interpolated. Panels (c)-(e) show the reference $S({\bf Q})$  for FM ($J = [-0.5,-0.5,J_3,J_4]$), $\Psi_4$ ($J = [0,0.5,J_3,J_4]$), and $\Psi_3/\Psi_2$ ($J = [0.5,-0.5,J_3,J_4]$). Panel (f) shows the structure factor for the center of the phase diagram, $J = [-0.05,0.0,J_3,J_4]$. Panel (g) shows the structure factor for the Heisenberg quantum spin liquid region $J = [0.5,0.5,J_3,J_4]$.}
	\label{fig:PhaseDiagramJ_YTO}
\end{figure*}

From the ED calculations, we used correlation analysis to build a phase diagram. Choosing a representative structure factor $S({\bf Q})$ in the $(hh\ell)$ scattering plane from deep inside each of the three ordered phases (FM, $\Psi_4$, and $\Psi_3/ \Psi_2$ \cite{Yan2017}), we built a phase diagram by using the Pearson $R$ coefficient to compute the pixel-by-pixel correlation with each of these representative phase  $S({\bf Q})$ (see the Supplemental Information for details \cite{SuppMat} in which is cited Ref. \cite{virtanen2020scipy}). 
We also computed the relative strength of the magnetic order (corresponding to the size of the static ordered moment) via the covariance between the calculated $S({\bf Q})$ and the representative $S({\bf Q})$. The results are shown in Fig. \ref{fig:PhaseDiagramJ_YTO}. 
Because the ED calculations are computationally expensive, we only calculated $S({\bf Q})$ for $\sim 450$ parameter sets, concentrating more calculations around the phase boundaries, interpolating between the computed points. 

From the ED results we also calculated the ground state energies across the phase diagram to define the phase boundaries. We calculated three line scans across the three phase boundaries with densely spaced points (two varying $J_1$, and one varying $J_2$), allowing us to calculate the first and second derivatives with respect to the exchange parameters.  These results are shown in Fig. \ref{fig:LineScans}, and reveal phase boundaries by discontinuities in the first or second derivatives of the energy.

\section{Results}

The results of the inelastic neutron experiment are striking. The spectrum in Fig. \ref{flo:measuredSpectra} smoothly evolves from a very diffuse spectrum at zero field to a sharp, well-defined magnon spectrum at 2~T. 
Such behavior has been reported before \cite{Ross_Hamiltonian,Thompson_2017}, but here we observe it using high-purity samples, where defects and disorder are negligible. The fact that significant broadening exists at 0.15~T (where a ferromagnetic monodomain is stabilized \cite{Saubert_2020}) falsifies the hypothesis that the broadening is from phase coexistence or domain effects. Rather, the broadening is intrinsic to the ferromagnetic ground state. What is more, these results indicate that a [001] magnetic field smoothly tunes \yto{} from 
a regime with 
well-defined magnons 
to one where the magnon picture breaks down.

At 2~T, the experimental data and all the theoretical simulations are in excellent agreement (though perhaps a slightly better description is found in NLSWT than LSWT), showing that magnon interactions are perturbative and that spin wave theory offers an accurate description at high fields. However, as the field decreases, the experimental data increasingly deviate from LSWT predictions. The primary discrepancies are (i) significant broadening of the magnon modes, and (ii) the magnon band between $X$ and $\Gamma$ is much flatter and much lower energy in experiment than in LSWT. 
NLSWT successfully accounts for the flattening of the $X \rightarrow \Gamma$ magnon mode and the 
 magnon peak broadening. 
However, there are still important differences with experiment. Namely, (i) the low-field experimental spectrum has much less spectral weight in the magnon branches and much more weight in the continuum, (ii) broadening at low-fields spans nearly the entire single-magnon spectrum, and (iii) at zero-field the NLSWT predicts sharp low-energy magnon modes whereas none are observed in experiment (although the zero-field calculation is weakly unstable near the $K$ wavevector, see Appendix \ref{app:NLSWT-stability}). 
Clearly, \yto{} low-field magnetism cannot be fully modeled with perturbative corrections.

This conclusion is bolstered by comparison with the non-perturbative THED spin-wave calculations in Fig. \ref{flo:measuredSpectra}. 
{At magnetic fields ($B \gtrsim 1$~T), THED reproduces both the high-energy magnon damping and the flattening of the lowest magnon band, consistent with the success of perturbative NLSWT in this regime.}
As the field decreases, however, the sharp magnon spectral features progressively lose intensity, which is transferred to the two-magnon continuum. This intensity redistribution, characterized by diminished single-magnon signals and enhanced two-magnon spectral weight, aligns very well with experimental observations. 
However, the match is still imperfect. In particular, (i) the intense flat feature at $K$ is missing in the low-field THED simulations [Fig. \ref{flo:measuredSpectra}(w)], and (ii) below 1~T the simulated low-energy magnon bands at all wavevectors are lower energy than experiment [Fig. \ref{flo:measuredSpectra}(w)-(z)]. This discrepancy can be attributed to our calculation's truncated Hilbert space limited to one- and two-magnon states, which means only the single-magnon energies experience a downward renormalization, while the two-magnon energies do not. Consequently, the lowest magnon modes remain outside the two-magnon continuum at all fields and remain sharp, unlike in experiment.

Including three-magnon states in the truncated Hilbert space would correct this discrepancy by renormalizing the two-magnon continuum downward. In particular, the emergence of a nearly flat band at low energies is expected to generate a renormalized two-magnon continuum across a wide energy range, covering the full energy scale of the single-magnon modes and thereby enabling the kinematic conditions for magnon decay observed experimentally across all energies. However, substantial contribution of three-magnon states is highly unconventional and such a calculation is beyond the scope of the present work. 
To go beyond the magnon picture entirely, we turn to ED simulations.

The ED simulated phase diagram in Fig. \ref{fig:PhaseDiagramJ_YTO} makes situation even more interesting. 
Figure \ref{fig:PhaseDiagramJ_YTO}(a) shows the phase correlations with the classical phase boundaries (from Ref. \cite{Yan2017}) superimposed. For the most part, the phases follow the classical bounds, although there are some differences. Interestingly, the Ref. \cite{Thompson_2017} \yto{} Hamiltonian is in the FM phase classically, but appears to be in the $\Psi_3/\Psi_2$ phase quantum mechanically. (This is consistent with previous semi-classical~\cite{Jaubert2015}, series expansion~\cite{Jaubert2015}, exact-diagonalization~\cite{Jaubert2015},  nonlinear spin wave calculations~\cite{rau2019magnon},  and more recent pseudo-fermion functional renormalization group (pf-FRG) calculations \cite{gresista2025quantum} showing that quantum effects move the boundary from the classical values.)

However, a striking departure from the classical phase diagram is the strength of the magnetic order in Fig. \ref{fig:PhaseDiagramJ_YTO}(b). This reveals significant static moment suppression along the boundary between FM and AFM order (both $\Psi_3/\Psi_2$ and $\Psi_4$). 
Classical simulations show a first-order boundary between the FM and $\Psi_3/\Psi_2$ phases \cite{Robert2015,Scheie_2022_Dynamical} rather than a regime of suppressed order.

\begin{figure*}
	\centering
	\includegraphics[width=\textwidth]{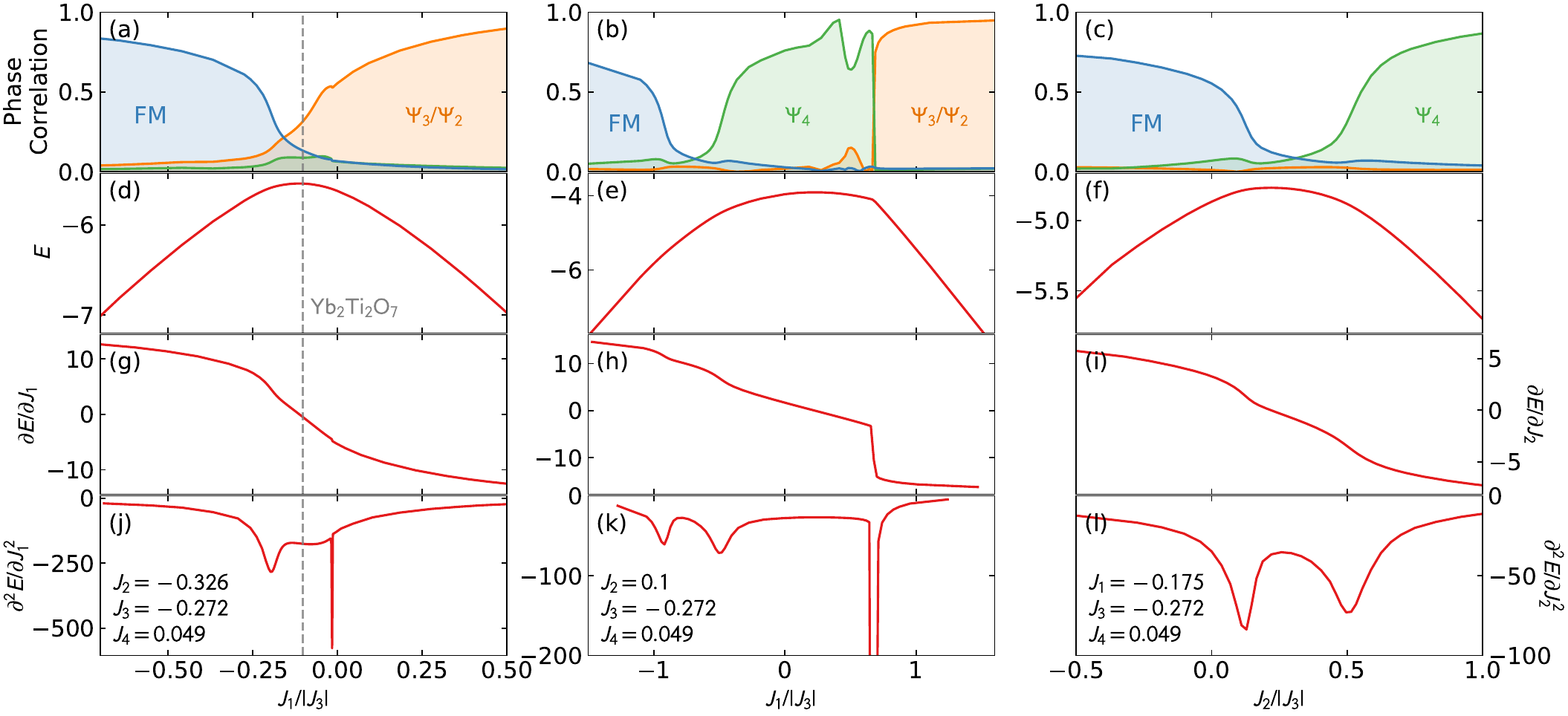}
	\caption{Line scans obtained by ED calculations using 32 sites through the phase boundaries in the $J_3=-0.272$~meV, $J_4=0.049$~meV phase diagram (Fig. \ref{fig:PhaseDiagramJ_YTO}). The top row (a)-(c) shows phase correlation with the three ordered phases. The second row (d)-(f) shows the ground state energy (in arbitrary units) across the line scan. The third row (g)-(i) show the first derivative in energy, while the fourth row (j)-(l) shows the second derivative in energy. The boundary between $\Psi_4$ and $\Psi_3/\Psi_2$ is very clearly first-order with an abrupt transition and a discontinuity in the first derivative. However, the boundary between FM and both AFM phases involves two distinct transitions with an intermediate phase in between. The FM-$\Psi_3/\Psi_2$ boundary appears to involve one second order transition and one weakly first-order transition, whereas the FM-$\Psi_4$ boundary involves two second order transitions. The discontinuities of the phase boundaries are also visible in the phase correlation scans.}
	\label{fig:LineScans}
\end{figure*}

The line scans in Fig. \ref{fig:LineScans} reveal more details about the phase boundaries. 
These calculations show that the phase boundaries between the FM and AFM phases split into \textit{two}, with a double minimum in the second derivative in energy. 
In the case of the FM-$\Psi_3/\Psi_2$ boundary (left column), it appears that one boundary is second-order and one is weakly first-order, with a slight discontinuity visible in the first derivative. For the FM-$\Psi_4$ boundary (right column), both transitions appear to be continuous and second-order. 
(Meanwhile the $\Psi_4$-$\Psi_3/\Psi_2$ transition is sharp, discontinuous, and clearly first-order.) 
Examining the phase correlations (Fig. \ref{fig:LineScans} top row), one can see the discontinuities in the phase correlations themselves, and the intermediate region between the two phase boundaries has strongly suppressed phase correlation. This indicates a finite phase without magnetic order on the boundary between FM and AFM. 

This extensive phase diagram allows us to go beyond merely describing \yto{}. These calculations reveal the onset of the quantum phase near the Heisenberg limit  $J_1=J_2>|J_3|$ [upper right corner of Fig. \ref{fig:PhaseDiagramJ_YTO}(b)]. 
The Heisenberg $S=1/2$ pyrochlore model is known to have a non-magnetically-ordered ground state (though the precise ground state is debated \cite{Canals_1998,Muller_2019,Hagymasi_2021,Astrakhantsev_2021,Hering_2022,Schafer_2023,pohle2023ground}). However the largest region of quantum disorder is the FM-AFM phase boundary, which has a distinct emergent quantum phase.

\section{Discussion}

Our results shed light on the \yto{} spectrum and pose some intriguing puzzles.
At high fields, the \yto{} spin dynamics are well described by a weakly interacting magnon picture, but near zero field even the infinite-order two-magnon simulations fail to describe the data. 
The need for magnon interactions to infinite order and beyond two-magnons makes one question whether it is meaningful to be talking about magnons at all in zero-field \yto{}. The magnon picture appears to completely break down in low-field. 

\begin{figure}
	\centering
	\includegraphics[width=0.49\textwidth]{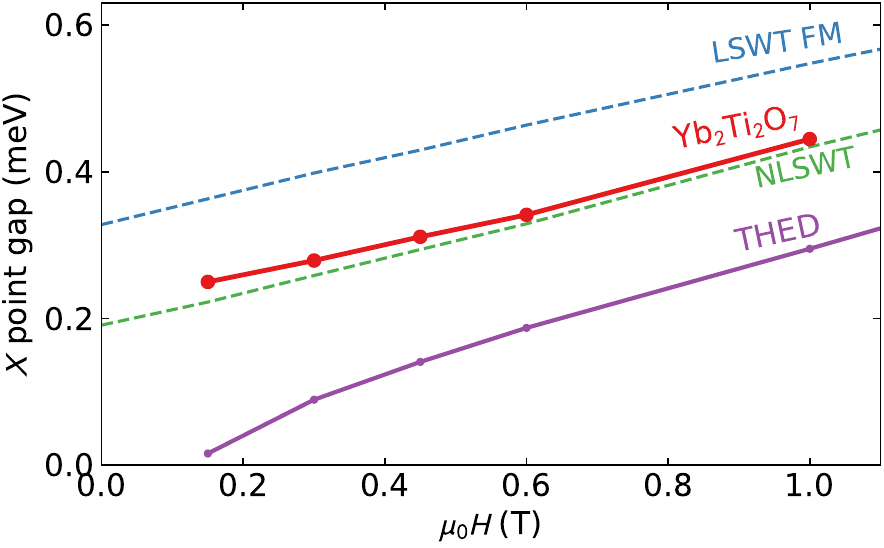}
	\caption{The gap at the $X$ point to the intense flat mode versus [001] magnetic field  for \yto{}, the THED calculation, the NLSWT calculation, and the LSWT model.}
	\label{fig:X-gap}
\end{figure}

Accompanying this magnon breakdown at low fields is the flat $X \rightarrow \Gamma$ magnon mode whose energy progressively decreases as the field decreases to zero. 
Figure \ref{fig:X-gap} shows the field evolution of this quasi-flat mode's energy compared with the theoretical calculations. 
The experimental data for \yto{} shows a linear dependence but with a positive energy intercept after linearly extrapolating to zero field. In comparison, the LSWT predicts a higher energy band, which is pushed down to a lower energy when magnon-magnon interaction is taken into account, as revealed by both NLSWT and THED calculations. The NLSWT has a remarkable agreement with the data, while THED overestimates the downward energy renormalizaiton. This mismatch arises from different treatment of the anomalous self-energy corrections between the two theoretical frameworks. This analysis suggests that \yto{} seems to be immune to the potential instability induced by the softening of the quasi-flat band. 
However, in zero field spectrum (Fig. \ref{flo:measuredSpectra}) the flat band has vanished---either because it has no intensity, has broadened beyond detection, or collapsed into the elastic line. 

In general, low-energy flat bands are harbingers of strong correlations and exotic physics \cite{Rhim01012021,checkelsky2024flat}, and the presence of a quasi-flat band here provides the necessary kinematic condition for the intrinsic broadening of magnon excitations at the full energy scales. In the pyrochlores, the low-energy $X \rightarrow \Gamma $ flat magnon mode indicates proximity to an extensive degeneracy. 
Classically, a ``pinch-line'' spin liquid exists at the meeting point of the FM and two AFM pyrochlore phases \cite{Benton2016,Yan2017}, and a ``spin-nematic'' phase exists on the boundary between the FM and $\Psi_4$ phases \cite{Francini_2025,chung2024mapping} where the magnon energies go to zero along certain reciprocal space directions \cite{gresista2025quantum,Francini_2025}. 
To clarify the situation at zero field, we adopted a fully quantum treatment, the ED study of the ground state.

The ED simulations show that quantum effects create an emergent quantum phase on the boundary with FM order with suppressed magnetic order. (This is reminiscent of the 2D triangular quantum spin liquid, where a sharp boundary between competing classical orders becomes an extended spin liquid phase in the quantum limit \cite{PhysRevB.92.041105,PhysRevB.92.140403,PhysRevB.93.144411,PhysRevB.94.121111,PhysRevB.95.035141,PhysRevB.96.075116,PhysRevLett.123.207203}.) 
A similar intermediate phase on the pyrochlore phase diagram was independently found with pf-FRG calculations (both on the FM-AFM boundary \cite{gresista2025quantum} and elsewhere on the pyrochlore phase diagram \cite{LozanoGomez_2024}), which strongly suggests that our conclusions persist beyond the 32-site system we calculated.  
We also see evidence of similarities for most parts of the phase diagram with smaller system sizes as well, see Appendix \ref{app:FiniteSize}.

The obvious next question is: what is the nature of the emergent quantum phase? The analysis here is insufficient to distinguish a quantum spin liquid from other forms of quantum non-magnetic-order (like valence bond crystals \cite{Fouet_2003}). However, we note that the intermediate phase displays the ``rods'' of scattering along $\{111\}$ reciprocal space directions, shown in Fig. \ref{fig:PhaseDiagramYTO} (and in further detail in the Supplemental Information \cite{SuppMat}). 

\begin{figure*}
	\centering
	\includegraphics[width=\textwidth]{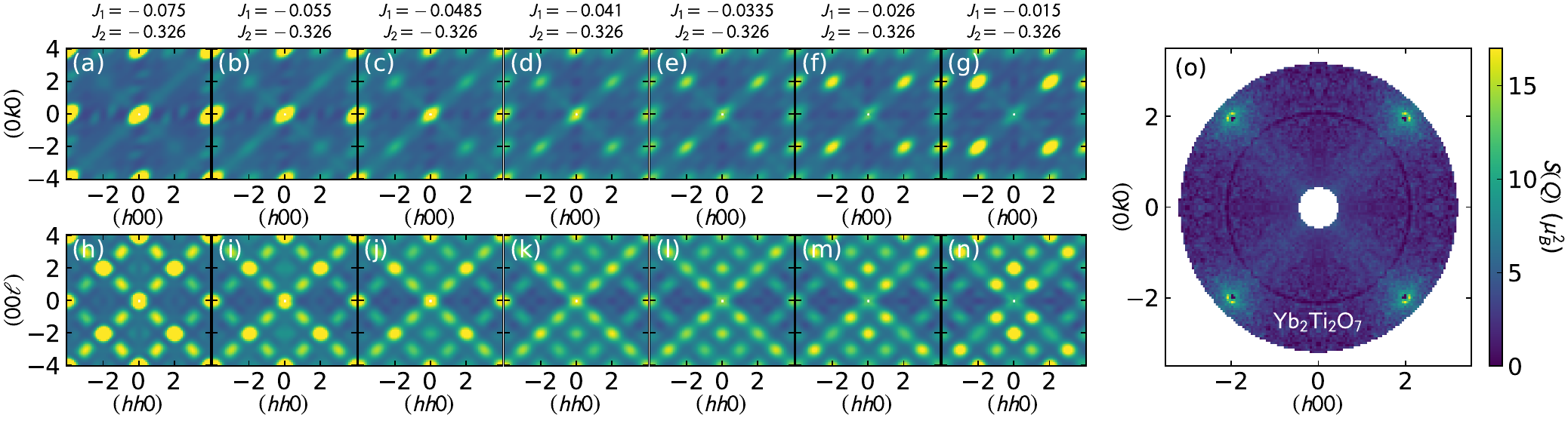}
	\caption{Evolution of $S(Q)$ tuning across the FM - $\Psi_3/\Psi_2$ phase boundary with $J_2=-0.326$,  $J_3=-0.272$, $J_4=0.049$ (from the fitted Yb$_2$Ti$_2$O$_7$ exchange Hamiltonian \cite{Thompson_2017}). Panels (a)-(g) show the $(hk0)$ scattering plane, while panels (h)-(n) show the $(hh\ell)$ scattering plane. (Note the asymmetric elongation in the $(hk0)$ structure factor is due to the ED supercell geometry, see Supplemental Information \cite{SuppMat}.) Panel (o) shows the experimental energy-integrated zero-field Yb$_2$Ti$_2$O$_7$ spectrum at 0.1~K, which resembles the scattering pattern intermediate to the two phases.}
	\label{fig:PhaseDiagramYTO}
\end{figure*}

Diffuse $\{111\}$ scattering rods, similar to what is calculated here, have been measured in previous neutron experiments on 
 \yto{} \cite{Bonville2004,Ross2009,Thompson_2011,Chang_2012_Higgs,Bowman2019,Scheie_2022_Dynamical} which indicate reduced dimensional correlations to 2D Kagome planes within the pyrochlore lattice. 
 This suggests a long-range collective effect rather than short-range singlets. 
(Note that these intermediate quantum phases are far from the quantum spin ice region, which is at $J_1 = -J_2 = J_3 = J_4 < 0$ \cite{Ross_Hamiltonian}.) Furthermore, the intermediate quantum phase is continuously connected to the center degenerate point, which is a classical spin liquid \cite{Benton2016} and exhibits symmetry-breaking ``nematic'' disorder in classical spin simulations \cite{Francini_2025}. 

Intriguingly, the most comprehensive spin Hamiltonian fit \cite{Thompson_2017} ($J_1=-0.028$~meV, $J_2=-0.326$~meV) puts \yto{} within the emergent quantum phase (Fig. \ref{fig:LineScans}). Of course experimental uncertainty and finite size calculations may change the precise location of the phase boundaries and the \yto{} Hamiltonian. 
Indeed, experiments unambiguously demonstrate that our samples possess a ferromagnetic ground state~\cite{SeyedPaper,Scheie2020}. Taken together, our study suggests two possible scenarios. First, the emergent quantum phase retains vestiges of ferromagnetism. Notably, the experimental 0.1~K energy-integrated $(hk0)$ scattering in Fig.~\ref{fig:PhaseDiagramYTO} shows a striking agreement with the ED calculated $S({\bf Q})$ within this intermediate phase. Alternatively, \yto{} may lie firmly within the FM phase but in close proximity to the emergent phase revealed by ED. Such proximity could be sufficient to induce emergent phenomena at finite energy scales, even within an ordered ferromagnetic ground state~\cite{Chern2019}.

The picture emerging from this study begins to explain many things about \yto{}, and suggests a new view of pyrochlore physics. \yto{} magnon broadening is intrinsic to its Hamiltonian rather than an extrinsic effect of domains. It appears to be associated with very low energy flat bands in the spectrum. This in turn creates a phase diagram remarkably similar to quantum critical electronic phases, in five respects. 
First, the competition between the FM and AFM phases suppresses the order parameters, producing an extended zero temperature phase (analogous to unconventional superconductivity \cite{sachdev2025foot}). Second, extensive degeneracy in the ground state is present with flat bands near zero energy, {closely paralleling the physics of heavy-fermion materials~\cite{aynajian2012visualizing,zhou2013visualizing}}. Third, scale-free fluctuations are observed at finite temperatures \cite{Scheie_2022_Dynamical}. Fourth, dimensional reduction (witnessed by rods of scattering) occurs near the phase boundary (c.f. CeCu$_6$ \cite{Stockert_1998}). Fifth, the magnon quasiparticle picture seems to break down \cite{Thompson_2017,Scheie2020}. 
These five features show striking resemblance to correlated electron quantum criticality. 
Importantly for experimental studies, \yto{} can be continuously tuned with a modest magnetic field from semiclassical physics to its nontrivial quantum ground state.

The emergent quantum phase here is unusual, because quantum fluctuations typically lift degeneracies by order-by-disorder \cite{Green_2018_Review}. Here we observe quantum fluctuations doing the opposite: stabilizing a degeneracy at a first-order phase boundary, and creating quantum critical behavior \cite{Raines_2024}.

Our experiments focused on only one model system, \yto{}. However, this dipolar phase diagram covers many other pyrochlore materials as well \cite{hallas2018experimental,Sarkis_2020,wulferding2023collective,xu2025ramification}. 
Furthermore, similar systematic calculations may also be useful for multipolar magnets like the cerium-based pyrochlores~\cite{Gaudet_CZO_2019, Gao_CZO_2019, Smith_CZO_2022, Bhardwaj_CZO_2022, Hosoi_PRL_2022, Poree_2025, Poree_CHO_2023, Bhardwaj_CHO_2025}, which also have a rich phase diagram~\cite{Hosoi_PRL_2022}.  
This offers an opportunity to study exotic phases of many-body systems with a precisely-known underlying Hamiltonian---a luxury not always available for other quantum critical systems like unconventional superconductors.

\section{Conclusions}

We have experimentally and theoretically explored the quantum limit of the anisotropic $S=1/2$ pyrochlore model, showing an emergent quantum disordered phase on the boundary between the FM phase and the two AFM phases. In this phase the magnetic order is strongly suppressed and scattering rods emerge which indicate reduced dimensional correlations. The scattering pattern computed from theory (ED) in this phase strongly resembles the experimental Yb$_2$Ti$_2$O$_7$ scattering pattern, which according to the spin wave model may lie within this the emergent quantum phase. 

The results of this study strongly suggest that the phenomenology of quantum criticality applies to Yb$_2$Ti$_2$O$_7$, as an insulating system with scale-free fluctuations and quasiparticle breakdown. This offers a remarkable opportunity to continuously tune from well-defined magnons to magnon breakdown in a well-understood and clean system. Our study stimulates the further exploration of the nature of quantum criticality along the FM/AFM phase boundary. 
The theoretical methods employed in this work each have unique limitations which restricts the extent to which we can reproduce the experimental low-field  Yb$_2$Ti$_2$O$_7$ spectrum. 
However, we believe this system could be a benchmark for many-body quantum simulation methods (including quantum computers) \cite{bartschi2024potential}. 

The most important result of this study is placing Yb$_2$Ti$_2$O$_7$ in context: in fact there are many regions where such exotic physics can be expected from pyrochlores, along the entire phase boundary between FM and AFM order. 
This puts forward the pyrochlore lattice generically (and \yto{} specifically) as a case where emergent physics--- potentially insulating quantum criticality---can be studied in a system with a well-understood underlying Hamiltonian.

\subsection*{Acknowledgments}
A portion of this research used resources at the Spallation Neutron Source, a DOE Office of Science User Facility operated by the Oak Ridge National Laboratory. The beam time was allocated to CNCS on proposal number IPTS-24263.1. 
AS acknowledges support from the U.S. Department of Energy, Office of Basic Energy Sciences, Division of Materials Science and Engineering under project ``Quantum Fluctuations in Narrow-Band Systems.'' 
Sample synthesis was supported through the Institute for Quantum Matter at Johns Hopkins University, by the U.S. Department of Energy, Division of Basic Energy Sciences, Grant DE-FG02-08ER46544. 
H.J.C. acknowledges funding from National Science Foundation Grant No. DMR 2046570 and the National High Magnetic Field Laboratory (NHMFL). NHMFL is supported by the National Science Foundation through NSF/DMR-1644779 and DMR-2128556 and the state of Florida. 
J.G.R. acknowledges funding from the Natural Sciences and Engineering Research Council of Canada (NSERC). 
A.B. and H.J.C. thank the Research Computing Center (RCC) and the Planck cluster at Florida State University for computational resources. This work also used Bridges-2 at Pittsburgh Supercomputing Center through allocation PHY240324 (Towards predictive modeling of strongly correlated quantum matter) from the Advanced Cyberinfrastructure Coordination Ecosystem: Services and Support (ACCESS) program, which is supported by U.S. National Science Foundation grants $\#$2138259, $\#$2138286, $\#$2138307, $\#$2137603, and $\#$2138296. 
We acknowledge helpful discussions with Nic Shannon and Cristian Batista.


\appendix

\section{Neutron experiment details \label{app:experiment}}

\begin{figure}
	\centering\includegraphics[width=0.22\textwidth]{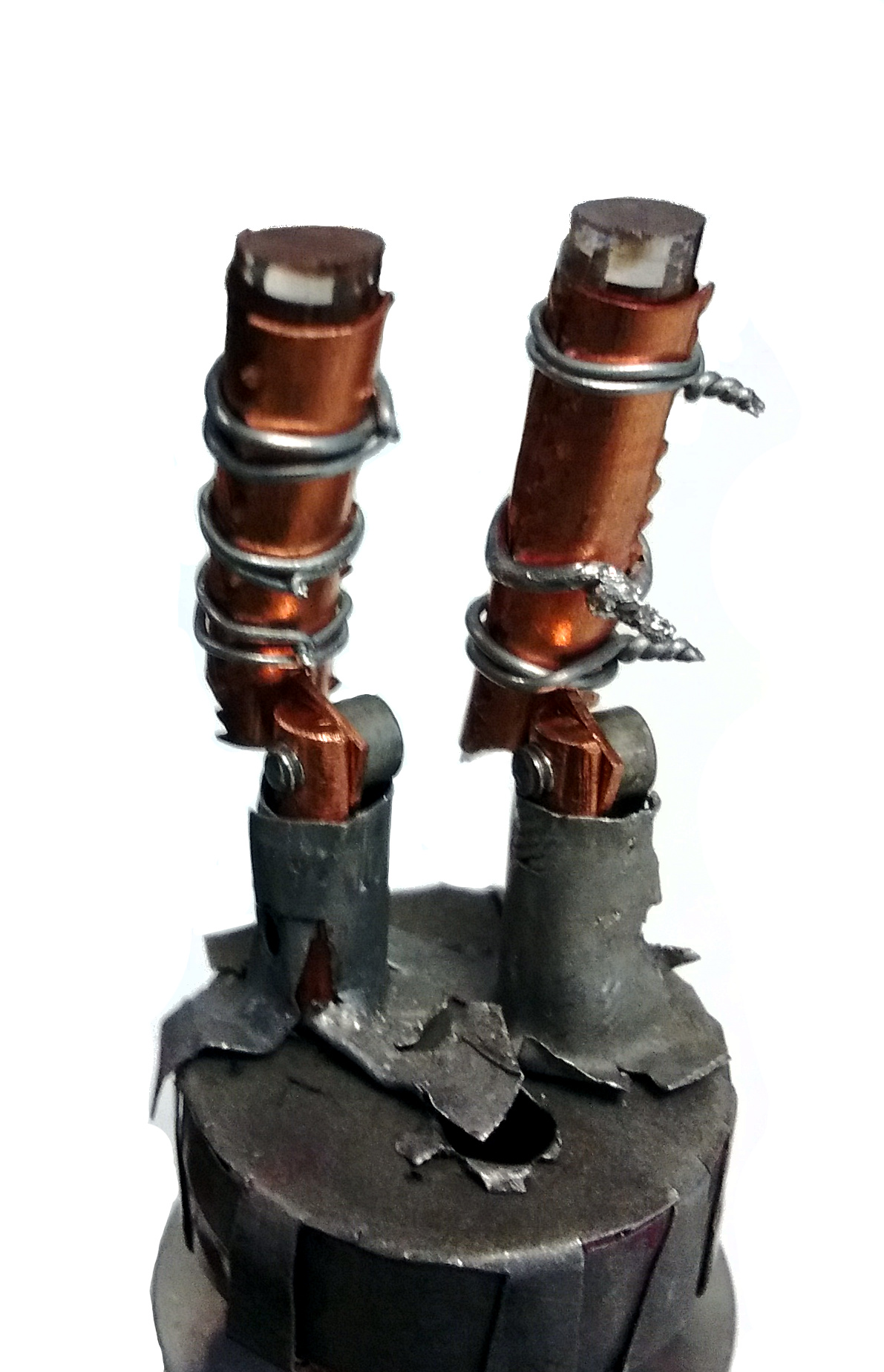}
	
	\caption{\yto{} sample used in the CNCS experiment: two coaligned crystals wrapped in copper foil and tied to a copper sample holder.}
	
	\label{flo:crystal}
\end{figure}

The sample used for this experiment, shown in Fig. \ref{flo:crystal}, was the same crystals as was used in ref. \cite{Scheie2020} reoriented so that $[001]$ is vertical. Total sample mass was 4.8~g. 
Data were measured rotating over $180^{\circ}$ in $0.5^{\circ}$ steps. 
We collected data using $E_i=3.32$~meV neutrons (disc chopper frequency 300 Hz, for an elastic line energy resolution $\Delta \hbar \omega = 0.11$~meV, taken from the FWHM of the experimental data) at several magnetic fields between 0~T and 2~T. 
We also briefly collected data with $E_i = 2.49$~meV neutrons (disc chopper frequency 240 Hz, for an elastic line energy resolution $\Delta \hbar \omega = 0.05$~meV from the FWHM of experimental data) at angles around $(220)$ in order to measure the gap at $(220)$. 

Data were background-subtracted by a measured 12~K \yto{} background. Prior to subtraction, inelastic 12~K data (above 0.15~meV) were subtracted by the median 12~K intensity in the scattering plane, which empirically works well in removing artifacts whilst preserving the magnetic scattering features. 
In the high-symmetry cuts plotted in Fig. \ref{flo:measuredSpectra}, the out-of-plane binning was $\pm 0.1$ reciprocal lattice units (RLU) along $[001]$ and  $\pm 0.05$~RLU along the transverse in-plane direction. 

The higher resolution data around the (220) wavevector with  $E_i = 2.49$~meV are shown in Fig.\ref{fig:LowEcuts} (for full colormaps, see the Supplemental Information \cite{SuppMat}). These data have a noisier signal than $E_i=3.32$~meV due to the lower flux, but the resonance at 0.11~meV is visible and outside error bars at the lowest temperatures.

\begin{figure}
	\centering\includegraphics[width=0.42\textwidth]{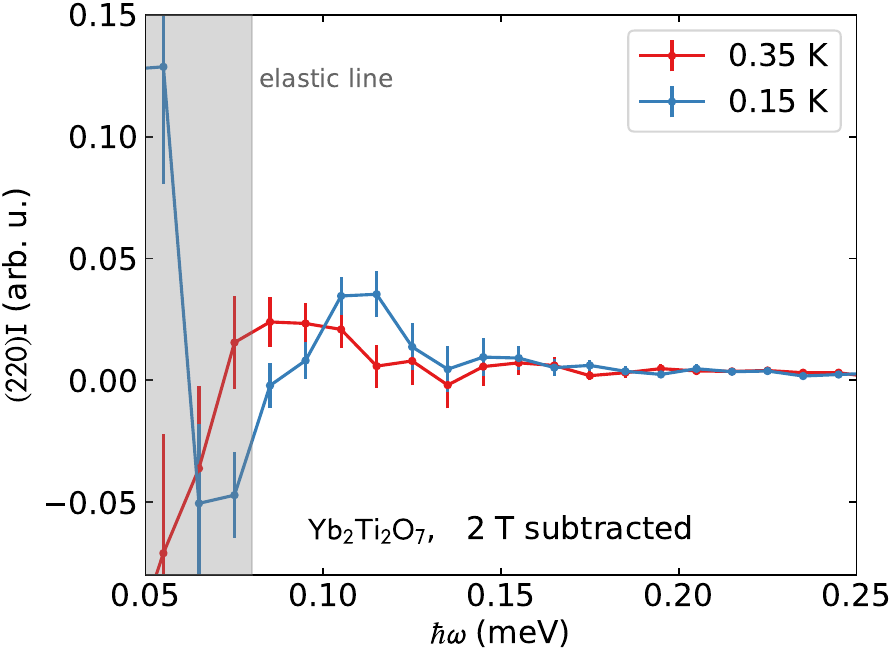}
	
	\caption{\yto{} scattering measured with $E_i = 2.49$~meV neutrons at $(220)$, with a 2~T background subtracted. Here we show the highest measured and the lowest measured temperatures in this configuration, which clearly reveals the presence of a 0.11 meV resonance at 0.15~K. The grey region indicates an oversubtraction from the 2T background data, which has nonzero Bragg intensity at $(220)$.}
	\label{fig:LowEcuts}
\end{figure}

\section{Absolute unit conversion and entanglement}

\begin{figure}
	\centering\includegraphics[width=0.44\textwidth]{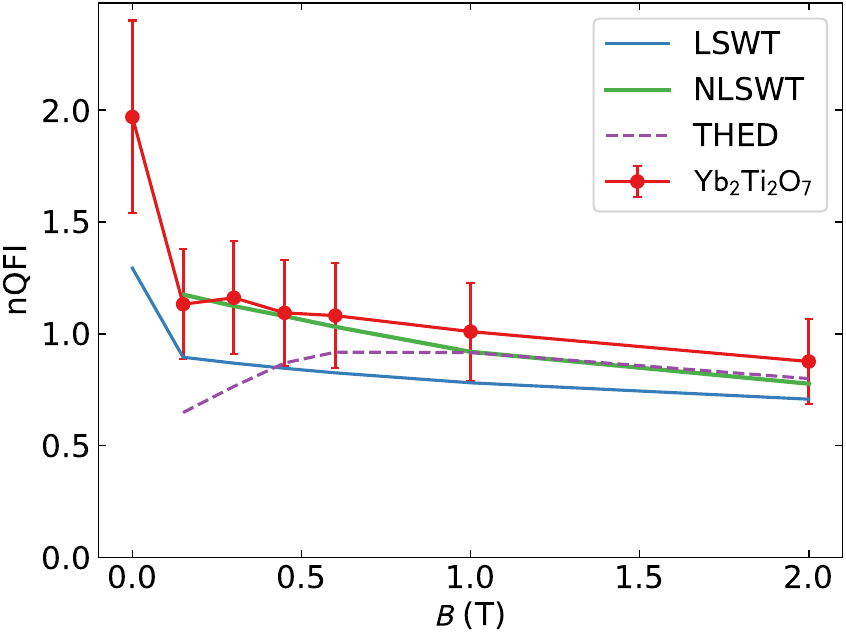}
	
	\caption{Quantum Fisher Information (QFI) of \yto{} at 0.1~K as a function of magnetic [001] field. The reported values are the maximal QFI along the high-symmetry cuts. At $B=0$ QFI is maximal at $K$, but at nonzero fields QFI is maximal at the midpoint between $K$ and $\Gamma$.}
	
	\label{flo:QFIvB}
\end{figure}

We are able to calculate the Quantum Fisher Information (QFI) of \yto{} by normalizing the inelastic scattering to be in absolute units using the intensity of the 2~T spin wave modes. This is done by integrating over the inelastic scattering via 
\begin{align}
	\begin{split}
		{\rm nQFI}[{\bf Q},T] =  \frac{1}{2 S^2} & 
		\int_{0}^\infty   \mathrm{d}(\hbar \omega) \bigg[ \tanh \left( \frac{\hbar \omega}{2 k_BT }\right) \\ 
		&\left( 1-e^{-\hbar\omega/k_B T} \right) \tilde{S}({\bf Q},\omega) 
		\bigg] \label{eq:bignQFI}
	\end{split}
\end{align}
where $S$ is the quantum spin number (here we normalize the Yb$^{3+}$ effective crystal field doublet to $S=1/2$), $T$ is temperature, and $\tilde{S}({\bf Q},\omega)$ is the unpolarized neutron structure factor \cite{laurell2024witnessing,scheie2024tutorial}. 
nQFI provides a lower bound to entanglement depth such that $\mathrm{nQFI} > m$ indicates a system with at least $m+1$ partite entanglement  \cite{Hauke2016,PhysRevB.103.224434}. 
The calculated nQFI versus field is shown in Fig. \ref{flo:QFIvB}. 

\begin{figure}
	\centering
	\includegraphics[width=0.45\textwidth]{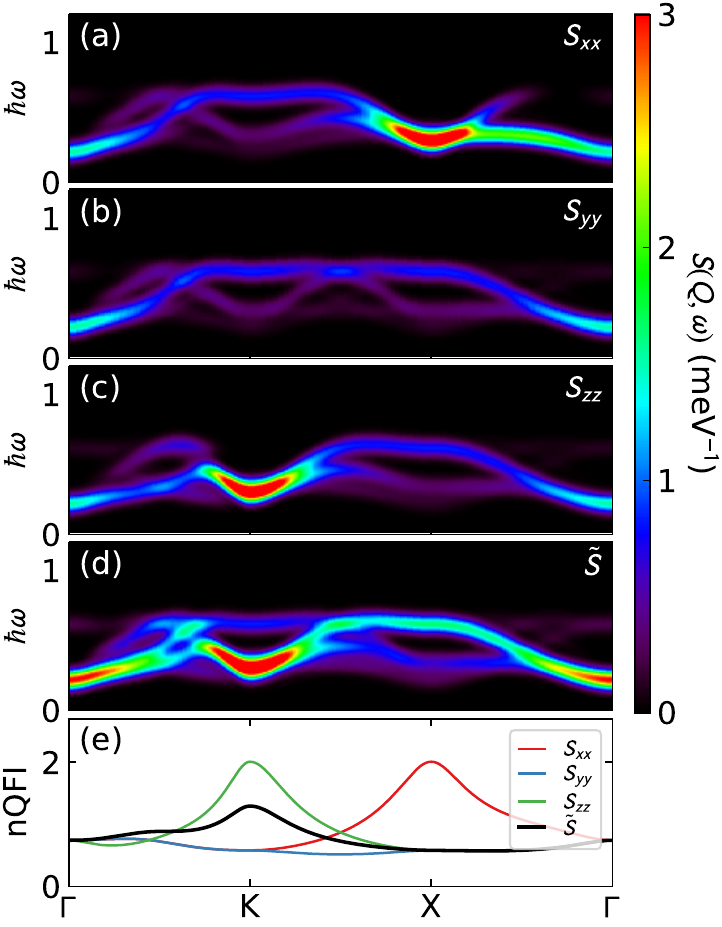}
	\caption{LSWT calculated $B=0$ scattering along different polarization channels (a)-(c) and the  $\tilde{S}$ with polarization factor applied 
		(d), and the resulting nQFI calculated from these spectra. The $S_{zz}$ channel nQFI is 1.55 times the $\tilde{S}$ nQFI (calculated with the isotropic approximation).}
	\label{fig:polarization}
\end{figure}

At $B=0$ the largest intensity (and correspondingly the largest QFI) is concentrated around $K$, giving $\mathrm{nQFI} = 1.9 \pm 0.4$. This witnesses at least bipartite entanglement per spin, consistent with a nontrivial quantum state. 
This value is almost certainly an underestimate, for three reasons. First, due to large uncertainties near the elastic line, we excluded all scattering below 0.07~meV, which would otherwise participate in the integral in Eq. \eqref{eq:bignQFI}. Second resolution effects are probably at play, which tends to broaden signals and suppress QFI values \cite{scheie2024tutorial}. Third, in Eq. \eqref{eq:bignQFI} we have assumed the isotropic approximation $S_{xx}({\bf Q},\omega) = S_{yy}({\bf Q},\omega) = S_{zz}({\bf Q},\omega) = \frac{1}{2} \tilde{S}({\bf Q},\omega)$, however the linear spin wave model shows that this approximation suppresses QFI by 36\% (see Fig. \ref{fig:polarization}). 

The inelastic scattering data were normalized to absolute units by the integrated intensity of the 2~T magnons at the $K$ point to LSWT calculations (similar to how phonon intensities are used in Ref. \cite{Xu_AbsUnits}).
Rather than normalizing to $\rm \mu_B/meV$ (for a ${\bf M}= \bf{g} \cdot {\bf S}$ model where $\bf{g}$ is the $g$-tensor), we  normalized to $\rm meV^{-1}$ for an effective $S=1/2$ model such that the sum rule $\int_{BZ}dq d\omega S(q,\omega) = S(S+1)$ is satisfied. This is necessary in order to apply the quantum entanglement measures that have been defined for $S=1/2$ \cite{scheie2024tutorial}.

Aside from an overall scale factor, the differences between the LSWT simulations are minor: mainly that the weaker modes have more intensity with the $g$-tensor included (see Supplemental Information \cite{SuppMat} for more plots). Thus we confidently use the 2~T simulations to normalize the experimental data to the $S=1/2$ model. (This normalization comes with a statistical uncertainty of 22\%, which is a typical value for absolute unit conversion \cite{Xu_AbsUnits}). 

Normalized QFI  (nQFI) of the experimental and simulated scattering is calculated via Eq. (6) of Ref. \cite{scheie2024tutorial}. 
Because the polarized scattering $S_{\alpha \alpha}$ was not measured, we have assumed the isotropic approximation
\begin{equation}
	S_{\alpha \alpha} = \frac{1}{2} \tilde{S} 
	\label{eq:isoapprox}
\end{equation}
to get Eq. \eqref{eq:bignQFI}.

As noted above, we have assumed the isotropic $S$ approximation for the QFI calculation as the most conservative way to estimate QFI from unpolarized scattering data. 
Although we cannot separate the experimental polarization components, we can use LSWT to estimate the influence of the different polarization channels. 
Figure \ref{fig:polarization} shows the effect in LSWT of the different polarization channels. In this model, the $S_{zz}$ channel would have given an nQFI over 50\% higher than the unpolarized $\tilde{S}$  
data. Taking this as an estimate for experimental polarization correction (like Ref. \cite{Laurell2021Quantifying}) gives an estimated experimental $B=0$ $\rm QFI \geq 2.9(6)$, witnessing $\geq 3$-partite entanglement in \yto{}. We suspect that in the real \yto{} polarization factor correction would be even greater, as the gap in  \yto{} is much smaller than in LSWT and intensities tend to diverge as gaps close. 

\section{Entropy from specific heat}

\begin{figure}
    \centering
    \includegraphics[width=0.95\linewidth]{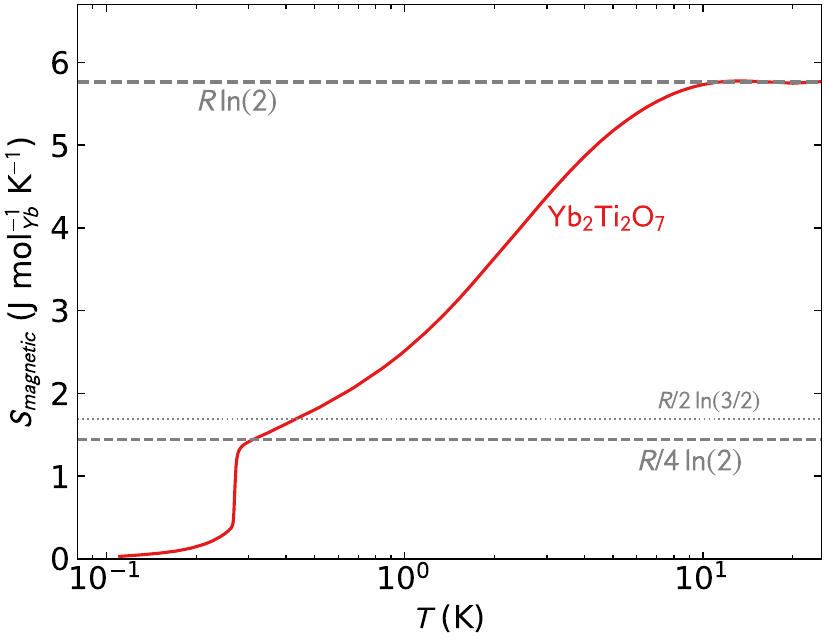}
    \caption{\yto{} entropy from zero-field heat capacity. From the paramagnetic phase to $T_c=270$~mK, the specific heat recovers $3/4 R\ln(2)$ entropy, in accord with ordered 2D kagome planes with intervening disordered spins.}
    \label{fig:entropy}
\end{figure}

Figure \ref{fig:entropy} shows the integrated entropy from zero-field specific heat reported in Ref. \cite{SeyedPaper}. The entropy of the short-ranged correlated phase just above $T_c=270$~mK goes below the spin ice Pauling entropy of $R/2 \ln(3/2)$ \cite{ramirez1999zero} and reaches $R/4 \ln(2)$. There is not a clear saturation of entropy, but this value is consistent with decoupled 2D kagome planes on the pyrochlore lattice wherein 1/4 of the spins are disordered and fluctuating. 

\section{Effects of Finite Size}\label{app:FiniteSize}
In order to assess the effects of finite system size, we have evaluated the ground-state energy with ED, by using two system sizes, $N=16 \times 1^3 = 16 $ sites (cubic unit cell with a 16 site basis) and $N = 4 \times 2^3 = 32$ (FCC unit cell with a 4 site basis). These geometries have been employed in previous work on pyrochlore magnets~\cite{Changlani2017quantum}, and have been further discussed in the Supplemental Information \cite{SuppMat}.
We used the energies 
to numerically compute their 
derivatives and second derivatives with respect to varying a parameter in the Hamiltonian. 

\begin{figure*}[]
	\centering
	\includegraphics[width=\linewidth]{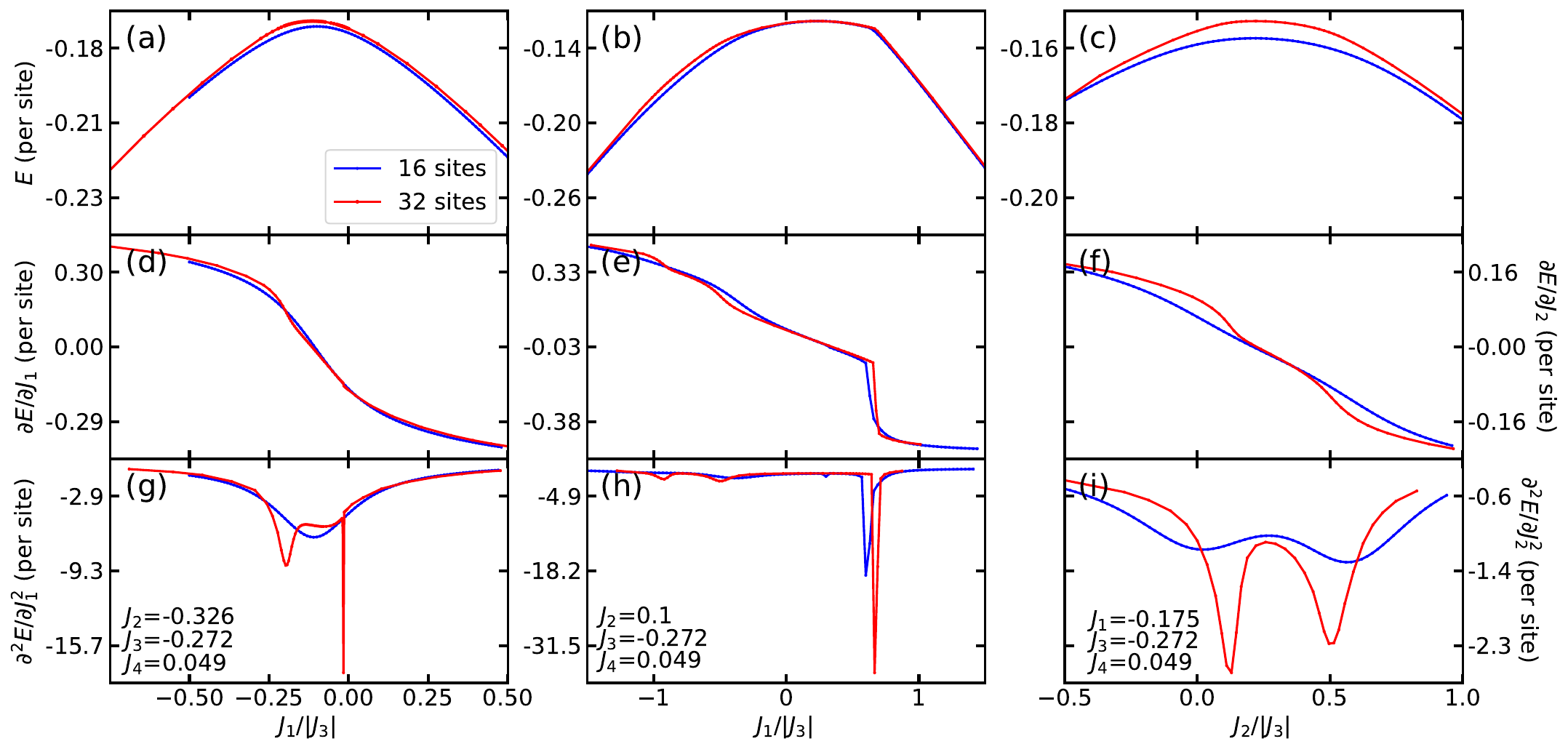}
	\caption{Comparison of energy scans obtained by using 16 sites (blue) and 32 sites (red) system. The first row compares the ground state energy (per particle) across the line scans. The second and third rows compare the derivative and the second derivative in Energy respectively.}
	\label{fig:16vs32}
\end{figure*}

Fig.~\ref{fig:16vs32}(a)-(i) shows our ED results for both sizes for three different line scans (see also Fig.~\ref{fig:LineScans}). We find that the energy per site 
agree reasonably well with each other for all the three line scans. We also observed than the $N=16$ system (like the $N=32$ site cluster) captures the first order phase transition at the $\Psi_4$ and $\Psi_3/\Psi_2$ boundary. However, the smaller sized system does not show any signature of any phase transition near the FM and $\Psi_3/\Psi_2$ boundary when moving along the line with $J_2$ fixed at -0.326, see Fig.~\ref{fig:16vs32}(g). As mentioned in the main text, the $N=32$ site cluster shows a transition in the form of a second order transition near this boundary. 

\section{$S({\bf Q})$ across phase boundaries}

\begin{figure*}
	\centering
	\includegraphics[width=0.9\textwidth]{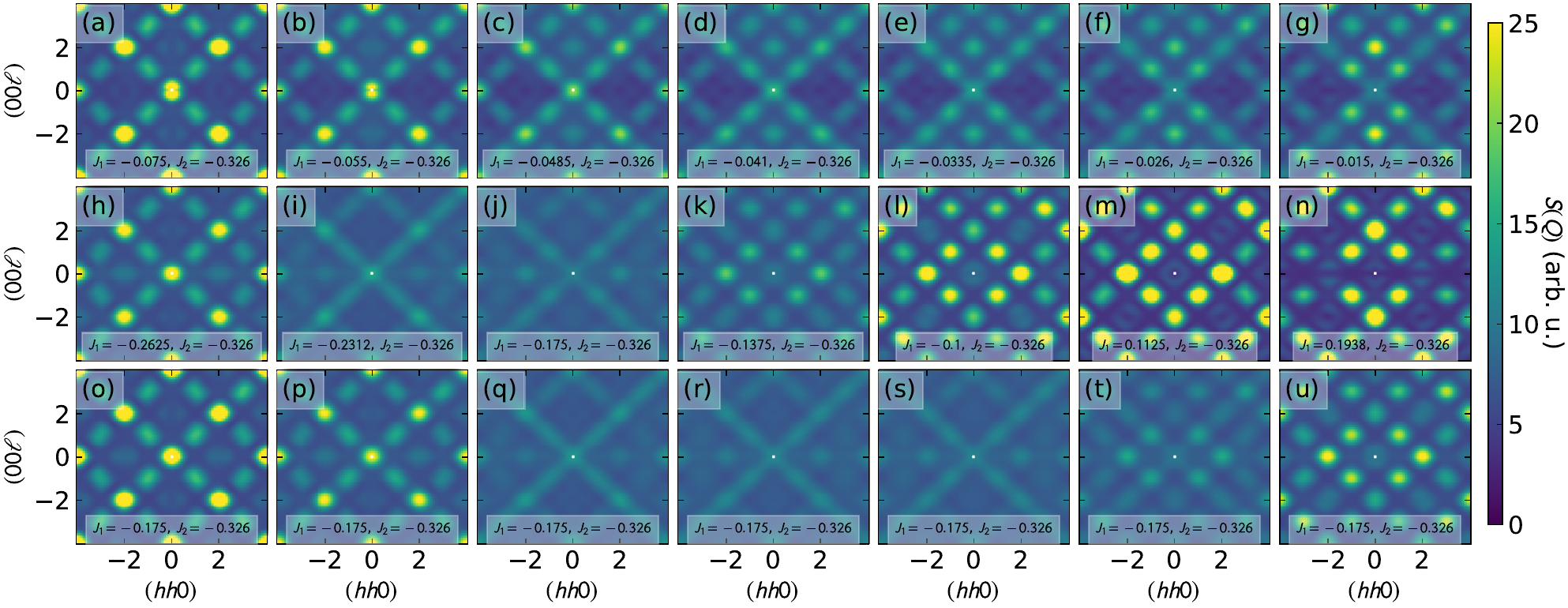}
	\caption{Calculated $S({\bf Q})$ along the lines in main text Fig. 4, tuning from FM $\rightarrow$ $\Psi_2/\Psi_3$ (a)-(g), FM $\rightarrow$ $\Psi_4$ $\rightarrow$ $\Psi_2/\Psi_3$ (h)-(n), FM $\rightarrow$ $\Psi_4$ (o)-(u). Note the existence of rod-like scattering in the crossover between FM and AFM scattering.}
	\label{fig:Sqw-along-lines}
\end{figure*}

\begin{figure}
	\centering
	\includegraphics[width=\linewidth]{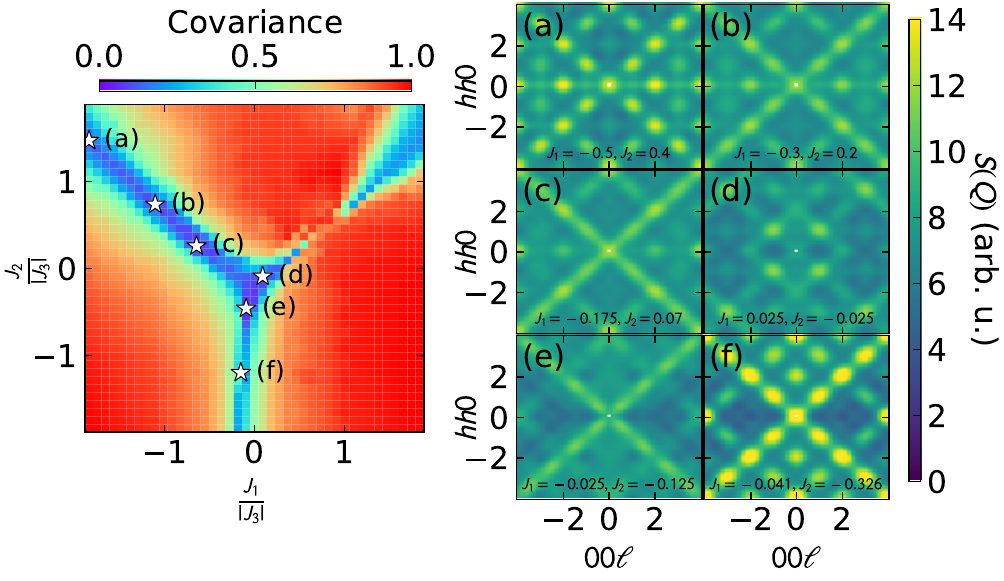}
	\caption{Calculated $S({\bf Q})$ along the FM-AFM boundary, showing the points on the covariance phase diagram (left) and the computed $S({\bf Q})$ (right). Note the rod-like scattering is most prevalent nearest the center degenerate point. At the degenerate point (d) the scattering takes on a different pattern, more akin to a superposition of the rod-like scattering and the Heisenberg QSL scattering.}
	\label{fig:Sqw-along-boundary}
\end{figure}

In this section we show the computed $S({\bf Q})$ along and across the phase boundaries in the phase diagram. 
Figure \ref{fig:Sqw-along-lines} shows the structure factor along the line scans explored in the main text, showing the onset of rod-like scattering at the phase boundary. 
Figure \ref{fig:Sqw-along-boundary} shows the evolution of the structure factor along the entire FM-AFM phase boundary, showing the universality of the rod-like features---although they are the clearest and most rod-like nearest the degenerate point at the middle.
That said, the scattering right at the degenerate point is less rod-like, and is more of a superposition of the Heisenberg QSL scattering and the FM-AFM rods. 

\section{NLSWT stability}\label{app:NLSWT-stability}

\begin{figure*}
	\centering
	\includegraphics[width=0.88\textwidth]{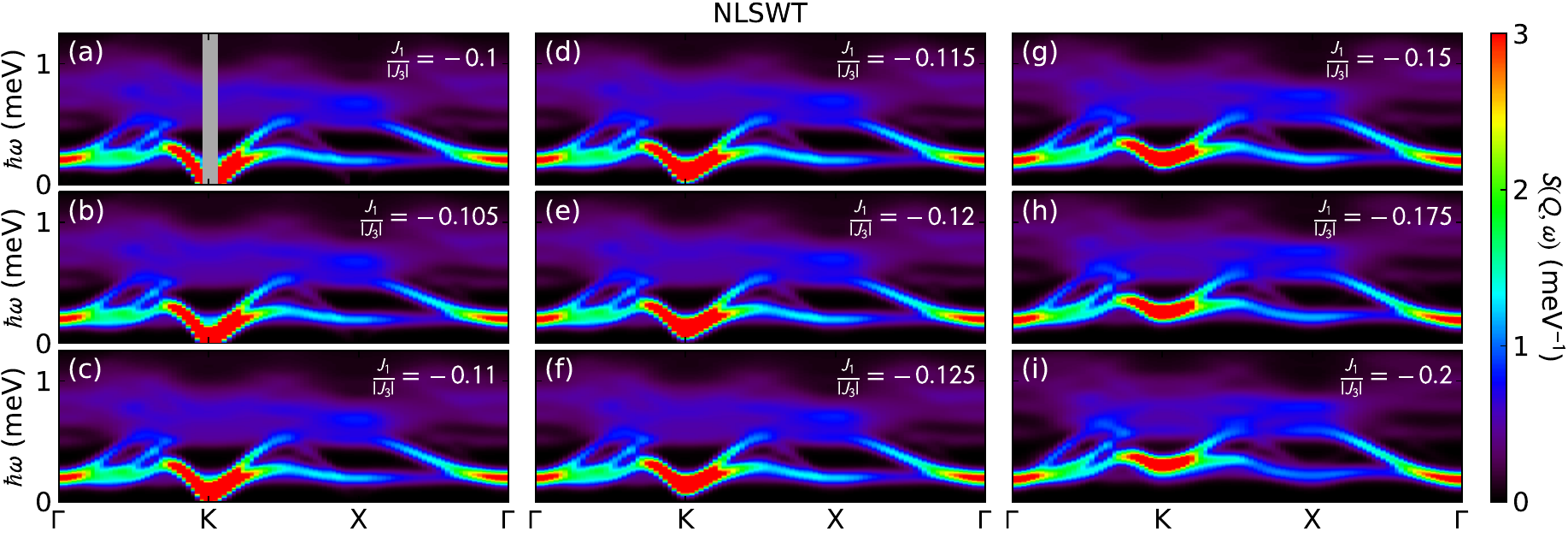}
	\caption{Zero-field nonlinear spin wave theory (NLSWT) calculations of the \yto{} Hamiltonian from Ref. \cite{Thompson_2017}, where $J_1$ is tuned systematically from the fitted $J_1/|J_3| = -0.1$~meV in panel (a) further into the FM phase in panels (b)-(i). A very slight modification yields numerical stability, and gaps the excitations at $K$.}
	\label{fig:TuneJ1}
\end{figure*}

As noted in the main text the NLSWT calculations are unstable around the $K$ wavevector at $B=0$, at least for the parameters in Ref. \cite{Thompson_2017}. 
In Fig. \ref{fig:TuneJ1}, we modify the Hamiltonian by tuning $J_1$ further into the FM phase. We find that a tiny modification allows numerical stability, and also gaps the excitations at $K$. 
This gap does not reproduce the intense flat mode shown in Fig. \ref{fig:LowEcuts} or in Ref. \cite{Antonio2017}, so there are still clear deficiencies in the NLSWT calculation even with parameters tuned so the FM phase is stable.

\begin{thebibliography}{108}%
\makeatletter
\providecommand \@ifxundefined [1]{%
 \@ifx{#1\undefined}
}%
\providecommand \@ifnum [1]{%
 \ifnum #1\expandafter \@firstoftwo
 \else \expandafter \@secondoftwo
 \fi
}%
\providecommand \@ifx [1]{%
 \ifx #1\expandafter \@firstoftwo
 \else \expandafter \@secondoftwo
 \fi
}%
\providecommand \natexlab [1]{#1}%
\providecommand \enquote  [1]{``#1''}%
\providecommand \bibnamefont  [1]{#1}%
\providecommand \bibfnamefont [1]{#1}%
\providecommand \citenamefont [1]{#1}%
\providecommand \href@noop [0]{\@secondoftwo}%
\providecommand \href [0]{\begingroup \@sanitize@url \@href}%
\providecommand \@href[1]{\@@startlink{#1}\@@href}%
\providecommand \@@href[1]{\endgroup#1\@@endlink}%
\providecommand \@sanitize@url [0]{\catcode `\\12\catcode `\$12\catcode `\&12\catcode `\#12\catcode `\^12\catcode `\_12\catcode `\%12\relax}%
\providecommand \@@startlink[1]{}%
\providecommand \@@endlink[0]{}%
\providecommand \url  [0]{\begingroup\@sanitize@url \@url }%
\providecommand \@url [1]{\endgroup\@href {#1}{\urlprefix }}%
\providecommand \urlprefix  [0]{URL }%
\providecommand \Eprint [0]{\href }%
\providecommand \doibase [0]{https://doi.org/}%
\providecommand \selectlanguage [0]{\@gobble}%
\providecommand \bibinfo  [0]{\@secondoftwo}%
\providecommand \bibfield  [0]{\@secondoftwo}%
\providecommand \translation [1]{[#1]}%
\providecommand \BibitemOpen [0]{}%
\providecommand \bibitemStop [0]{}%
\providecommand \bibitemNoStop [0]{.\EOS\space}%
\providecommand \EOS [0]{\spacefactor3000\relax}%
\providecommand \BibitemShut  [1]{\csname bibitem#1\endcsname}%
\let\auto@bib@innerbib\@empty
\bibitem [{\citenamefont {Altland}\ and\ \citenamefont {Simons}(2010)}]{altland2010condensed}%
  \BibitemOpen
  \bibfield  {author} {\bibinfo {author} {\bibfnamefont {A.}~\bibnamefont {Altland}}\ and\ \bibinfo {author} {\bibfnamefont {B.~D.}\ \bibnamefont {Simons}},\ }\href@noop {} {\emph {\bibinfo {title} {Condensed matter field theory}}}\ (\bibinfo  {publisher} {Cambridge university press},\ \bibinfo {year} {2010})\BibitemShut {NoStop}%
\bibitem [{\citenamefont {Pitaevskii}(1959)}]{pitaevskii1959properties}%
  \BibitemOpen
  \bibfield  {author} {\bibinfo {author} {\bibfnamefont {L.}~\bibnamefont {Pitaevskii}},\ }\href@noop {} {\bibfield  {journal} {\bibinfo  {journal} {Sov. Phys. JETP}\ }\textbf {\bibinfo {volume} {9}},\ \bibinfo {pages} {830} (\bibinfo {year} {1959})}\BibitemShut {NoStop}%
\bibitem [{\citenamefont {Nichitiu}\ \emph {et~al.}(2024)\citenamefont {Nichitiu}, \citenamefont {Brown},\ and\ \citenamefont {Zaliznyak}}]{Nichitiu_2024_He4}%
  \BibitemOpen
  \bibfield  {author} {\bibinfo {author} {\bibfnamefont {M.~D.}\ \bibnamefont {Nichitiu}}, \bibinfo {author} {\bibfnamefont {C.}~\bibnamefont {Brown}},\ and\ \bibinfo {author} {\bibfnamefont {I.~A.}\ \bibnamefont {Zaliznyak}},\ }\href {https://doi.org/10.1103/PhysRevB.109.L060502} {\bibfield  {journal} {\bibinfo  {journal} {Phys. Rev. B}\ }\textbf {\bibinfo {volume} {109}},\ \bibinfo {pages} {L060502} (\bibinfo {year} {2024})}\BibitemShut {NoStop}%
\bibitem [{\citenamefont {Laughlin}(1999)}]{Laughlin_Nobel}%
  \BibitemOpen
  \bibfield  {author} {\bibinfo {author} {\bibfnamefont {R.~B.}\ \bibnamefont {Laughlin}},\ }\href {https://doi.org/10.1103/RevModPhys.71.863} {\bibfield  {journal} {\bibinfo  {journal} {Rev. Mod. Phys.}\ }\textbf {\bibinfo {volume} {71}},\ \bibinfo {pages} {863} (\bibinfo {year} {1999})}\BibitemShut {NoStop}%
\bibitem [{\citenamefont {Broholm}\ \emph {et~al.}(2020)\citenamefont {Broholm}, \citenamefont {Cava}, \citenamefont {Kivelson}, \citenamefont {Nocera}, \citenamefont {Norman},\ and\ \citenamefont {Senthil}}]{broholm2020quantum}%
  \BibitemOpen
  \bibfield  {author} {\bibinfo {author} {\bibfnamefont {C.}~\bibnamefont {Broholm}}, \bibinfo {author} {\bibfnamefont {R.~J.}\ \bibnamefont {Cava}}, \bibinfo {author} {\bibfnamefont {S.}~\bibnamefont {Kivelson}}, \bibinfo {author} {\bibfnamefont {D.}~\bibnamefont {Nocera}}, \bibinfo {author} {\bibfnamefont {M.}~\bibnamefont {Norman}},\ and\ \bibinfo {author} {\bibfnamefont {T.}~\bibnamefont {Senthil}},\ }\href {https://doi.org/10.1126/science.aay0668} {\bibfield  {journal} {\bibinfo  {journal} {Science}\ }\textbf {\bibinfo {volume} {367}},\ \bibinfo {pages} {eaay0668} (\bibinfo {year} {2020})}\BibitemShut {NoStop}%
\bibitem [{\citenamefont {Stone}\ \emph {et~al.}(2006)\citenamefont {Stone}, \citenamefont {Zaliznyak}, \citenamefont {Hong}, \citenamefont {Broholm},\ and\ \citenamefont {Reich}}]{stone2006quasiparticle}%
  \BibitemOpen
  \bibfield  {author} {\bibinfo {author} {\bibfnamefont {M.~B.}\ \bibnamefont {Stone}}, \bibinfo {author} {\bibfnamefont {I.~A.}\ \bibnamefont {Zaliznyak}}, \bibinfo {author} {\bibfnamefont {T.}~\bibnamefont {Hong}}, \bibinfo {author} {\bibfnamefont {C.~L.}\ \bibnamefont {Broholm}},\ and\ \bibinfo {author} {\bibfnamefont {D.~H.}\ \bibnamefont {Reich}},\ }\href {https://doi.org/10.1038/nature04593} {\bibfield  {journal} {\bibinfo  {journal} {Nature}\ }\textbf {\bibinfo {volume} {440}},\ \bibinfo {pages} {187} (\bibinfo {year} {2006})}\BibitemShut {NoStop}%
\bibitem [{\citenamefont {Hu}\ \emph {et~al.}(2024)\citenamefont {Hu}, \citenamefont {Chen},\ and\ \citenamefont {Si}}]{hu2024quantum}%
  \BibitemOpen
  \bibfield  {author} {\bibinfo {author} {\bibfnamefont {H.}~\bibnamefont {Hu}}, \bibinfo {author} {\bibfnamefont {L.}~\bibnamefont {Chen}},\ and\ \bibinfo {author} {\bibfnamefont {Q.}~\bibnamefont {Si}},\ }\href {https://doi.org/10.1038/s41567-024-02679-7} {\bibfield  {journal} {\bibinfo  {journal} {Nature Physics}\ ,\ \bibinfo {pages} {1}} (\bibinfo {year} {2024})}\BibitemShut {NoStop}%
\bibitem [{\citenamefont {Abanov}\ \emph {et~al.}(2003)\citenamefont {Abanov}, \citenamefont {Chubukov},\ and\ \citenamefont {Schmalian}}]{abanov2003quantum}%
  \BibitemOpen
  \bibfield  {author} {\bibinfo {author} {\bibfnamefont {A.}~\bibnamefont {Abanov}}, \bibinfo {author} {\bibfnamefont {A.~V.}\ \bibnamefont {Chubukov}},\ and\ \bibinfo {author} {\bibfnamefont {J.}~\bibnamefont {Schmalian}},\ }\href {https://doi.org/10.1080/0001873021000057123} {\bibfield  {journal} {\bibinfo  {journal} {Advances in Physics}\ }\textbf {\bibinfo {volume} {52}},\ \bibinfo {pages} {119} (\bibinfo {year} {2003})}\BibitemShut {NoStop}%
\bibitem [{\citenamefont {Abanov}\ and\ \citenamefont {Chubukov}(2020)}]{abanov2020interplay}%
  \BibitemOpen
  \bibfield  {author} {\bibinfo {author} {\bibfnamefont {A.}~\bibnamefont {Abanov}}\ and\ \bibinfo {author} {\bibfnamefont {A.~V.}\ \bibnamefont {Chubukov}},\ }\href {https://doi.org/10.1103/PhysRevB.102.024524} {\bibfield  {journal} {\bibinfo  {journal} {Phys. Rev. B}\ }\textbf {\bibinfo {volume} {102}},\ \bibinfo {pages} {024524} (\bibinfo {year} {2020})}\BibitemShut {NoStop}%
\bibitem [{\citenamefont {Eberlein}\ \emph {et~al.}(2016)\citenamefont {Eberlein}, \citenamefont {Mandal},\ and\ \citenamefont {Sachdev}}]{eberlein2016hyperscaling}%
  \BibitemOpen
  \bibfield  {author} {\bibinfo {author} {\bibfnamefont {A.}~\bibnamefont {Eberlein}}, \bibinfo {author} {\bibfnamefont {I.}~\bibnamefont {Mandal}},\ and\ \bibinfo {author} {\bibfnamefont {S.}~\bibnamefont {Sachdev}},\ }\href {https://doi.org/10.1103/PhysRevB.94.045133} {\bibfield  {journal} {\bibinfo  {journal} {Phys. Rev. B}\ }\textbf {\bibinfo {volume} {94}},\ \bibinfo {pages} {045133} (\bibinfo {year} {2016})}\BibitemShut {NoStop}%
\bibitem [{\citenamefont {Hartnoll}\ \emph {et~al.}(2011)\citenamefont {Hartnoll}, \citenamefont {Hofman}, \citenamefont {Metlitski},\ and\ \citenamefont {Sachdev}}]{hartnoll2011quantum}%
  \BibitemOpen
  \bibfield  {author} {\bibinfo {author} {\bibfnamefont {S.~A.}\ \bibnamefont {Hartnoll}}, \bibinfo {author} {\bibfnamefont {D.~M.}\ \bibnamefont {Hofman}}, \bibinfo {author} {\bibfnamefont {M.~A.}\ \bibnamefont {Metlitski}},\ and\ \bibinfo {author} {\bibfnamefont {S.}~\bibnamefont {Sachdev}},\ }\href {https://doi.org/10.1103/PhysRevB.84.125115} {\bibfield  {journal} {\bibinfo  {journal} {Phys. Rev. B}\ }\textbf {\bibinfo {volume} {84}},\ \bibinfo {pages} {125115} (\bibinfo {year} {2011})}\BibitemShut {NoStop}%
\bibitem [{\citenamefont {Bl{\"o}te}\ \emph {et~al.}(1969)\citenamefont {Bl{\"o}te}, \citenamefont {Wielinga},\ and\ \citenamefont {Huiskamp}}]{Blote1969}%
  \BibitemOpen
  \bibfield  {author} {\bibinfo {author} {\bibfnamefont {H.}~\bibnamefont {Bl{\"o}te}}, \bibinfo {author} {\bibfnamefont {R.}~\bibnamefont {Wielinga}},\ and\ \bibinfo {author} {\bibfnamefont {W.}~\bibnamefont {Huiskamp}},\ }\href {https://doi.org/https://doi.org/10.1016/0031-8914(69)90187-6} {\bibfield  {journal} {\bibinfo  {journal} {Physica}\ }\textbf {\bibinfo {volume} {43}},\ \bibinfo {pages} {549} (\bibinfo {year} {1969})}\BibitemShut {NoStop}%
\bibitem [{\citenamefont {Hodges}\ \emph {et~al.}(2001)\citenamefont {Hodges}, \citenamefont {Bonville}, \citenamefont {Forget}, \citenamefont {Rams}, \citenamefont {Krolas},\ and\ \citenamefont {Dhalenne}}]{HodgesCEF}%
  \BibitemOpen
  \bibfield  {author} {\bibinfo {author} {\bibfnamefont {J.~A.}\ \bibnamefont {Hodges}}, \bibinfo {author} {\bibfnamefont {P.}~\bibnamefont {Bonville}}, \bibinfo {author} {\bibfnamefont {A.}~\bibnamefont {Forget}}, \bibinfo {author} {\bibfnamefont {M.}~\bibnamefont {Rams}}, \bibinfo {author} {\bibfnamefont {K.}~\bibnamefont {Krolas}},\ and\ \bibinfo {author} {\bibfnamefont {G.}~\bibnamefont {Dhalenne}},\ }\href {http://stacks.iop.org/0953-8984/13/i=41/a=318} {\bibfield  {journal} {\bibinfo  {journal} {Journal of Physics: Condensed Matter}\ }\textbf {\bibinfo {volume} {13}},\ \bibinfo {pages} {9301} (\bibinfo {year} {2001})}\BibitemShut {NoStop}%
\bibitem [{\citenamefont {Onoda}(2011)}]{Onoda_2011}%
  \BibitemOpen
  \bibfield  {author} {\bibinfo {author} {\bibfnamefont {S.}~\bibnamefont {Onoda}},\ }\href {https://doi.org/10.1088/1742-6596/320/1/012065} {\bibfield  {journal} {\bibinfo  {journal} {Journal of Physics: Conference Series}\ }\textbf {\bibinfo {volume} {320}},\ \bibinfo {pages} {012065} (\bibinfo {year} {2011})}\BibitemShut {NoStop}%
\bibitem [{\citenamefont {Yasui}\ \emph {et~al.}(2003)\citenamefont {Yasui}, \citenamefont {Soda}, \citenamefont {Iikubo}, \citenamefont {Ito}, \citenamefont {Sato}, \citenamefont {Hamaguchi}, \citenamefont {Matsushita}, \citenamefont {Wada}, \citenamefont {Takeuchi}, \citenamefont {Aso},\ and\ \citenamefont {Kakurai}}]{FM_order2003}%
  \BibitemOpen
  \bibfield  {author} {\bibinfo {author} {\bibfnamefont {Y.}~\bibnamefont {Yasui}}, \bibinfo {author} {\bibfnamefont {M.}~\bibnamefont {Soda}}, \bibinfo {author} {\bibfnamefont {S.}~\bibnamefont {Iikubo}}, \bibinfo {author} {\bibfnamefont {M.}~\bibnamefont {Ito}}, \bibinfo {author} {\bibfnamefont {M.}~\bibnamefont {Sato}}, \bibinfo {author} {\bibfnamefont {N.}~\bibnamefont {Hamaguchi}}, \bibinfo {author} {\bibfnamefont {T.}~\bibnamefont {Matsushita}}, \bibinfo {author} {\bibfnamefont {N.}~\bibnamefont {Wada}}, \bibinfo {author} {\bibfnamefont {T.}~\bibnamefont {Takeuchi}}, \bibinfo {author} {\bibfnamefont {N.}~\bibnamefont {Aso}},\ and\ \bibinfo {author} {\bibfnamefont {K.}~\bibnamefont {Kakurai}},\ }\href {https://doi.org/10.1143/JPSJ.72.3014} {\bibfield  {journal} {\bibinfo  {journal} {Journal of the Physical Society of Japan}\ }\textbf {\bibinfo {volume} {72}},\ \bibinfo {pages} {3014} (\bibinfo {year} {2003})}\BibitemShut {NoStop}%
\bibitem [{\citenamefont {Chang}\ \emph {et~al.}(2012)\citenamefont {Chang}, \citenamefont {Onoda}, \citenamefont {Su}, \citenamefont {Kao}, \citenamefont {Tsuei}, \citenamefont {Yasui}, \citenamefont {Kakurai},\ and\ \citenamefont {Lees}}]{Chang_2012_Higgs}%
  \BibitemOpen
  \bibfield  {author} {\bibinfo {author} {\bibfnamefont {L.-J.}\ \bibnamefont {Chang}}, \bibinfo {author} {\bibfnamefont {S.}~\bibnamefont {Onoda}}, \bibinfo {author} {\bibfnamefont {Y.}~\bibnamefont {Su}}, \bibinfo {author} {\bibfnamefont {Y.-J.}\ \bibnamefont {Kao}}, \bibinfo {author} {\bibfnamefont {K.-D.}\ \bibnamefont {Tsuei}}, \bibinfo {author} {\bibfnamefont {Y.}~\bibnamefont {Yasui}}, \bibinfo {author} {\bibfnamefont {K.}~\bibnamefont {Kakurai}},\ and\ \bibinfo {author} {\bibfnamefont {M.~R.}\ \bibnamefont {Lees}},\ }\href {https://doi.org/10.1038/ncomms1989} {\bibfield  {journal} {\bibinfo  {journal} {Nature Communications}\ }\textbf {\bibinfo {volume} {3}},\ \bibinfo {pages} {992} (\bibinfo {year} {2012})}\BibitemShut {NoStop}%
\bibitem [{\citenamefont {Shafieizadeh}\ \emph {et~al.}(2018)\citenamefont {Shafieizadeh}, \citenamefont {Xin}, \citenamefont {Koohpayeh}, \citenamefont {Huang},\ and\ \citenamefont {Zhou}}]{Shafieizadeh2018}%
  \BibitemOpen
  \bibfield  {author} {\bibinfo {author} {\bibfnamefont {Z.}~\bibnamefont {Shafieizadeh}}, \bibinfo {author} {\bibfnamefont {Y.}~\bibnamefont {Xin}}, \bibinfo {author} {\bibfnamefont {S.~M.}\ \bibnamefont {Koohpayeh}}, \bibinfo {author} {\bibfnamefont {Q.}~\bibnamefont {Huang}},\ and\ \bibinfo {author} {\bibfnamefont {H.}~\bibnamefont {Zhou}},\ }\href {https://doi.org/10.1038/s41598-018-35283-w} {\bibfield  {journal} {\bibinfo  {journal} {Scientific Reports}\ }\textbf {\bibinfo {volume} {8}},\ \bibinfo {pages} {17202} (\bibinfo {year} {2018})}\BibitemShut {NoStop}%
\bibitem [{\citenamefont {Yaouanc}\ \emph {et~al.}(2016)\citenamefont {Yaouanc}, \citenamefont {de~R\'eotier}, \citenamefont {Keller}, \citenamefont {Roessli},\ and\ \citenamefont {Forget}}]{Yaouanc_order}%
  \BibitemOpen
  \bibfield  {author} {\bibinfo {author} {\bibfnamefont {A.}~\bibnamefont {Yaouanc}}, \bibinfo {author} {\bibfnamefont {P.~D.}\ \bibnamefont {de~R\'eotier}}, \bibinfo {author} {\bibfnamefont {L.}~\bibnamefont {Keller}}, \bibinfo {author} {\bibfnamefont {B.}~\bibnamefont {Roessli}},\ and\ \bibinfo {author} {\bibfnamefont {A.}~\bibnamefont {Forget}},\ }\href {http://stacks.iop.org/0953-8984/28/i=42/a=426002} {\bibfield  {journal} {\bibinfo  {journal} {Journal of Physics: Condensed Matter}\ }\textbf {\bibinfo {volume} {28}},\ \bibinfo {pages} {426002} (\bibinfo {year} {2016})}\BibitemShut {NoStop}%
\bibitem [{\citenamefont {Scheie}\ \emph {et~al.}(2017)\citenamefont {Scheie}, \citenamefont {Kindervater}, \citenamefont {S\"aubert}, \citenamefont {Duvinage}, \citenamefont {Pfleiderer}, \citenamefont {Changlani}, \citenamefont {Zhang}, \citenamefont {Harriger}, \citenamefont {Arpino}, \citenamefont {Koohpayeh}, \citenamefont {Tchernyshyov},\ and\ \citenamefont {Broholm}}]{Scheie2017}%
  \BibitemOpen
  \bibfield  {author} {\bibinfo {author} {\bibfnamefont {A.}~\bibnamefont {Scheie}}, \bibinfo {author} {\bibfnamefont {J.}~\bibnamefont {Kindervater}}, \bibinfo {author} {\bibfnamefont {S.}~\bibnamefont {S\"aubert}}, \bibinfo {author} {\bibfnamefont {C.}~\bibnamefont {Duvinage}}, \bibinfo {author} {\bibfnamefont {C.}~\bibnamefont {Pfleiderer}}, \bibinfo {author} {\bibfnamefont {H.~J.}\ \bibnamefont {Changlani}}, \bibinfo {author} {\bibfnamefont {S.}~\bibnamefont {Zhang}}, \bibinfo {author} {\bibfnamefont {L.}~\bibnamefont {Harriger}}, \bibinfo {author} {\bibfnamefont {K.}~\bibnamefont {Arpino}}, \bibinfo {author} {\bibfnamefont {S.~M.}\ \bibnamefont {Koohpayeh}}, \bibinfo {author} {\bibfnamefont {O.}~\bibnamefont {Tchernyshyov}},\ and\ \bibinfo {author} {\bibfnamefont {C.}~\bibnamefont {Broholm}},\ }\href {https://doi.org/10.1103/PhysRevLett.119.127201} {\bibfield  {journal} {\bibinfo  {journal} {Phys. Rev. Lett.}\ }\textbf {\bibinfo {volume} {119}},\ \bibinfo {pages} {127201} (\bibinfo {year}
  {2017})}\BibitemShut {NoStop}%
\bibitem [{\citenamefont {S\"aubert}\ \emph {et~al.}(2020)\citenamefont {S\"aubert}, \citenamefont {Scheie}, \citenamefont {Duvinage}, \citenamefont {Kindervater}, \citenamefont {Zhang}, \citenamefont {Changlani}, \citenamefont {Xu}, \citenamefont {Koohpayeh}, \citenamefont {Tchernyshyov}, \citenamefont {Broholm},\ and\ \citenamefont {Pfleiderer}}]{Saubert_2020}%
  \BibitemOpen
  \bibfield  {author} {\bibinfo {author} {\bibfnamefont {S.}~\bibnamefont {S\"aubert}}, \bibinfo {author} {\bibfnamefont {A.}~\bibnamefont {Scheie}}, \bibinfo {author} {\bibfnamefont {C.}~\bibnamefont {Duvinage}}, \bibinfo {author} {\bibfnamefont {J.}~\bibnamefont {Kindervater}}, \bibinfo {author} {\bibfnamefont {S.}~\bibnamefont {Zhang}}, \bibinfo {author} {\bibfnamefont {H.~J.}\ \bibnamefont {Changlani}}, \bibinfo {author} {\bibfnamefont {G.}~\bibnamefont {Xu}}, \bibinfo {author} {\bibfnamefont {S.~M.}\ \bibnamefont {Koohpayeh}}, \bibinfo {author} {\bibfnamefont {O.}~\bibnamefont {Tchernyshyov}}, \bibinfo {author} {\bibfnamefont {C.~L.}\ \bibnamefont {Broholm}},\ and\ \bibinfo {author} {\bibfnamefont {C.}~\bibnamefont {Pfleiderer}},\ }\href {https://doi.org/10.1103/PhysRevB.101.174434} {\bibfield  {journal} {\bibinfo  {journal} {Phys. Rev. B}\ }\textbf {\bibinfo {volume} {101}},\ \bibinfo {pages} {174434} (\bibinfo {year} {2020})}\BibitemShut {NoStop}%
\bibitem [{\citenamefont {Yaouanc}\ \emph {et~al.}(2011)\citenamefont {Yaouanc}, \citenamefont {Dalmas~de R\'eotier}, \citenamefont {Marin},\ and\ \citenamefont {Glazkov}}]{SampleDependence_HC}%
  \BibitemOpen
  \bibfield  {author} {\bibinfo {author} {\bibfnamefont {A.}~\bibnamefont {Yaouanc}}, \bibinfo {author} {\bibfnamefont {P.}~\bibnamefont {Dalmas~de R\'eotier}}, \bibinfo {author} {\bibfnamefont {C.}~\bibnamefont {Marin}},\ and\ \bibinfo {author} {\bibfnamefont {V.}~\bibnamefont {Glazkov}},\ }\href {https://doi.org/10.1103/PhysRevB.84.172408} {\bibfield  {journal} {\bibinfo  {journal} {Phys. Rev. B}\ }\textbf {\bibinfo {volume} {84}},\ \bibinfo {pages} {172408} (\bibinfo {year} {2011})}\BibitemShut {NoStop}%
\bibitem [{\citenamefont {Ross}\ \emph {et~al.}(2011{\natexlab{a}})\citenamefont {Ross}, \citenamefont {Yaraskavitch}, \citenamefont {Laver}, \citenamefont {Gardner}, \citenamefont {Quilliam}, \citenamefont {Meng}, \citenamefont {Kycia}, \citenamefont {Singh}, \citenamefont {Proffen}, \citenamefont {Dabkowska},\ and\ \citenamefont {Gaulin}}]{SampleDependence_Ross}%
  \BibitemOpen
  \bibfield  {author} {\bibinfo {author} {\bibfnamefont {K.~A.}\ \bibnamefont {Ross}}, \bibinfo {author} {\bibfnamefont {L.~R.}\ \bibnamefont {Yaraskavitch}}, \bibinfo {author} {\bibfnamefont {M.}~\bibnamefont {Laver}}, \bibinfo {author} {\bibfnamefont {J.~S.}\ \bibnamefont {Gardner}}, \bibinfo {author} {\bibfnamefont {J.~A.}\ \bibnamefont {Quilliam}}, \bibinfo {author} {\bibfnamefont {S.}~\bibnamefont {Meng}}, \bibinfo {author} {\bibfnamefont {J.~B.}\ \bibnamefont {Kycia}}, \bibinfo {author} {\bibfnamefont {D.~K.}\ \bibnamefont {Singh}}, \bibinfo {author} {\bibfnamefont {T.}~\bibnamefont {Proffen}}, \bibinfo {author} {\bibfnamefont {H.~A.}\ \bibnamefont {Dabkowska}},\ and\ \bibinfo {author} {\bibfnamefont {B.~D.}\ \bibnamefont {Gaulin}},\ }\href {https://doi.org/10.1103/PhysRevB.84.174442} {\bibfield  {journal} {\bibinfo  {journal} {Phys. Rev. B}\ }\textbf {\bibinfo {volume} {84}},\ \bibinfo {pages} {174442} (\bibinfo {year} {2011}{\natexlab{a}})}\BibitemShut {NoStop}%
\bibitem [{\citenamefont {Ross}\ \emph {et~al.}(2012)\citenamefont {Ross}, \citenamefont {Proffen}, \citenamefont {Dabkowska}, \citenamefont {Quilliam}, \citenamefont {Yaraskavitch}, \citenamefont {Kycia},\ and\ \citenamefont {Gaulin}}]{Ross_stuffing}%
  \BibitemOpen
  \bibfield  {author} {\bibinfo {author} {\bibfnamefont {K.~A.}\ \bibnamefont {Ross}}, \bibinfo {author} {\bibfnamefont {T.}~\bibnamefont {Proffen}}, \bibinfo {author} {\bibfnamefont {H.~A.}\ \bibnamefont {Dabkowska}}, \bibinfo {author} {\bibfnamefont {J.~A.}\ \bibnamefont {Quilliam}}, \bibinfo {author} {\bibfnamefont {L.~R.}\ \bibnamefont {Yaraskavitch}}, \bibinfo {author} {\bibfnamefont {J.~B.}\ \bibnamefont {Kycia}},\ and\ \bibinfo {author} {\bibfnamefont {B.~D.}\ \bibnamefont {Gaulin}},\ }\href {https://doi.org/10.1103/PhysRevB.86.174424} {\bibfield  {journal} {\bibinfo  {journal} {Phys. Rev. B}\ }\textbf {\bibinfo {volume} {86}},\ \bibinfo {pages} {174424} (\bibinfo {year} {2012})}\BibitemShut {NoStop}%
\bibitem [{\citenamefont {Gaudet}\ \emph {et~al.}(2015)\citenamefont {Gaudet}, \citenamefont {Maharaj}, \citenamefont {Sala}, \citenamefont {Kermarrec}, \citenamefont {Ross}, \citenamefont {Dabkowska}, \citenamefont {Kolesnikov}, \citenamefont {Granroth},\ and\ \citenamefont {Gaulin}}]{GaudetRoss_CEF}%
  \BibitemOpen
  \bibfield  {author} {\bibinfo {author} {\bibfnamefont {J.}~\bibnamefont {Gaudet}}, \bibinfo {author} {\bibfnamefont {D.~D.}\ \bibnamefont {Maharaj}}, \bibinfo {author} {\bibfnamefont {G.}~\bibnamefont {Sala}}, \bibinfo {author} {\bibfnamefont {E.}~\bibnamefont {Kermarrec}}, \bibinfo {author} {\bibfnamefont {K.~A.}\ \bibnamefont {Ross}}, \bibinfo {author} {\bibfnamefont {H.~A.}\ \bibnamefont {Dabkowska}}, \bibinfo {author} {\bibfnamefont {A.~I.}\ \bibnamefont {Kolesnikov}}, \bibinfo {author} {\bibfnamefont {G.~E.}\ \bibnamefont {Granroth}},\ and\ \bibinfo {author} {\bibfnamefont {B.~D.}\ \bibnamefont {Gaulin}},\ }\href {https://doi.org/10.1103/PhysRevB.92.134420} {\bibfield  {journal} {\bibinfo  {journal} {Phys. Rev. B}\ }\textbf {\bibinfo {volume} {92}},\ \bibinfo {pages} {134420} (\bibinfo {year} {2015})}\BibitemShut {NoStop}%
\bibitem [{\citenamefont {Mostaed}\ \emph {et~al.}(2017)\citenamefont {Mostaed}, \citenamefont {Balakrishnan}, \citenamefont {Lees}, \citenamefont {Yasui}, \citenamefont {Chang},\ and\ \citenamefont {Beanland}}]{Mostaed2017}%
  \BibitemOpen
  \bibfield  {author} {\bibinfo {author} {\bibfnamefont {A.}~\bibnamefont {Mostaed}}, \bibinfo {author} {\bibfnamefont {G.}~\bibnamefont {Balakrishnan}}, \bibinfo {author} {\bibfnamefont {M.~R.}\ \bibnamefont {Lees}}, \bibinfo {author} {\bibfnamefont {Y.}~\bibnamefont {Yasui}}, \bibinfo {author} {\bibfnamefont {L.-J.}\ \bibnamefont {Chang}},\ and\ \bibinfo {author} {\bibfnamefont {R.}~\bibnamefont {Beanland}},\ }\href {https://doi.org/10.1103/PhysRevB.95.094431} {\bibfield  {journal} {\bibinfo  {journal} {Phys. Rev. B}\ }\textbf {\bibinfo {volume} {95}},\ \bibinfo {pages} {094431} (\bibinfo {year} {2017})}\BibitemShut {NoStop}%
\bibitem [{\citenamefont {D'Ortenzio}\ \emph {et~al.}(2013)\citenamefont {D'Ortenzio}, \citenamefont {Dabkowska}, \citenamefont {Dunsiger}, \citenamefont {Gaulin}, \citenamefont {Gingras}, \citenamefont {Goko}, \citenamefont {Kycia}, \citenamefont {Liu}, \citenamefont {Medina}, \citenamefont {Munsie}, \citenamefont {Pomaranski}, \citenamefont {Ross}, \citenamefont {Uemura}, \citenamefont {Williams},\ and\ \citenamefont {Luke}}]{DOrtenzio_noGsOrder}%
  \BibitemOpen
  \bibfield  {author} {\bibinfo {author} {\bibfnamefont {R.~M.}\ \bibnamefont {D'Ortenzio}}, \bibinfo {author} {\bibfnamefont {H.~A.}\ \bibnamefont {Dabkowska}}, \bibinfo {author} {\bibfnamefont {S.~R.}\ \bibnamefont {Dunsiger}}, \bibinfo {author} {\bibfnamefont {B.~D.}\ \bibnamefont {Gaulin}}, \bibinfo {author} {\bibfnamefont {M.~J.~P.}\ \bibnamefont {Gingras}}, \bibinfo {author} {\bibfnamefont {T.}~\bibnamefont {Goko}}, \bibinfo {author} {\bibfnamefont {J.~B.}\ \bibnamefont {Kycia}}, \bibinfo {author} {\bibfnamefont {L.}~\bibnamefont {Liu}}, \bibinfo {author} {\bibfnamefont {T.}~\bibnamefont {Medina}}, \bibinfo {author} {\bibfnamefont {T.~J.}\ \bibnamefont {Munsie}}, \bibinfo {author} {\bibfnamefont {D.}~\bibnamefont {Pomaranski}}, \bibinfo {author} {\bibfnamefont {K.~A.}\ \bibnamefont {Ross}}, \bibinfo {author} {\bibfnamefont {Y.~J.}\ \bibnamefont {Uemura}}, \bibinfo {author} {\bibfnamefont {T.~J.}\ \bibnamefont {Williams}},\ and\ \bibinfo {author} {\bibfnamefont {G.~M.}\ \bibnamefont {Luke}},\ }\href
  {https://doi.org/10.1103/PhysRevB.88.134428} {\bibfield  {journal} {\bibinfo  {journal} {Phys. Rev. B}\ }\textbf {\bibinfo {volume} {88}},\ \bibinfo {pages} {134428} (\bibinfo {year} {2013})}\BibitemShut {NoStop}%
\bibitem [{\citenamefont {Bowman}\ \emph {et~al.}(2019)\citenamefont {Bowman}, \citenamefont {Cemal}, \citenamefont {Lehner}, \citenamefont {Wildes}, \citenamefont {Mangin-Thro}, \citenamefont {Nilsen}, \citenamefont {Gutmann}, \citenamefont {Voneshen}, \citenamefont {Prabhakaran}, \citenamefont {Boothroyd}, \citenamefont {Porter}, \citenamefont {Castelnovo}, \citenamefont {Refson},\ and\ \citenamefont {Goff}}]{Bowman2019}%
  \BibitemOpen
  \bibfield  {author} {\bibinfo {author} {\bibfnamefont {D.~F.}\ \bibnamefont {Bowman}}, \bibinfo {author} {\bibfnamefont {E.}~\bibnamefont {Cemal}}, \bibinfo {author} {\bibfnamefont {T.}~\bibnamefont {Lehner}}, \bibinfo {author} {\bibfnamefont {A.~R.}\ \bibnamefont {Wildes}}, \bibinfo {author} {\bibfnamefont {L.}~\bibnamefont {Mangin-Thro}}, \bibinfo {author} {\bibfnamefont {G.~J.}\ \bibnamefont {Nilsen}}, \bibinfo {author} {\bibfnamefont {M.~J.}\ \bibnamefont {Gutmann}}, \bibinfo {author} {\bibfnamefont {D.~J.}\ \bibnamefont {Voneshen}}, \bibinfo {author} {\bibfnamefont {D.}~\bibnamefont {Prabhakaran}}, \bibinfo {author} {\bibfnamefont {A.~T.}\ \bibnamefont {Boothroyd}}, \bibinfo {author} {\bibfnamefont {D.~G.}\ \bibnamefont {Porter}}, \bibinfo {author} {\bibfnamefont {C.}~\bibnamefont {Castelnovo}}, \bibinfo {author} {\bibfnamefont {K.}~\bibnamefont {Refson}},\ and\ \bibinfo {author} {\bibfnamefont {J.~P.}\ \bibnamefont {Goff}},\ }\href {https://doi.org/10.1038/s41467-019-08598-z} {\bibfield  {journal}
  {\bibinfo  {journal} {Nature Communications}\ }\textbf {\bibinfo {volume} {10}},\ \bibinfo {pages} {637} (\bibinfo {year} {2019})}\BibitemShut {NoStop}%
\bibitem [{\citenamefont {Arpino}\ \emph {et~al.}(2017)\citenamefont {Arpino}, \citenamefont {Trump}, \citenamefont {Scheie}, \citenamefont {McQueen},\ and\ \citenamefont {Koohpayeh}}]{SeyedPaper}%
  \BibitemOpen
  \bibfield  {author} {\bibinfo {author} {\bibfnamefont {K.~E.}\ \bibnamefont {Arpino}}, \bibinfo {author} {\bibfnamefont {B.~A.}\ \bibnamefont {Trump}}, \bibinfo {author} {\bibfnamefont {A.~O.}\ \bibnamefont {Scheie}}, \bibinfo {author} {\bibfnamefont {T.~M.}\ \bibnamefont {McQueen}},\ and\ \bibinfo {author} {\bibfnamefont {S.~M.}\ \bibnamefont {Koohpayeh}},\ }\href {https://doi.org/10.1103/PhysRevB.95.094407} {\bibfield  {journal} {\bibinfo  {journal} {Phys. Rev. B}\ }\textbf {\bibinfo {volume} {95}},\ \bibinfo {pages} {094407} (\bibinfo {year} {2017})}\BibitemShut {NoStop}%
\bibitem [{\citenamefont {Ross}\ \emph {et~al.}(2009)\citenamefont {Ross}, \citenamefont {Ruff}, \citenamefont {Adams}, \citenamefont {Gardner}, \citenamefont {Dabkowska}, \citenamefont {Qiu}, \citenamefont {Copley},\ and\ \citenamefont {Gaulin}}]{Ross2009}%
  \BibitemOpen
  \bibfield  {author} {\bibinfo {author} {\bibfnamefont {K.~A.}\ \bibnamefont {Ross}}, \bibinfo {author} {\bibfnamefont {J.~P.~C.}\ \bibnamefont {Ruff}}, \bibinfo {author} {\bibfnamefont {C.~P.}\ \bibnamefont {Adams}}, \bibinfo {author} {\bibfnamefont {J.~S.}\ \bibnamefont {Gardner}}, \bibinfo {author} {\bibfnamefont {H.~A.}\ \bibnamefont {Dabkowska}}, \bibinfo {author} {\bibfnamefont {Y.}~\bibnamefont {Qiu}}, \bibinfo {author} {\bibfnamefont {J.~R.~D.}\ \bibnamefont {Copley}},\ and\ \bibinfo {author} {\bibfnamefont {B.~D.}\ \bibnamefont {Gaulin}},\ }\href {https://doi.org/10.1103/PhysRevLett.103.227202} {\bibfield  {journal} {\bibinfo  {journal} {Phys. Rev. Lett.}\ }\textbf {\bibinfo {volume} {103}},\ \bibinfo {pages} {227202} (\bibinfo {year} {2009})}\BibitemShut {NoStop}%
\bibitem [{\citenamefont {Ross}\ \emph {et~al.}(2011{\natexlab{b}})\citenamefont {Ross}, \citenamefont {Savary}, \citenamefont {Gaulin},\ and\ \citenamefont {Balents}}]{Ross_Hamiltonian}%
  \BibitemOpen
  \bibfield  {author} {\bibinfo {author} {\bibfnamefont {K.~A.}\ \bibnamefont {Ross}}, \bibinfo {author} {\bibfnamefont {L.}~\bibnamefont {Savary}}, \bibinfo {author} {\bibfnamefont {B.~D.}\ \bibnamefont {Gaulin}},\ and\ \bibinfo {author} {\bibfnamefont {L.}~\bibnamefont {Balents}},\ }\href {https://doi.org/10.1103/PhysRevX.1.021002} {\bibfield  {journal} {\bibinfo  {journal} {Phys. Rev. X}\ }\textbf {\bibinfo {volume} {1}},\ \bibinfo {pages} {021002} (\bibinfo {year} {2011}{\natexlab{b}})}\BibitemShut {NoStop}%
\bibitem [{\citenamefont {Thompson}\ \emph {et~al.}(2017)\citenamefont {Thompson}, \citenamefont {McClarty}, \citenamefont {Prabhakaran}, \citenamefont {Cabrera}, \citenamefont {Guidi},\ and\ \citenamefont {Coldea}}]{Thompson_2017}%
  \BibitemOpen
  \bibfield  {author} {\bibinfo {author} {\bibfnamefont {J.~D.}\ \bibnamefont {Thompson}}, \bibinfo {author} {\bibfnamefont {P.~A.}\ \bibnamefont {McClarty}}, \bibinfo {author} {\bibfnamefont {D.}~\bibnamefont {Prabhakaran}}, \bibinfo {author} {\bibfnamefont {I.}~\bibnamefont {Cabrera}}, \bibinfo {author} {\bibfnamefont {T.}~\bibnamefont {Guidi}},\ and\ \bibinfo {author} {\bibfnamefont {R.}~\bibnamefont {Coldea}},\ }\href {https://doi.org/10.1103/PhysRevLett.119.057203} {\bibfield  {journal} {\bibinfo  {journal} {Phys. Rev. Lett.}\ }\textbf {\bibinfo {volume} {119}},\ \bibinfo {pages} {057203} (\bibinfo {year} {2017})}\BibitemShut {NoStop}%
\bibitem [{\citenamefont {Scheie}\ \emph {et~al.}(2020)\citenamefont {Scheie}, \citenamefont {Kindervater}, \citenamefont {Zhang}, \citenamefont {Changlani}, \citenamefont {Sala}, \citenamefont {Ehlers}, \citenamefont {Heinemann}, \citenamefont {Tucker}, \citenamefont {Koohpayeh},\ and\ \citenamefont {Broholm}}]{Scheie2020}%
  \BibitemOpen
  \bibfield  {author} {\bibinfo {author} {\bibfnamefont {A.}~\bibnamefont {Scheie}}, \bibinfo {author} {\bibfnamefont {J.}~\bibnamefont {Kindervater}}, \bibinfo {author} {\bibfnamefont {S.}~\bibnamefont {Zhang}}, \bibinfo {author} {\bibfnamefont {H.~J.}\ \bibnamefont {Changlani}}, \bibinfo {author} {\bibfnamefont {G.}~\bibnamefont {Sala}}, \bibinfo {author} {\bibfnamefont {G.}~\bibnamefont {Ehlers}}, \bibinfo {author} {\bibfnamefont {A.}~\bibnamefont {Heinemann}}, \bibinfo {author} {\bibfnamefont {G.~S.}\ \bibnamefont {Tucker}}, \bibinfo {author} {\bibfnamefont {S.~M.}\ \bibnamefont {Koohpayeh}},\ and\ \bibinfo {author} {\bibfnamefont {C.}~\bibnamefont {Broholm}},\ }\href {https://doi.org/10.1073/pnas.2008791117} {\bibfield  {journal} {\bibinfo  {journal} {Proceedings of the National Academy of Sciences}\ }\textbf {\bibinfo {volume} {117}},\ \bibinfo {pages} {27245} (\bibinfo {year} {2020})}\BibitemShut {NoStop}%
\bibitem [{\citenamefont {Fishman}\ \emph {et~al.}(2018)\citenamefont {Fishman}, \citenamefont {Fernandez-Baca},\ and\ \citenamefont {R{\~o}{\~o}m}}]{fishman2018spin}%
  \BibitemOpen
  \bibfield  {author} {\bibinfo {author} {\bibfnamefont {R.~S.}\ \bibnamefont {Fishman}}, \bibinfo {author} {\bibfnamefont {J.~A.}\ \bibnamefont {Fernandez-Baca}},\ and\ \bibinfo {author} {\bibfnamefont {T.}~\bibnamefont {R{\~o}{\~o}m}},\ }\href@noop {} {\emph {\bibinfo {title} {Spin-Wave Theory and its Applications to Neutron Scattering and THz Spectroscopy}}}\ (\bibinfo  {publisher} {Morgan \& Claypool Publishers},\ \bibinfo {year} {2018})\BibitemShut {NoStop}%
\bibitem [{\citenamefont {Tokiwa}\ \emph {et~al.}(2016)\citenamefont {Tokiwa}, \citenamefont {Yamashita}, \citenamefont {Udagawa}, \citenamefont {Kittaka}, \citenamefont {Sakakibara}, \citenamefont {Terazawa}, \citenamefont {Shimoyama}, \citenamefont {Terashima}, \citenamefont {Yasui}, \citenamefont {Shibauchi},\ and\ \citenamefont {Matsuda}}]{MonopolesConductivity}%
  \BibitemOpen
  \bibfield  {author} {\bibinfo {author} {\bibfnamefont {Y.}~\bibnamefont {Tokiwa}}, \bibinfo {author} {\bibfnamefont {T.}~\bibnamefont {Yamashita}}, \bibinfo {author} {\bibfnamefont {M.}~\bibnamefont {Udagawa}}, \bibinfo {author} {\bibfnamefont {S.}~\bibnamefont {Kittaka}}, \bibinfo {author} {\bibfnamefont {T.}~\bibnamefont {Sakakibara}}, \bibinfo {author} {\bibfnamefont {D.}~\bibnamefont {Terazawa}}, \bibinfo {author} {\bibfnamefont {Y.}~\bibnamefont {Shimoyama}}, \bibinfo {author} {\bibfnamefont {T.}~\bibnamefont {Terashima}}, \bibinfo {author} {\bibfnamefont {Y.}~\bibnamefont {Yasui}}, \bibinfo {author} {\bibfnamefont {T.}~\bibnamefont {Shibauchi}},\ and\ \bibinfo {author} {\bibfnamefont {Y.}~\bibnamefont {Matsuda}},\ }\href {http://dx.doi.org/10.1038/ncomms10807} {\bibfield  {journal} {\bibinfo  {journal} {Nature Communications}\ }\textbf {\bibinfo {volume} {7}},\ \bibinfo {pages} {10807} (\bibinfo {year} {2016})}\BibitemShut {NoStop}%
\bibitem [{\citenamefont {Hirschberger}\ \emph {et~al.}(2019)\citenamefont {Hirschberger}, \citenamefont {Czajka}, \citenamefont {Koohpayeh}, \citenamefont {Wang},\ and\ \citenamefont {Ong}}]{hirschberger2019enhanced}%
  \BibitemOpen
  \bibfield  {author} {\bibinfo {author} {\bibfnamefont {M.}~\bibnamefont {Hirschberger}}, \bibinfo {author} {\bibfnamefont {P.}~\bibnamefont {Czajka}}, \bibinfo {author} {\bibfnamefont {S.}~\bibnamefont {Koohpayeh}}, \bibinfo {author} {\bibfnamefont {W.}~\bibnamefont {Wang}},\ and\ \bibinfo {author} {\bibfnamefont {N.~P.}\ \bibnamefont {Ong}},\ }\href {https://doi.org/10.48550/arXiv.1903.00595} {\bibfield  {journal} {\bibinfo  {journal} {arXiv preprint arXiv:1903.00595}\ } (\bibinfo {year} {2019})}\BibitemShut {NoStop}%
\bibitem [{\citenamefont {Scheie}\ \emph {et~al.}(2022)\citenamefont {Scheie}, \citenamefont {Benton}, \citenamefont {Taillefumier}, \citenamefont {Jaubert}, \citenamefont {Sala}, \citenamefont {Jalarvo}, \citenamefont {Koohpayeh},\ and\ \citenamefont {Shannon}}]{Scheie_2022_Dynamical}%
  \BibitemOpen
  \bibfield  {author} {\bibinfo {author} {\bibfnamefont {A.}~\bibnamefont {Scheie}}, \bibinfo {author} {\bibfnamefont {O.}~\bibnamefont {Benton}}, \bibinfo {author} {\bibfnamefont {M.}~\bibnamefont {Taillefumier}}, \bibinfo {author} {\bibfnamefont {L.~D.~C.}\ \bibnamefont {Jaubert}}, \bibinfo {author} {\bibfnamefont {G.}~\bibnamefont {Sala}}, \bibinfo {author} {\bibfnamefont {N.}~\bibnamefont {Jalarvo}}, \bibinfo {author} {\bibfnamefont {S.~M.}\ \bibnamefont {Koohpayeh}},\ and\ \bibinfo {author} {\bibfnamefont {N.}~\bibnamefont {Shannon}},\ }\href {https://doi.org/10.1103/PhysRevLett.129.217202} {\bibfield  {journal} {\bibinfo  {journal} {Phys. Rev. Lett.}\ }\textbf {\bibinfo {volume} {129}},\ \bibinfo {pages} {217202} (\bibinfo {year} {2022})}\BibitemShut {NoStop}%
\bibitem [{\citenamefont {Yan}\ \emph {et~al.}(2017)\citenamefont {Yan}, \citenamefont {Benton}, \citenamefont {Jaubert},\ and\ \citenamefont {Shannon}}]{Yan2017}%
  \BibitemOpen
  \bibfield  {author} {\bibinfo {author} {\bibfnamefont {H.}~\bibnamefont {Yan}}, \bibinfo {author} {\bibfnamefont {O.}~\bibnamefont {Benton}}, \bibinfo {author} {\bibfnamefont {L.}~\bibnamefont {Jaubert}},\ and\ \bibinfo {author} {\bibfnamefont {N.}~\bibnamefont {Shannon}},\ }\href {https://doi.org/10.1103/PhysRevB.95.094422} {\bibfield  {journal} {\bibinfo  {journal} {Phys. Rev. B}\ }\textbf {\bibinfo {volume} {95}},\ \bibinfo {pages} {094422} (\bibinfo {year} {2017})}\BibitemShut {NoStop}%
\bibitem [{\citenamefont {Hayre}\ \emph {et~al.}(2013)\citenamefont {Hayre}, \citenamefont {Ross}, \citenamefont {Applegate}, \citenamefont {Lin}, \citenamefont {Singh}, \citenamefont {Gaulin},\ and\ \citenamefont {Gingras}}]{Hayre2013}%
  \BibitemOpen
  \bibfield  {author} {\bibinfo {author} {\bibfnamefont {N.~R.}\ \bibnamefont {Hayre}}, \bibinfo {author} {\bibfnamefont {K.~A.}\ \bibnamefont {Ross}}, \bibinfo {author} {\bibfnamefont {R.}~\bibnamefont {Applegate}}, \bibinfo {author} {\bibfnamefont {T.}~\bibnamefont {Lin}}, \bibinfo {author} {\bibfnamefont {R.~R.~P.}\ \bibnamefont {Singh}}, \bibinfo {author} {\bibfnamefont {B.~D.}\ \bibnamefont {Gaulin}},\ and\ \bibinfo {author} {\bibfnamefont {M.~J.~P.}\ \bibnamefont {Gingras}},\ }\href {https://doi.org/10.1103/PhysRevB.87.184423} {\bibfield  {journal} {\bibinfo  {journal} {Phys. Rev. B}\ }\textbf {\bibinfo {volume} {87}},\ \bibinfo {pages} {184423} (\bibinfo {year} {2013})}\BibitemShut {NoStop}%
\bibitem [{\citenamefont {Hamachi}\ \emph {et~al.}(2016)\citenamefont {Hamachi}, \citenamefont {Yasui}, \citenamefont {Araki}, \citenamefont {Kittaka},\ and\ \citenamefont {Sakakibara}}]{Hamachi_2016}%
  \BibitemOpen
  \bibfield  {author} {\bibinfo {author} {\bibfnamefont {N.}~\bibnamefont {Hamachi}}, \bibinfo {author} {\bibfnamefont {Y.}~\bibnamefont {Yasui}}, \bibinfo {author} {\bibfnamefont {K.}~\bibnamefont {Araki}}, \bibinfo {author} {\bibfnamefont {S.}~\bibnamefont {Kittaka}},\ and\ \bibinfo {author} {\bibfnamefont {T.}~\bibnamefont {Sakakibara}},\ }\href {https://doi.org/http://dx.doi.org/10.1063/1.4944337} {\bibfield  {journal} {\bibinfo  {journal} {AIP Advances}\ }\textbf {\bibinfo {volume} {6}},\ \bibinfo {eid} {055707} (\bibinfo {year} {2016})}\BibitemShut {NoStop}%
\bibitem [{\citenamefont {Pan}\ \emph {et~al.}(2016)\citenamefont {Pan}, \citenamefont {Laurita}, \citenamefont {Ross}, \citenamefont {Gaulin},\ and\ \citenamefont {Armitage}}]{Armitage_monopoles_2016}%
  \BibitemOpen
  \bibfield  {author} {\bibinfo {author} {\bibfnamefont {L.}~\bibnamefont {Pan}}, \bibinfo {author} {\bibfnamefont {N.~J.}\ \bibnamefont {Laurita}}, \bibinfo {author} {\bibfnamefont {K.~A.}\ \bibnamefont {Ross}}, \bibinfo {author} {\bibfnamefont {B.~D.}\ \bibnamefont {Gaulin}},\ and\ \bibinfo {author} {\bibfnamefont {N.~P.}\ \bibnamefont {Armitage}},\ }\href {http://dx.doi.org/10.1038/nphys3608} {\bibfield  {journal} {\bibinfo  {journal} {Nat Phys}\ }\textbf {\bibinfo {volume} {12}},\ \bibinfo {pages} {361} (\bibinfo {year} {2016})}\BibitemShut {NoStop}%
\bibitem [{\citenamefont {Applegate}\ \emph {et~al.}(2012)\citenamefont {Applegate}, \citenamefont {Hayre}, \citenamefont {Singh}, \citenamefont {Lin}, \citenamefont {Day},\ and\ \citenamefont {Gingras}}]{ApplegateSpinIce}%
  \BibitemOpen
  \bibfield  {author} {\bibinfo {author} {\bibfnamefont {R.}~\bibnamefont {Applegate}}, \bibinfo {author} {\bibfnamefont {N.~R.}\ \bibnamefont {Hayre}}, \bibinfo {author} {\bibfnamefont {R.~R.~P.}\ \bibnamefont {Singh}}, \bibinfo {author} {\bibfnamefont {T.}~\bibnamefont {Lin}}, \bibinfo {author} {\bibfnamefont {A.~G.~R.}\ \bibnamefont {Day}},\ and\ \bibinfo {author} {\bibfnamefont {M.~J.~P.}\ \bibnamefont {Gingras}},\ }\href {https://doi.org/10.1103/PhysRevLett.109.097205} {\bibfield  {journal} {\bibinfo  {journal} {Phys. Rev. Lett.}\ }\textbf {\bibinfo {volume} {109}},\ \bibinfo {pages} {097205} (\bibinfo {year} {2012})}\BibitemShut {NoStop}%
\bibitem [{\citenamefont {Gingras}\ and\ \citenamefont {McClarty}(2014)}]{SpinIce_review}%
  \BibitemOpen
  \bibfield  {author} {\bibinfo {author} {\bibfnamefont {M.~J.~P.}\ \bibnamefont {Gingras}}\ and\ \bibinfo {author} {\bibfnamefont {P.~A.}\ \bibnamefont {McClarty}},\ }\href {http://stacks.iop.org/0034-4885/77/i=5/a=056501} {\bibfield  {journal} {\bibinfo  {journal} {Reports on Progress in Physics}\ }\textbf {\bibinfo {volume} {77}},\ \bibinfo {pages} {056501} (\bibinfo {year} {2014})}\BibitemShut {NoStop}%
\bibitem [{\citenamefont {Robert}\ \emph {et~al.}(2015)\citenamefont {Robert}, \citenamefont {Lhotel}, \citenamefont {Remenyi}, \citenamefont {Sahling}, \citenamefont {Mirebeau}, \citenamefont {Decorse}, \citenamefont {Canals},\ and\ \citenamefont {Petit}}]{Robert2015}%
  \BibitemOpen
  \bibfield  {author} {\bibinfo {author} {\bibfnamefont {J.}~\bibnamefont {Robert}}, \bibinfo {author} {\bibfnamefont {E.}~\bibnamefont {Lhotel}}, \bibinfo {author} {\bibfnamefont {G.}~\bibnamefont {Remenyi}}, \bibinfo {author} {\bibfnamefont {S.}~\bibnamefont {Sahling}}, \bibinfo {author} {\bibfnamefont {I.}~\bibnamefont {Mirebeau}}, \bibinfo {author} {\bibfnamefont {C.}~\bibnamefont {Decorse}}, \bibinfo {author} {\bibfnamefont {B.}~\bibnamefont {Canals}},\ and\ \bibinfo {author} {\bibfnamefont {S.}~\bibnamefont {Petit}},\ }\href {https://doi.org/10.1103/PhysRevB.92.064425} {\bibfield  {journal} {\bibinfo  {journal} {Phys. Rev. B}\ }\textbf {\bibinfo {volume} {92}},\ \bibinfo {pages} {064425} (\bibinfo {year} {2015})}\BibitemShut {NoStop}%
\bibitem [{\citenamefont {Rau}\ \emph {et~al.}(2019)\citenamefont {Rau}, \citenamefont {Moessner},\ and\ \citenamefont {McClarty}}]{rau2019magnon}%
  \BibitemOpen
  \bibfield  {author} {\bibinfo {author} {\bibfnamefont {J.~G.}\ \bibnamefont {Rau}}, \bibinfo {author} {\bibfnamefont {R.}~\bibnamefont {Moessner}},\ and\ \bibinfo {author} {\bibfnamefont {P.~A.}\ \bibnamefont {McClarty}},\ }\href {https://doi.org/10.1103/PhysRevB.100.104423} {\bibfield  {journal} {\bibinfo  {journal} {Phys. Rev. B}\ }\textbf {\bibinfo {volume} {100}},\ \bibinfo {pages} {104423} (\bibinfo {year} {2019})}\BibitemShut {NoStop}%
\bibitem [{\citenamefont {Jaubert}\ \emph {et~al.}(2015)\citenamefont {Jaubert}, \citenamefont {Benton}, \citenamefont {Rau}, \citenamefont {Oitmaa}, \citenamefont {Singh}, \citenamefont {Shannon},\ and\ \citenamefont {Gingras}}]{Jaubert2015}%
  \BibitemOpen
  \bibfield  {author} {\bibinfo {author} {\bibfnamefont {L.~D.~C.}\ \bibnamefont {Jaubert}}, \bibinfo {author} {\bibfnamefont {O.}~\bibnamefont {Benton}}, \bibinfo {author} {\bibfnamefont {J.~G.}\ \bibnamefont {Rau}}, \bibinfo {author} {\bibfnamefont {J.}~\bibnamefont {Oitmaa}}, \bibinfo {author} {\bibfnamefont {R.~R.~P.}\ \bibnamefont {Singh}}, \bibinfo {author} {\bibfnamefont {N.}~\bibnamefont {Shannon}},\ and\ \bibinfo {author} {\bibfnamefont {M.~J.~P.}\ \bibnamefont {Gingras}},\ }\href {https://doi.org/10.1103/PhysRevLett.115.267208} {\bibfield  {journal} {\bibinfo  {journal} {Phys. Rev. Lett.}\ }\textbf {\bibinfo {volume} {115}},\ \bibinfo {pages} {267208} (\bibinfo {year} {2015})}\BibitemShut {NoStop}%
\bibitem [{Sup()}]{SuppMat}%
  \BibitemOpen
  \href@noop {} {}\bibinfo {note} {See Supplemental Material at [URL will be inserted by publisher] for more details of the experiments and calculations.}\BibitemShut {Stop}%
\bibitem [{\citenamefont {Ehlers}\ \emph {et~al.}(2011)\citenamefont {Ehlers}, \citenamefont {Podlesnyak}, \citenamefont {Niedziela}, \citenamefont {Iverson},\ and\ \citenamefont {Sokol}}]{CNCS}%
  \BibitemOpen
  \bibfield  {author} {\bibinfo {author} {\bibfnamefont {G.}~\bibnamefont {Ehlers}}, \bibinfo {author} {\bibfnamefont {A.~A.}\ \bibnamefont {Podlesnyak}}, \bibinfo {author} {\bibfnamefont {J.~L.}\ \bibnamefont {Niedziela}}, \bibinfo {author} {\bibfnamefont {E.~B.}\ \bibnamefont {Iverson}},\ and\ \bibinfo {author} {\bibfnamefont {P.~E.}\ \bibnamefont {Sokol}},\ }\href {https://doi.org/10.1063/1.3626935} {\bibfield  {journal} {\bibinfo  {journal} {Review of Scientific Instruments}\ }\textbf {\bibinfo {volume} {82}},\ \bibinfo {pages} {085108} (\bibinfo {year} {2011})}\BibitemShut {NoStop}%
\bibitem [{\citenamefont {Curnoe}(2007)}]{Curnoe_2007}%
  \BibitemOpen
  \bibfield  {author} {\bibinfo {author} {\bibfnamefont {S.~H.}\ \bibnamefont {Curnoe}},\ }\href {https://doi.org/10.1103/PhysRevB.75.212404} {\bibfield  {journal} {\bibinfo  {journal} {Phys. Rev. B}\ }\textbf {\bibinfo {volume} {75}},\ \bibinfo {pages} {212404} (\bibinfo {year} {2007})}\BibitemShut {NoStop}%
\bibitem [{\citenamefont {Toth}\ and\ \citenamefont {Lake}(2015)}]{SpinW}%
  \BibitemOpen
  \bibfield  {author} {\bibinfo {author} {\bibfnamefont {S.}~\bibnamefont {Toth}}\ and\ \bibinfo {author} {\bibfnamefont {B.}~\bibnamefont {Lake}},\ }\href {https://doi.org/10.1088/0953-8984/27/16/166002} {\bibfield  {journal} {\bibinfo  {journal} {Journal of Physics: Condensed Matter}\ }\textbf {\bibinfo {volume} {27}},\ \bibinfo {pages} {166002} (\bibinfo {year} {2015})}\BibitemShut {NoStop}%
\bibitem [{\citenamefont {Zhitomirsky}\ and\ \citenamefont {Chernyshev}(2013)}]{zhito2013}%
  \BibitemOpen
  \bibfield  {author} {\bibinfo {author} {\bibfnamefont {M.~E.}\ \bibnamefont {Zhitomirsky}}\ and\ \bibinfo {author} {\bibfnamefont {A.~L.}\ \bibnamefont {Chernyshev}},\ }\href {https://doi.org/10.1103/RevModPhys.85.219} {\bibfield  {journal} {\bibinfo  {journal} {Rev. Mod. Phys.}\ }\textbf {\bibinfo {volume} {85}},\ \bibinfo {pages} {219} (\bibinfo {year} {2013})}\BibitemShut {NoStop}%
\bibitem [{\citenamefont {Mourigal}\ \emph {et~al.}(2013)\citenamefont {Mourigal}, \citenamefont {Fuhrman}, \citenamefont {Chernyshev},\ and\ \citenamefont {Zhitomirsky}}]{mourigal2013}%
  \BibitemOpen
  \bibfield  {author} {\bibinfo {author} {\bibfnamefont {M.}~\bibnamefont {Mourigal}}, \bibinfo {author} {\bibfnamefont {W.~T.}\ \bibnamefont {Fuhrman}}, \bibinfo {author} {\bibfnamefont {A.~L.}\ \bibnamefont {Chernyshev}},\ and\ \bibinfo {author} {\bibfnamefont {M.~E.}\ \bibnamefont {Zhitomirsky}},\ }\href {https://doi.org/10.1103/PhysRevB.88.094407} {\bibfield  {journal} {\bibinfo  {journal} {Phys. Rev. B}\ }\textbf {\bibinfo {volume} {88}},\ \bibinfo {pages} {094407} (\bibinfo {year} {2013})}\BibitemShut {NoStop}%
\bibitem [{\citenamefont {Legros}\ \emph {et~al.}(2021)\citenamefont {Legros}, \citenamefont {Zhang}, \citenamefont {Bai}, \citenamefont {Zhang}, \citenamefont {Dun}, \citenamefont {Phelan}, \citenamefont {Batista}, \citenamefont {Mourigal},\ and\ \citenamefont {Armitage}}]{Legros_2021}%
  \BibitemOpen
  \bibfield  {author} {\bibinfo {author} {\bibfnamefont {A.}~\bibnamefont {Legros}}, \bibinfo {author} {\bibfnamefont {S.-S.}\ \bibnamefont {Zhang}}, \bibinfo {author} {\bibfnamefont {X.}~\bibnamefont {Bai}}, \bibinfo {author} {\bibfnamefont {H.}~\bibnamefont {Zhang}}, \bibinfo {author} {\bibfnamefont {Z.}~\bibnamefont {Dun}}, \bibinfo {author} {\bibfnamefont {W.~A.}\ \bibnamefont {Phelan}}, \bibinfo {author} {\bibfnamefont {C.~D.}\ \bibnamefont {Batista}}, \bibinfo {author} {\bibfnamefont {M.}~\bibnamefont {Mourigal}},\ and\ \bibinfo {author} {\bibfnamefont {N.~P.}\ \bibnamefont {Armitage}},\ }\href {https://doi.org/10.1103/PhysRevLett.127.267201} {\bibfield  {journal} {\bibinfo  {journal} {Phys. Rev. Lett.}\ }\textbf {\bibinfo {volume} {127}},\ \bibinfo {pages} {267201} (\bibinfo {year} {2021})}\BibitemShut {NoStop}%
\bibitem [{\citenamefont {Zhang}\ \emph {et~al.}(2025)\citenamefont {Zhang}, \citenamefont {Huang}, \citenamefont {Scheie}, \citenamefont {Zhu}, \citenamefont {Xie}, \citenamefont {Murai}, \citenamefont {Ohira-Kawamura}, \citenamefont {Zheludev}, \citenamefont {L{\"a}uchli},\ and\ \citenamefont {Batista}}]{zhang2025nonperturbative}%
  \BibitemOpen
  \bibfield  {author} {\bibinfo {author} {\bibfnamefont {H.}~\bibnamefont {Zhang}}, \bibinfo {author} {\bibfnamefont {T.}~\bibnamefont {Huang}}, \bibinfo {author} {\bibfnamefont {A.~O.}\ \bibnamefont {Scheie}}, \bibinfo {author} {\bibfnamefont {M.}~\bibnamefont {Zhu}}, \bibinfo {author} {\bibfnamefont {T.}~\bibnamefont {Xie}}, \bibinfo {author} {\bibfnamefont {N.}~\bibnamefont {Murai}}, \bibinfo {author} {\bibfnamefont {S.}~\bibnamefont {Ohira-Kawamura}}, \bibinfo {author} {\bibfnamefont {A.}~\bibnamefont {Zheludev}}, \bibinfo {author} {\bibfnamefont {A.~M.}\ \bibnamefont {L{\"a}uchli}},\ and\ \bibinfo {author} {\bibfnamefont {C.~D.}\ \bibnamefont {Batista}},\ }\bibfield  {journal} {\bibinfo  {journal} {arXiv preprint arXiv:2508.21142}\ }\href {https://doi.org/10.48550/arXiv.2508.21142} {10.48550/arXiv.2508.21142} (\bibinfo {year} {2025})\BibitemShut {NoStop}%
\bibitem [{\citenamefont {Changlani}(2017)}]{Changlani2017quantum}%
  \BibitemOpen
  \bibfield  {author} {\bibinfo {author} {\bibfnamefont {H.~J.}\ \bibnamefont {Changlani}},\ }\bibfield  {journal} {\bibinfo  {journal} {arXiv preprint arXiv:1710.02234}\ }\href {https://doi.org/10.48550/arXiv.1710.02234} {10.48550/arXiv.1710.02234} (\bibinfo {year} {2017})\BibitemShut {NoStop}%
\bibitem [{\citenamefont {Virtanen}\ \emph {et~al.}(2020)\citenamefont {Virtanen}, \citenamefont {Gommers}, \citenamefont {Oliphant}, \citenamefont {Haberland}, \citenamefont {Reddy}, \citenamefont {Cournapeau}, \citenamefont {Burovski}, \citenamefont {Peterson}, \citenamefont {Weckesser}, \citenamefont {Bright} \emph {et~al.}}]{virtanen2020scipy}%
  \BibitemOpen
  \bibfield  {author} {\bibinfo {author} {\bibfnamefont {P.}~\bibnamefont {Virtanen}}, \bibinfo {author} {\bibfnamefont {R.}~\bibnamefont {Gommers}}, \bibinfo {author} {\bibfnamefont {T.~E.}\ \bibnamefont {Oliphant}}, \bibinfo {author} {\bibfnamefont {M.}~\bibnamefont {Haberland}}, \bibinfo {author} {\bibfnamefont {T.}~\bibnamefont {Reddy}}, \bibinfo {author} {\bibfnamefont {D.}~\bibnamefont {Cournapeau}}, \bibinfo {author} {\bibfnamefont {E.}~\bibnamefont {Burovski}}, \bibinfo {author} {\bibfnamefont {P.}~\bibnamefont {Peterson}}, \bibinfo {author} {\bibfnamefont {W.}~\bibnamefont {Weckesser}}, \bibinfo {author} {\bibfnamefont {J.}~\bibnamefont {Bright}}, \emph {et~al.},\ }\href {https://doi.org/10.1038/s41592-019-0686-2} {\bibfield  {journal} {\bibinfo  {journal} {Nature methods}\ }\textbf {\bibinfo {volume} {17}},\ \bibinfo {pages} {261} (\bibinfo {year} {2020})}\BibitemShut {NoStop}%
\bibitem [{\citenamefont {Gresista}\ \emph {et~al.}(2025)\citenamefont {Gresista}, \citenamefont {Lozano-G\'omez}, \citenamefont {Vojta}, \citenamefont {Trebst},\ and\ \citenamefont {Iqbal}}]{gresista2025quantum}%
  \BibitemOpen
  \bibfield  {author} {\bibinfo {author} {\bibfnamefont {L.}~\bibnamefont {Gresista}}, \bibinfo {author} {\bibfnamefont {D.}~\bibnamefont {Lozano-G\'omez}}, \bibinfo {author} {\bibfnamefont {M.}~\bibnamefont {Vojta}}, \bibinfo {author} {\bibfnamefont {S.}~\bibnamefont {Trebst}},\ and\ \bibinfo {author} {\bibfnamefont {Y.}~\bibnamefont {Iqbal}},\ }\href {https://doi.org/10.48550/arXiv.2503.03749} {\bibfield  {journal} {\bibinfo  {journal} {arXiv preprint arXiv:2503.03749}\ } (\bibinfo {year} {2025})}\BibitemShut {NoStop}%
\bibitem [{\citenamefont {Canals}\ and\ \citenamefont {Lacroix}(1998)}]{Canals_1998}%
  \BibitemOpen
  \bibfield  {author} {\bibinfo {author} {\bibfnamefont {B.}~\bibnamefont {Canals}}\ and\ \bibinfo {author} {\bibfnamefont {C.}~\bibnamefont {Lacroix}},\ }\href {https://doi.org/10.1103/PhysRevLett.80.2933} {\bibfield  {journal} {\bibinfo  {journal} {Phys. Rev. Lett.}\ }\textbf {\bibinfo {volume} {80}},\ \bibinfo {pages} {2933} (\bibinfo {year} {1998})}\BibitemShut {NoStop}%
\bibitem [{\citenamefont {M\"uller}\ \emph {et~al.}(2019)\citenamefont {M\"uller}, \citenamefont {Lohmann}, \citenamefont {Richter},\ and\ \citenamefont {Derzhko}}]{Muller_2019}%
  \BibitemOpen
  \bibfield  {author} {\bibinfo {author} {\bibfnamefont {P.}~\bibnamefont {M\"uller}}, \bibinfo {author} {\bibfnamefont {A.}~\bibnamefont {Lohmann}}, \bibinfo {author} {\bibfnamefont {J.}~\bibnamefont {Richter}},\ and\ \bibinfo {author} {\bibfnamefont {O.}~\bibnamefont {Derzhko}},\ }\href {https://doi.org/10.1103/PhysRevB.100.024424} {\bibfield  {journal} {\bibinfo  {journal} {Phys. Rev. B}\ }\textbf {\bibinfo {volume} {100}},\ \bibinfo {pages} {024424} (\bibinfo {year} {2019})}\BibitemShut {NoStop}%
\bibitem [{\citenamefont {Hagym\'asi}\ \emph {et~al.}(2021)\citenamefont {Hagym\'asi}, \citenamefont {Sch\"afer}, \citenamefont {Moessner},\ and\ \citenamefont {Luitz}}]{Hagymasi_2021}%
  \BibitemOpen
  \bibfield  {author} {\bibinfo {author} {\bibfnamefont {I.}~\bibnamefont {Hagym\'asi}}, \bibinfo {author} {\bibfnamefont {R.}~\bibnamefont {Sch\"afer}}, \bibinfo {author} {\bibfnamefont {R.}~\bibnamefont {Moessner}},\ and\ \bibinfo {author} {\bibfnamefont {D.~J.}\ \bibnamefont {Luitz}},\ }\href {https://doi.org/10.1103/PhysRevLett.126.117204} {\bibfield  {journal} {\bibinfo  {journal} {Phys. Rev. Lett.}\ }\textbf {\bibinfo {volume} {126}},\ \bibinfo {pages} {117204} (\bibinfo {year} {2021})}\BibitemShut {NoStop}%
\bibitem [{\citenamefont {Astrakhantsev}\ \emph {et~al.}(2021)\citenamefont {Astrakhantsev}, \citenamefont {Westerhout}, \citenamefont {Tiwari}, \citenamefont {Choo}, \citenamefont {Chen}, \citenamefont {Fischer}, \citenamefont {Carleo},\ and\ \citenamefont {Neupert}}]{Astrakhantsev_2021}%
  \BibitemOpen
  \bibfield  {author} {\bibinfo {author} {\bibfnamefont {N.}~\bibnamefont {Astrakhantsev}}, \bibinfo {author} {\bibfnamefont {T.}~\bibnamefont {Westerhout}}, \bibinfo {author} {\bibfnamefont {A.}~\bibnamefont {Tiwari}}, \bibinfo {author} {\bibfnamefont {K.}~\bibnamefont {Choo}}, \bibinfo {author} {\bibfnamefont {A.}~\bibnamefont {Chen}}, \bibinfo {author} {\bibfnamefont {M.~H.}\ \bibnamefont {Fischer}}, \bibinfo {author} {\bibfnamefont {G.}~\bibnamefont {Carleo}},\ and\ \bibinfo {author} {\bibfnamefont {T.}~\bibnamefont {Neupert}},\ }\href {https://doi.org/10.1103/PhysRevX.11.041021} {\bibfield  {journal} {\bibinfo  {journal} {Phys. Rev. X}\ }\textbf {\bibinfo {volume} {11}},\ \bibinfo {pages} {041021} (\bibinfo {year} {2021})}\BibitemShut {NoStop}%
\bibitem [{\citenamefont {Hering}\ \emph {et~al.}(2022)\citenamefont {Hering}, \citenamefont {Noculak}, \citenamefont {Ferrari}, \citenamefont {Iqbal},\ and\ \citenamefont {Reuther}}]{Hering_2022}%
  \BibitemOpen
  \bibfield  {author} {\bibinfo {author} {\bibfnamefont {M.}~\bibnamefont {Hering}}, \bibinfo {author} {\bibfnamefont {V.}~\bibnamefont {Noculak}}, \bibinfo {author} {\bibfnamefont {F.}~\bibnamefont {Ferrari}}, \bibinfo {author} {\bibfnamefont {Y.}~\bibnamefont {Iqbal}},\ and\ \bibinfo {author} {\bibfnamefont {J.}~\bibnamefont {Reuther}},\ }\href {https://doi.org/10.1103/PhysRevB.105.054426} {\bibfield  {journal} {\bibinfo  {journal} {Phys. Rev. B}\ }\textbf {\bibinfo {volume} {105}},\ \bibinfo {pages} {054426} (\bibinfo {year} {2022})}\BibitemShut {NoStop}%
\bibitem [{\citenamefont {Sch\"afer}\ \emph {et~al.}(2023)\citenamefont {Sch\"afer}, \citenamefont {Placke}, \citenamefont {Benton},\ and\ \citenamefont {Moessner}}]{Schafer_2023}%
  \BibitemOpen
  \bibfield  {author} {\bibinfo {author} {\bibfnamefont {R.}~\bibnamefont {Sch\"afer}}, \bibinfo {author} {\bibfnamefont {B.}~\bibnamefont {Placke}}, \bibinfo {author} {\bibfnamefont {O.}~\bibnamefont {Benton}},\ and\ \bibinfo {author} {\bibfnamefont {R.}~\bibnamefont {Moessner}},\ }\href {https://doi.org/10.1103/PhysRevLett.131.096702} {\bibfield  {journal} {\bibinfo  {journal} {Phys. Rev. Lett.}\ }\textbf {\bibinfo {volume} {131}},\ \bibinfo {pages} {096702} (\bibinfo {year} {2023})}\BibitemShut {NoStop}%
\bibitem [{\citenamefont {Pohle}\ \emph {et~al.}(2023)\citenamefont {Pohle}, \citenamefont {Yamaji},\ and\ \citenamefont {Imada}}]{pohle2023ground}%
  \BibitemOpen
  \bibfield  {author} {\bibinfo {author} {\bibfnamefont {R.}~\bibnamefont {Pohle}}, \bibinfo {author} {\bibfnamefont {Y.}~\bibnamefont {Yamaji}},\ and\ \bibinfo {author} {\bibfnamefont {M.}~\bibnamefont {Imada}},\ }\href {https://doi.org/10.48550/arXiv.2311.11561} {\bibfield  {journal} {\bibinfo  {journal} {arXiv preprint arXiv:2311.11561}\ } (\bibinfo {year} {2023})}\BibitemShut {NoStop}%
\bibitem [{\citenamefont {Rhim}\ and\ \citenamefont {and}(2021)}]{Rhim01012021}%
  \BibitemOpen
  \bibfield  {author} {\bibinfo {author} {\bibfnamefont {J.-W.}\ \bibnamefont {Rhim}}\ and\ \bibinfo {author} {\bibfnamefont {B.-J.~Y.}\ \bibnamefont {and}},\ }\href {https://doi.org/10.1080/23746149.2021.1901606} {\bibfield  {journal} {\bibinfo  {journal} {Advances in Physics: X}\ }\textbf {\bibinfo {volume} {6}},\ \bibinfo {pages} {1901606} (\bibinfo {year} {2021})}\BibitemShut {NoStop}%
\bibitem [{\citenamefont {Checkelsky}\ \emph {et~al.}(2024)\citenamefont {Checkelsky}, \citenamefont {Bernevig}, \citenamefont {Coleman}, \citenamefont {Si},\ and\ \citenamefont {Paschen}}]{checkelsky2024flat}%
  \BibitemOpen
  \bibfield  {author} {\bibinfo {author} {\bibfnamefont {J.~G.}\ \bibnamefont {Checkelsky}}, \bibinfo {author} {\bibfnamefont {B.~A.}\ \bibnamefont {Bernevig}}, \bibinfo {author} {\bibfnamefont {P.}~\bibnamefont {Coleman}}, \bibinfo {author} {\bibfnamefont {Q.}~\bibnamefont {Si}},\ and\ \bibinfo {author} {\bibfnamefont {S.}~\bibnamefont {Paschen}},\ }\href {https://doi.org/10.1038/s41578-023-00644-z} {\bibfield  {journal} {\bibinfo  {journal} {Nature Reviews Materials}\ }\textbf {\bibinfo {volume} {9}},\ \bibinfo {pages} {509} (\bibinfo {year} {2024})}\BibitemShut {NoStop}%
\bibitem [{\citenamefont {Benton}\ \emph {et~al.}(2016)\citenamefont {Benton}, \citenamefont {Jaubert}, \citenamefont {Yan},\ and\ \citenamefont {Shannon}}]{Benton2016}%
  \BibitemOpen
  \bibfield  {author} {\bibinfo {author} {\bibfnamefont {O.}~\bibnamefont {Benton}}, \bibinfo {author} {\bibfnamefont {L.~D.~C.}\ \bibnamefont {Jaubert}}, \bibinfo {author} {\bibfnamefont {H.}~\bibnamefont {Yan}},\ and\ \bibinfo {author} {\bibfnamefont {N.}~\bibnamefont {Shannon}},\ }\href {https://doi.org/10.1038/ncomms11572} {\bibfield  {journal} {\bibinfo  {journal} {Nature Communications}\ }\textbf {\bibinfo {volume} {7}},\ \bibinfo {pages} {11572} (\bibinfo {year} {2016})}\BibitemShut {NoStop}%
\bibitem [{\citenamefont {Francini}\ \emph {et~al.}(2025)\citenamefont {Francini}, \citenamefont {Janssen},\ and\ \citenamefont {Lozano-G\'omez}}]{Francini_2025}%
  \BibitemOpen
  \bibfield  {author} {\bibinfo {author} {\bibfnamefont {N.}~\bibnamefont {Francini}}, \bibinfo {author} {\bibfnamefont {L.}~\bibnamefont {Janssen}},\ and\ \bibinfo {author} {\bibfnamefont {D.}~\bibnamefont {Lozano-G\'omez}},\ }\href {https://doi.org/10.1103/PhysRevB.111.085140} {\bibfield  {journal} {\bibinfo  {journal} {Phys. Rev. B}\ }\textbf {\bibinfo {volume} {111}},\ \bibinfo {pages} {085140} (\bibinfo {year} {2025})}\BibitemShut {NoStop}%
\bibitem [{\citenamefont {Chung}(2024)}]{chung2024mapping}%
  \BibitemOpen
  \bibfield  {author} {\bibinfo {author} {\bibfnamefont {K.~T.~K.}\ \bibnamefont {Chung}},\ }\href {https://doi.org/10.48550/arXiv.2411.03429} {\bibfield  {journal} {\bibinfo  {journal} {arXiv preprint arXiv:2411.03429}\ } (\bibinfo {year} {2024})}\BibitemShut {NoStop}%
\bibitem [{\citenamefont {Zhu}\ and\ \citenamefont {White}(2015)}]{PhysRevB.92.041105}%
  \BibitemOpen
  \bibfield  {author} {\bibinfo {author} {\bibfnamefont {Z.}~\bibnamefont {Zhu}}\ and\ \bibinfo {author} {\bibfnamefont {S.~R.}\ \bibnamefont {White}},\ }\href {https://doi.org/10.1103/PhysRevB.92.041105} {\bibfield  {journal} {\bibinfo  {journal} {Phys. Rev. B}\ }\textbf {\bibinfo {volume} {92}},\ \bibinfo {pages} {041105} (\bibinfo {year} {2015})}\BibitemShut {NoStop}%
\bibitem [{\citenamefont {Hu}\ \emph {et~al.}(2015)\citenamefont {Hu}, \citenamefont {Gong}, \citenamefont {Zhu},\ and\ \citenamefont {Sheng}}]{PhysRevB.92.140403}%
  \BibitemOpen
  \bibfield  {author} {\bibinfo {author} {\bibfnamefont {W.-J.}\ \bibnamefont {Hu}}, \bibinfo {author} {\bibfnamefont {S.-S.}\ \bibnamefont {Gong}}, \bibinfo {author} {\bibfnamefont {W.}~\bibnamefont {Zhu}},\ and\ \bibinfo {author} {\bibfnamefont {D.~N.}\ \bibnamefont {Sheng}},\ }\href {https://doi.org/10.1103/PhysRevB.92.140403} {\bibfield  {journal} {\bibinfo  {journal} {Phys. Rev. B}\ }\textbf {\bibinfo {volume} {92}},\ \bibinfo {pages} {140403} (\bibinfo {year} {2015})}\BibitemShut {NoStop}%
\bibitem [{\citenamefont {Iqbal}\ \emph {et~al.}(2016)\citenamefont {Iqbal}, \citenamefont {Hu}, \citenamefont {Thomale}, \citenamefont {Poilblanc},\ and\ \citenamefont {Becca}}]{PhysRevB.93.144411}%
  \BibitemOpen
  \bibfield  {author} {\bibinfo {author} {\bibfnamefont {Y.}~\bibnamefont {Iqbal}}, \bibinfo {author} {\bibfnamefont {W.-J.}\ \bibnamefont {Hu}}, \bibinfo {author} {\bibfnamefont {R.}~\bibnamefont {Thomale}}, \bibinfo {author} {\bibfnamefont {D.}~\bibnamefont {Poilblanc}},\ and\ \bibinfo {author} {\bibfnamefont {F.}~\bibnamefont {Becca}},\ }\href {https://doi.org/10.1103/PhysRevB.93.144411} {\bibfield  {journal} {\bibinfo  {journal} {Phys. Rev. B}\ }\textbf {\bibinfo {volume} {93}},\ \bibinfo {pages} {144411} (\bibinfo {year} {2016})}\BibitemShut {NoStop}%
\bibitem [{\citenamefont {Saadatmand}\ and\ \citenamefont {McCulloch}(2016)}]{PhysRevB.94.121111}%
  \BibitemOpen
  \bibfield  {author} {\bibinfo {author} {\bibfnamefont {S.~N.}\ \bibnamefont {Saadatmand}}\ and\ \bibinfo {author} {\bibfnamefont {I.~P.}\ \bibnamefont {McCulloch}},\ }\href {https://doi.org/10.1103/PhysRevB.94.121111} {\bibfield  {journal} {\bibinfo  {journal} {Phys. Rev. B}\ }\textbf {\bibinfo {volume} {94}},\ \bibinfo {pages} {121111} (\bibinfo {year} {2016})}\BibitemShut {NoStop}%
\bibitem [{\citenamefont {Wietek}\ and\ \citenamefont {L\"auchli}(2017)}]{PhysRevB.95.035141}%
  \BibitemOpen
  \bibfield  {author} {\bibinfo {author} {\bibfnamefont {A.}~\bibnamefont {Wietek}}\ and\ \bibinfo {author} {\bibfnamefont {A.~M.}\ \bibnamefont {L\"auchli}},\ }\href {https://doi.org/10.1103/PhysRevB.95.035141} {\bibfield  {journal} {\bibinfo  {journal} {Phys. Rev. B}\ }\textbf {\bibinfo {volume} {95}},\ \bibinfo {pages} {035141} (\bibinfo {year} {2017})}\BibitemShut {NoStop}%
\bibitem [{\citenamefont {Gong}\ \emph {et~al.}(2017)\citenamefont {Gong}, \citenamefont {Zhu}, \citenamefont {Zhu}, \citenamefont {Sheng},\ and\ \citenamefont {Yang}}]{PhysRevB.96.075116}%
  \BibitemOpen
  \bibfield  {author} {\bibinfo {author} {\bibfnamefont {S.-S.}\ \bibnamefont {Gong}}, \bibinfo {author} {\bibfnamefont {W.}~\bibnamefont {Zhu}}, \bibinfo {author} {\bibfnamefont {J.-X.}\ \bibnamefont {Zhu}}, \bibinfo {author} {\bibfnamefont {D.~N.}\ \bibnamefont {Sheng}},\ and\ \bibinfo {author} {\bibfnamefont {K.}~\bibnamefont {Yang}},\ }\href {https://doi.org/10.1103/PhysRevB.96.075116} {\bibfield  {journal} {\bibinfo  {journal} {Phys. Rev. B}\ }\textbf {\bibinfo {volume} {96}},\ \bibinfo {pages} {075116} (\bibinfo {year} {2017})}\BibitemShut {NoStop}%
\bibitem [{\citenamefont {Hu}\ \emph {et~al.}(2019)\citenamefont {Hu}, \citenamefont {Zhu}, \citenamefont {Eggert},\ and\ \citenamefont {He}}]{PhysRevLett.123.207203}%
  \BibitemOpen
  \bibfield  {author} {\bibinfo {author} {\bibfnamefont {S.}~\bibnamefont {Hu}}, \bibinfo {author} {\bibfnamefont {W.}~\bibnamefont {Zhu}}, \bibinfo {author} {\bibfnamefont {S.}~\bibnamefont {Eggert}},\ and\ \bibinfo {author} {\bibfnamefont {Y.-C.}\ \bibnamefont {He}},\ }\href {https://doi.org/10.1103/PhysRevLett.123.207203} {\bibfield  {journal} {\bibinfo  {journal} {Phys. Rev. Lett.}\ }\textbf {\bibinfo {volume} {123}},\ \bibinfo {pages} {207203} (\bibinfo {year} {2019})}\BibitemShut {NoStop}%
\bibitem [{\citenamefont {Lozano-G\'omez}\ \emph {et~al.}(2024{\natexlab{a}})\citenamefont {Lozano-G\'omez}, \citenamefont {Noculak}, \citenamefont {Oitmaa}, \citenamefont {Singh}, \citenamefont {Iqbal}, \citenamefont {Reuther},\ and\ \citenamefont {Gingras}}]{LozanoGomez_2024}%
  \BibitemOpen
  \bibfield  {author} {\bibinfo {author} {\bibfnamefont {D.}~\bibnamefont {Lozano-G\'omez}}, \bibinfo {author} {\bibfnamefont {V.}~\bibnamefont {Noculak}}, \bibinfo {author} {\bibfnamefont {J.}~\bibnamefont {Oitmaa}}, \bibinfo {author} {\bibfnamefont {R.~R.~P.}\ \bibnamefont {Singh}}, \bibinfo {author} {\bibfnamefont {Y.}~\bibnamefont {Iqbal}}, \bibinfo {author} {\bibfnamefont {J.}~\bibnamefont {Reuther}},\ and\ \bibinfo {author} {\bibfnamefont {M.~J.~P.}\ \bibnamefont {Gingras}},\ }\href {https://doi.org/10.1073/pnas.2403487121} {\bibfield  {journal} {\bibinfo  {journal} {Proceedings of the National Academy of Sciences}\ }\textbf {\bibinfo {volume} {121}},\ \bibinfo {pages} {e2403487121} (\bibinfo {year} {2024}{\natexlab{a}})}\BibitemShut {NoStop}%
\bibitem [{\citenamefont {Fouet}\ \emph {et~al.}(2003)\citenamefont {Fouet}, \citenamefont {Mambrini}, \citenamefont {Sindzingre},\ and\ \citenamefont {Lhuillier}}]{Fouet_2003}%
  \BibitemOpen
  \bibfield  {author} {\bibinfo {author} {\bibfnamefont {J.-B.}\ \bibnamefont {Fouet}}, \bibinfo {author} {\bibfnamefont {M.}~\bibnamefont {Mambrini}}, \bibinfo {author} {\bibfnamefont {P.}~\bibnamefont {Sindzingre}},\ and\ \bibinfo {author} {\bibfnamefont {C.}~\bibnamefont {Lhuillier}},\ }\href {https://doi.org/10.1103/PhysRevB.67.054411} {\bibfield  {journal} {\bibinfo  {journal} {Phys. Rev. B}\ }\textbf {\bibinfo {volume} {67}},\ \bibinfo {pages} {054411} (\bibinfo {year} {2003})}\BibitemShut {NoStop}%
\bibitem [{\citenamefont {Bonville}\ \emph {et~al.}(2004)\citenamefont {Bonville}, \citenamefont {Hodges}, \citenamefont {Bertin}, \citenamefont {Bouchaud}, \citenamefont {Dalmas~de R{\'e}otier}, \citenamefont {Regnault}, \citenamefont {R{\o}nnow}, \citenamefont {Sanchez}, \citenamefont {Sosin},\ and\ \citenamefont {Yaouanc}}]{Bonville2004}%
  \BibitemOpen
  \bibfield  {author} {\bibinfo {author} {\bibfnamefont {P.}~\bibnamefont {Bonville}}, \bibinfo {author} {\bibfnamefont {J.~A.}\ \bibnamefont {Hodges}}, \bibinfo {author} {\bibfnamefont {E.}~\bibnamefont {Bertin}}, \bibinfo {author} {\bibfnamefont {J.-P.}\ \bibnamefont {Bouchaud}}, \bibinfo {author} {\bibfnamefont {P.}~\bibnamefont {Dalmas~de R{\'e}otier}}, \bibinfo {author} {\bibfnamefont {L.-P.}\ \bibnamefont {Regnault}}, \bibinfo {author} {\bibfnamefont {H.~M.}\ \bibnamefont {R{\o}nnow}}, \bibinfo {author} {\bibfnamefont {J.-P.}\ \bibnamefont {Sanchez}}, \bibinfo {author} {\bibfnamefont {S.}~\bibnamefont {Sosin}},\ and\ \bibinfo {author} {\bibfnamefont {A.}~\bibnamefont {Yaouanc}},\ }\href {https://doi.org/10.1023/B:HYPE.0000043235.21257.13} {\bibfield  {journal} {\bibinfo  {journal} {Hyperfine Interactions}\ }\textbf {\bibinfo {volume} {156}},\ \bibinfo {pages} {103} (\bibinfo {year} {2004})}\BibitemShut {NoStop}%
\bibitem [{\citenamefont {Thompson}\ \emph {et~al.}(2011)\citenamefont {Thompson}, \citenamefont {McClarty}, \citenamefont {R\o{}nnow}, \citenamefont {Regnault}, \citenamefont {Sorge},\ and\ \citenamefont {Gingras}}]{Thompson_2011}%
  \BibitemOpen
  \bibfield  {author} {\bibinfo {author} {\bibfnamefont {J.~D.}\ \bibnamefont {Thompson}}, \bibinfo {author} {\bibfnamefont {P.~A.}\ \bibnamefont {McClarty}}, \bibinfo {author} {\bibfnamefont {H.~M.}\ \bibnamefont {R\o{}nnow}}, \bibinfo {author} {\bibfnamefont {L.~P.}\ \bibnamefont {Regnault}}, \bibinfo {author} {\bibfnamefont {A.}~\bibnamefont {Sorge}},\ and\ \bibinfo {author} {\bibfnamefont {M.~J.~P.}\ \bibnamefont {Gingras}},\ }\href {https://doi.org/10.1103/PhysRevLett.106.187202} {\bibfield  {journal} {\bibinfo  {journal} {Phys. Rev. Lett.}\ }\textbf {\bibinfo {volume} {106}},\ \bibinfo {pages} {187202} (\bibinfo {year} {2011})}\BibitemShut {NoStop}%
\bibitem [{\citenamefont {Chern}\ and\ \citenamefont {Kim}(2019)}]{Chern2019}%
  \BibitemOpen
  \bibfield  {author} {\bibinfo {author} {\bibfnamefont {L.~E.}\ \bibnamefont {Chern}}\ and\ \bibinfo {author} {\bibfnamefont {Y.~B.}\ \bibnamefont {Kim}},\ }\href {https://doi.org/10.1038/s41598-019-47517-6} {\bibfield  {journal} {\bibinfo  {journal} {Scientific Reports}\ }\textbf {\bibinfo {volume} {9}},\ \bibinfo {pages} {10974} (\bibinfo {year} {2019})}\BibitemShut {NoStop}%
\bibitem [{\citenamefont {Sachdev}(2025)}]{sachdev2025foot}%
  \BibitemOpen
  \bibfield  {author} {\bibinfo {author} {\bibfnamefont {S.}~\bibnamefont {Sachdev}},\ }\href {https://doi.org/10.1016/j.physc.2025.1354707} {\bibfield  {journal} {\bibinfo  {journal} {Physica C: Superconductivity and its Applications}\ }\textbf {\bibinfo {volume} {633}},\ \bibinfo {pages} {1354707} (\bibinfo {year} {2025})}\BibitemShut {NoStop}%
\bibitem [{\citenamefont {Aynajian}\ \emph {et~al.}(2012)\citenamefont {Aynajian}, \citenamefont {da~Silva~Neto}, \citenamefont {Gyenis}, \citenamefont {Baumbach}, \citenamefont {Thompson}, \citenamefont {Fisk}, \citenamefont {Bauer},\ and\ \citenamefont {Yazdani}}]{aynajian2012visualizing}%
  \BibitemOpen
  \bibfield  {author} {\bibinfo {author} {\bibfnamefont {P.}~\bibnamefont {Aynajian}}, \bibinfo {author} {\bibfnamefont {E.~H.}\ \bibnamefont {da~Silva~Neto}}, \bibinfo {author} {\bibfnamefont {A.}~\bibnamefont {Gyenis}}, \bibinfo {author} {\bibfnamefont {R.~E.}\ \bibnamefont {Baumbach}}, \bibinfo {author} {\bibfnamefont {J.}~\bibnamefont {Thompson}}, \bibinfo {author} {\bibfnamefont {Z.}~\bibnamefont {Fisk}}, \bibinfo {author} {\bibfnamefont {E.~D.}\ \bibnamefont {Bauer}},\ and\ \bibinfo {author} {\bibfnamefont {A.}~\bibnamefont {Yazdani}},\ }\href@noop {} {\bibfield  {journal} {\bibinfo  {journal} {Nature}\ }\textbf {\bibinfo {volume} {486}},\ \bibinfo {pages} {201} (\bibinfo {year} {2012})}\BibitemShut {NoStop}%
\bibitem [{\citenamefont {Zhou}\ \emph {et~al.}(2013)\citenamefont {Zhou}, \citenamefont {Misra}, \citenamefont {da~Silva~Neto}, \citenamefont {Aynajian}, \citenamefont {Baumbach}, \citenamefont {Thompson}, \citenamefont {Bauer},\ and\ \citenamefont {Yazdani}}]{zhou2013visualizing}%
  \BibitemOpen
  \bibfield  {author} {\bibinfo {author} {\bibfnamefont {B.~B.}\ \bibnamefont {Zhou}}, \bibinfo {author} {\bibfnamefont {S.}~\bibnamefont {Misra}}, \bibinfo {author} {\bibfnamefont {E.~H.}\ \bibnamefont {da~Silva~Neto}}, \bibinfo {author} {\bibfnamefont {P.}~\bibnamefont {Aynajian}}, \bibinfo {author} {\bibfnamefont {R.~E.}\ \bibnamefont {Baumbach}}, \bibinfo {author} {\bibfnamefont {J.}~\bibnamefont {Thompson}}, \bibinfo {author} {\bibfnamefont {E.~D.}\ \bibnamefont {Bauer}},\ and\ \bibinfo {author} {\bibfnamefont {A.}~\bibnamefont {Yazdani}},\ }\href@noop {} {\bibfield  {journal} {\bibinfo  {journal} {Nature physics}\ }\textbf {\bibinfo {volume} {9}},\ \bibinfo {pages} {474} (\bibinfo {year} {2013})}\BibitemShut {NoStop}%
\bibitem [{\citenamefont {Stockert}\ \emph {et~al.}(1998)\citenamefont {Stockert}, \citenamefont {L\"ohneysen}, \citenamefont {Rosch}, \citenamefont {Pyka},\ and\ \citenamefont {Loewenhaupt}}]{Stockert_1998}%
  \BibitemOpen
  \bibfield  {author} {\bibinfo {author} {\bibfnamefont {O.}~\bibnamefont {Stockert}}, \bibinfo {author} {\bibfnamefont {H.~v.}\ \bibnamefont {L\"ohneysen}}, \bibinfo {author} {\bibfnamefont {A.}~\bibnamefont {Rosch}}, \bibinfo {author} {\bibfnamefont {N.}~\bibnamefont {Pyka}},\ and\ \bibinfo {author} {\bibfnamefont {M.}~\bibnamefont {Loewenhaupt}},\ }\href {https://doi.org/10.1103/PhysRevLett.80.5627} {\bibfield  {journal} {\bibinfo  {journal} {Phys. Rev. Lett.}\ }\textbf {\bibinfo {volume} {80}},\ \bibinfo {pages} {5627} (\bibinfo {year} {1998})}\BibitemShut {NoStop}%
\bibitem [{\citenamefont {Green}\ \emph {et~al.}(2018)\citenamefont {Green}, \citenamefont {Conduit},\ and\ \citenamefont {Krüger}}]{Green_2018_Review}%
  \BibitemOpen
  \bibfield  {author} {\bibinfo {author} {\bibfnamefont {A.~G.}\ \bibnamefont {Green}}, \bibinfo {author} {\bibfnamefont {G.}~\bibnamefont {Conduit}},\ and\ \bibinfo {author} {\bibfnamefont {F.}~\bibnamefont {Krüger}},\ }\href {https://doi.org/https://doi.org/10.1146/annurev-conmatphys-033117-053925} {\bibfield  {journal} {\bibinfo  {journal} {Annual Review of Condensed Matter Physics}\ }\textbf {\bibinfo {volume} {9}},\ \bibinfo {pages} {59} (\bibinfo {year} {2018})}\BibitemShut {NoStop}%
\bibitem [{\citenamefont {Raines}\ \emph {et~al.}(2024)\citenamefont {Raines}, \citenamefont {Glazman},\ and\ \citenamefont {Chubukov}}]{Raines_2024}%
  \BibitemOpen
  \bibfield  {author} {\bibinfo {author} {\bibfnamefont {Z.~M.}\ \bibnamefont {Raines}}, \bibinfo {author} {\bibfnamefont {L.~I.}\ \bibnamefont {Glazman}},\ and\ \bibinfo {author} {\bibfnamefont {A.~V.}\ \bibnamefont {Chubukov}},\ }\href {https://doi.org/10.1103/PhysRevLett.133.146501} {\bibfield  {journal} {\bibinfo  {journal} {Phys. Rev. Lett.}\ }\textbf {\bibinfo {volume} {133}},\ \bibinfo {pages} {146501} (\bibinfo {year} {2024})}\BibitemShut {NoStop}%
\bibitem [{\citenamefont {Hallas}\ \emph {et~al.}(2018)\citenamefont {Hallas}, \citenamefont {Gaudet},\ and\ \citenamefont {Gaulin}}]{hallas2018experimental}%
  \BibitemOpen
  \bibfield  {author} {\bibinfo {author} {\bibfnamefont {A.~M.}\ \bibnamefont {Hallas}}, \bibinfo {author} {\bibfnamefont {J.}~\bibnamefont {Gaudet}},\ and\ \bibinfo {author} {\bibfnamefont {B.~D.}\ \bibnamefont {Gaulin}},\ }\href {https://doi.org/https://doi.org/10.1146/annurev-conmatphys-031016-025218} {\bibfield  {journal} {\bibinfo  {journal} {Annual Review of Condensed Matter Physics}\ }\textbf {\bibinfo {volume} {9}},\ \bibinfo {pages} {105} (\bibinfo {year} {2018})}\BibitemShut {NoStop}%
\bibitem [{\citenamefont {Sarkis}\ \emph {et~al.}(2020)\citenamefont {Sarkis}, \citenamefont {Rau}, \citenamefont {Sanjeewa}, \citenamefont {Powell}, \citenamefont {Kolis}, \citenamefont {Marbey}, \citenamefont {Hill}, \citenamefont {Rodriguez-Rivera}, \citenamefont {Nair}, \citenamefont {Yahne}, \citenamefont {S\"aubert}, \citenamefont {Gingras},\ and\ \citenamefont {Ross}}]{Sarkis_2020}%
  \BibitemOpen
  \bibfield  {author} {\bibinfo {author} {\bibfnamefont {C.~L.}\ \bibnamefont {Sarkis}}, \bibinfo {author} {\bibfnamefont {J.~G.}\ \bibnamefont {Rau}}, \bibinfo {author} {\bibfnamefont {L.~D.}\ \bibnamefont {Sanjeewa}}, \bibinfo {author} {\bibfnamefont {M.}~\bibnamefont {Powell}}, \bibinfo {author} {\bibfnamefont {J.}~\bibnamefont {Kolis}}, \bibinfo {author} {\bibfnamefont {J.}~\bibnamefont {Marbey}}, \bibinfo {author} {\bibfnamefont {S.}~\bibnamefont {Hill}}, \bibinfo {author} {\bibfnamefont {J.~A.}\ \bibnamefont {Rodriguez-Rivera}}, \bibinfo {author} {\bibfnamefont {H.~S.}\ \bibnamefont {Nair}}, \bibinfo {author} {\bibfnamefont {D.~R.}\ \bibnamefont {Yahne}}, \bibinfo {author} {\bibfnamefont {S.}~\bibnamefont {S\"aubert}}, \bibinfo {author} {\bibfnamefont {M.~J.~P.}\ \bibnamefont {Gingras}},\ and\ \bibinfo {author} {\bibfnamefont {K.~A.}\ \bibnamefont {Ross}},\ }\href {https://doi.org/10.1103/PhysRevB.102.134418} {\bibfield  {journal} {\bibinfo  {journal} {Phys. Rev. B}\ }\textbf {\bibinfo {volume} {102}},\
  \bibinfo {pages} {134418} (\bibinfo {year} {2020})}\BibitemShut {NoStop}%
\bibitem [{\citenamefont {Wulferding}\ \emph {et~al.}(2023)\citenamefont {Wulferding}, \citenamefont {Kim}, \citenamefont {Kim}, \citenamefont {Yang}, \citenamefont {Lee}, \citenamefont {Song}, \citenamefont {Oh}, \citenamefont {Kim}, \citenamefont {Chern}, \citenamefont {Kim}, \citenamefont {Noh}, \citenamefont {Choi}, \citenamefont {Perkins}, \citenamefont {Kim},\ and\ \citenamefont {Park}}]{wulferding2023collective}%
  \BibitemOpen
  \bibfield  {author} {\bibinfo {author} {\bibfnamefont {D.}~\bibnamefont {Wulferding}}, \bibinfo {author} {\bibfnamefont {J.}~\bibnamefont {Kim}}, \bibinfo {author} {\bibfnamefont {M.~K.}\ \bibnamefont {Kim}}, \bibinfo {author} {\bibfnamefont {Y.}~\bibnamefont {Yang}}, \bibinfo {author} {\bibfnamefont {J.~H.}\ \bibnamefont {Lee}}, \bibinfo {author} {\bibfnamefont {D.}~\bibnamefont {Song}}, \bibinfo {author} {\bibfnamefont {D.}~\bibnamefont {Oh}}, \bibinfo {author} {\bibfnamefont {H.-S.}\ \bibnamefont {Kim}}, \bibinfo {author} {\bibfnamefont {L.~E.}\ \bibnamefont {Chern}}, \bibinfo {author} {\bibfnamefont {Y.~B.}\ \bibnamefont {Kim}}, \bibinfo {author} {\bibfnamefont {M.}~\bibnamefont {Noh}}, \bibinfo {author} {\bibfnamefont {H.}~\bibnamefont {Choi}}, \bibinfo {author} {\bibfnamefont {N.~B.}\ \bibnamefont {Perkins}}, \bibinfo {author} {\bibfnamefont {C.}~\bibnamefont {Kim}},\ and\ \bibinfo {author} {\bibfnamefont {S.~R.}\ \bibnamefont {Park}},\ }\href@noop {} {\bibfield  {journal} {\bibinfo  {journal} {npj
  Quantum Materials}\ }\textbf {\bibinfo {volume} {8}},\ \bibinfo {pages} {40} (\bibinfo {year} {2023})}\BibitemShut {NoStop}%
\bibitem [{\citenamefont {Xu}\ \emph {et~al.}(2025)\citenamefont {Xu}, \citenamefont {Yang}, \citenamefont {Teyssier}, \citenamefont {Ohtsuki}, \citenamefont {Qiu}, \citenamefont {Nakatsuji}, \citenamefont {van~der Marel}, \citenamefont {Perkins},\ and\ \citenamefont {Drichko}}]{xu2025ramification}%
  \BibitemOpen
  \bibfield  {author} {\bibinfo {author} {\bibfnamefont {Y.}~\bibnamefont {Xu}}, \bibinfo {author} {\bibfnamefont {Y.}~\bibnamefont {Yang}}, \bibinfo {author} {\bibfnamefont {J.}~\bibnamefont {Teyssier}}, \bibinfo {author} {\bibfnamefont {T.}~\bibnamefont {Ohtsuki}}, \bibinfo {author} {\bibfnamefont {Y.}~\bibnamefont {Qiu}}, \bibinfo {author} {\bibfnamefont {S.}~\bibnamefont {Nakatsuji}}, \bibinfo {author} {\bibfnamefont {D.}~\bibnamefont {van~der Marel}}, \bibinfo {author} {\bibfnamefont {N.~B.}\ \bibnamefont {Perkins}},\ and\ \bibinfo {author} {\bibfnamefont {N.}~\bibnamefont {Drichko}},\ }\href@noop {} {\bibfield  {journal} {\bibinfo  {journal} {npj Quantum Materials}\ }\textbf {\bibinfo {volume} {10}},\ \bibinfo {pages} {88} (\bibinfo {year} {2025})}\BibitemShut {NoStop}%
\bibitem [{\citenamefont {Gaudet}\ \emph {et~al.}(2019)\citenamefont {Gaudet}, \citenamefont {Smith}, \citenamefont {Dudemaine}, \citenamefont {Beare}, \citenamefont {Buhariwalla}, \citenamefont {Butch}, \citenamefont {Stone}, \citenamefont {Kolesnikov}, \citenamefont {Xu}, \citenamefont {Yahne}, \citenamefont {Ross}, \citenamefont {Marjerrison}, \citenamefont {Garrett}, \citenamefont {Luke}, \citenamefont {Bianchi},\ and\ \citenamefont {Gaulin}}]{Gaudet_CZO_2019}%
  \BibitemOpen
  \bibfield  {author} {\bibinfo {author} {\bibfnamefont {J.}~\bibnamefont {Gaudet}}, \bibinfo {author} {\bibfnamefont {E.~M.}\ \bibnamefont {Smith}}, \bibinfo {author} {\bibfnamefont {J.}~\bibnamefont {Dudemaine}}, \bibinfo {author} {\bibfnamefont {J.}~\bibnamefont {Beare}}, \bibinfo {author} {\bibfnamefont {C.~R.~C.}\ \bibnamefont {Buhariwalla}}, \bibinfo {author} {\bibfnamefont {N.~P.}\ \bibnamefont {Butch}}, \bibinfo {author} {\bibfnamefont {M.~B.}\ \bibnamefont {Stone}}, \bibinfo {author} {\bibfnamefont {A.~I.}\ \bibnamefont {Kolesnikov}}, \bibinfo {author} {\bibfnamefont {G.}~\bibnamefont {Xu}}, \bibinfo {author} {\bibfnamefont {D.~R.}\ \bibnamefont {Yahne}}, \bibinfo {author} {\bibfnamefont {K.~A.}\ \bibnamefont {Ross}}, \bibinfo {author} {\bibfnamefont {C.~A.}\ \bibnamefont {Marjerrison}}, \bibinfo {author} {\bibfnamefont {J.~D.}\ \bibnamefont {Garrett}}, \bibinfo {author} {\bibfnamefont {G.~M.}\ \bibnamefont {Luke}}, \bibinfo {author} {\bibfnamefont {A.~D.}\ \bibnamefont {Bianchi}},\ and\ \bibinfo
  {author} {\bibfnamefont {B.~D.}\ \bibnamefont {Gaulin}},\ }\href {https://doi.org/10.1103/PhysRevLett.122.187201} {\bibfield  {journal} {\bibinfo  {journal} {Phys. Rev. Lett.}\ }\textbf {\bibinfo {volume} {122}},\ \bibinfo {pages} {187201} (\bibinfo {year} {2019})}\BibitemShut {NoStop}%
\bibitem [{\citenamefont {Gao}\ \emph {et~al.}(2019)\citenamefont {Gao}, \citenamefont {Chen}, \citenamefont {Tam}, \citenamefont {Huang}, \citenamefont {Sasmal}, \citenamefont {Adroja}, \citenamefont {Ye}, \citenamefont {Cao}, \citenamefont {Sala}, \citenamefont {Stone}, \citenamefont {Baines}, \citenamefont {Verezhak}, \citenamefont {Hu}, \citenamefont {Chung}, \citenamefont {Xu}, \citenamefont {Cheong}, \citenamefont {Nallaiyan}, \citenamefont {Spagna}, \citenamefont {Maple}, \citenamefont {Nevidomskyy}, \citenamefont {Morosan}, \citenamefont {Chen},\ and\ \citenamefont {Dai}}]{Gao_CZO_2019}%
  \BibitemOpen
  \bibfield  {author} {\bibinfo {author} {\bibfnamefont {B.}~\bibnamefont {Gao}}, \bibinfo {author} {\bibfnamefont {T.}~\bibnamefont {Chen}}, \bibinfo {author} {\bibfnamefont {D.~W.}\ \bibnamefont {Tam}}, \bibinfo {author} {\bibfnamefont {C.-L.}\ \bibnamefont {Huang}}, \bibinfo {author} {\bibfnamefont {K.}~\bibnamefont {Sasmal}}, \bibinfo {author} {\bibfnamefont {D.~T.}\ \bibnamefont {Adroja}}, \bibinfo {author} {\bibfnamefont {F.}~\bibnamefont {Ye}}, \bibinfo {author} {\bibfnamefont {H.}~\bibnamefont {Cao}}, \bibinfo {author} {\bibfnamefont {G.}~\bibnamefont {Sala}}, \bibinfo {author} {\bibfnamefont {M.~B.}\ \bibnamefont {Stone}}, \bibinfo {author} {\bibfnamefont {C.}~\bibnamefont {Baines}}, \bibinfo {author} {\bibfnamefont {J.~A.~T.}\ \bibnamefont {Verezhak}}, \bibinfo {author} {\bibfnamefont {H.}~\bibnamefont {Hu}}, \bibinfo {author} {\bibfnamefont {J.-H.}\ \bibnamefont {Chung}}, \bibinfo {author} {\bibfnamefont {X.}~\bibnamefont {Xu}}, \bibinfo {author} {\bibfnamefont {S.-W.}\ \bibnamefont {Cheong}},
  \bibinfo {author} {\bibfnamefont {M.}~\bibnamefont {Nallaiyan}}, \bibinfo {author} {\bibfnamefont {S.}~\bibnamefont {Spagna}}, \bibinfo {author} {\bibfnamefont {M.~B.}\ \bibnamefont {Maple}}, \bibinfo {author} {\bibfnamefont {A.~H.}\ \bibnamefont {Nevidomskyy}}, \bibinfo {author} {\bibfnamefont {E.}~\bibnamefont {Morosan}}, \bibinfo {author} {\bibfnamefont {G.}~\bibnamefont {Chen}},\ and\ \bibinfo {author} {\bibfnamefont {P.}~\bibnamefont {Dai}},\ }\href {https://doi.org/10.1038/s41567-019-0577-6} {\bibfield  {journal} {\bibinfo  {journal} {Nature Physics}\ }\textbf {\bibinfo {volume} {15}},\ \bibinfo {pages} {1052} (\bibinfo {year} {2019})}\BibitemShut {NoStop}%
\bibitem [{\citenamefont {Smith}\ \emph {et~al.}(2022)\citenamefont {Smith}, \citenamefont {Benton}, \citenamefont {Yahne}, \citenamefont {Placke}, \citenamefont {Sch\"afer}, \citenamefont {Gaudet}, \citenamefont {Dudemaine}, \citenamefont {Fitterman}, \citenamefont {Beare}, \citenamefont {Wildes}, \citenamefont {Bhattacharya}, \citenamefont {DeLazzer}, \citenamefont {Buhariwalla}, \citenamefont {Butch}, \citenamefont {Movshovich}, \citenamefont {Garrett}, \citenamefont {Marjerrison}, \citenamefont {Clancy}, \citenamefont {Kermarrec}, \citenamefont {Luke}, \citenamefont {Bianchi}, \citenamefont {Ross},\ and\ \citenamefont {Gaulin}}]{Smith_CZO_2022}%
  \BibitemOpen
  \bibfield  {author} {\bibinfo {author} {\bibfnamefont {E.~M.}\ \bibnamefont {Smith}}, \bibinfo {author} {\bibfnamefont {O.}~\bibnamefont {Benton}}, \bibinfo {author} {\bibfnamefont {D.~R.}\ \bibnamefont {Yahne}}, \bibinfo {author} {\bibfnamefont {B.}~\bibnamefont {Placke}}, \bibinfo {author} {\bibfnamefont {R.}~\bibnamefont {Sch\"afer}}, \bibinfo {author} {\bibfnamefont {J.}~\bibnamefont {Gaudet}}, \bibinfo {author} {\bibfnamefont {J.}~\bibnamefont {Dudemaine}}, \bibinfo {author} {\bibfnamefont {A.}~\bibnamefont {Fitterman}}, \bibinfo {author} {\bibfnamefont {J.}~\bibnamefont {Beare}}, \bibinfo {author} {\bibfnamefont {A.~R.}\ \bibnamefont {Wildes}}, \bibinfo {author} {\bibfnamefont {S.}~\bibnamefont {Bhattacharya}}, \bibinfo {author} {\bibfnamefont {T.}~\bibnamefont {DeLazzer}}, \bibinfo {author} {\bibfnamefont {C.~R.~C.}\ \bibnamefont {Buhariwalla}}, \bibinfo {author} {\bibfnamefont {N.~P.}\ \bibnamefont {Butch}}, \bibinfo {author} {\bibfnamefont {R.}~\bibnamefont {Movshovich}}, \bibinfo {author}
  {\bibfnamefont {J.~D.}\ \bibnamefont {Garrett}}, \bibinfo {author} {\bibfnamefont {C.~A.}\ \bibnamefont {Marjerrison}}, \bibinfo {author} {\bibfnamefont {J.~P.}\ \bibnamefont {Clancy}}, \bibinfo {author} {\bibfnamefont {E.}~\bibnamefont {Kermarrec}}, \bibinfo {author} {\bibfnamefont {G.~M.}\ \bibnamefont {Luke}}, \bibinfo {author} {\bibfnamefont {A.~D.}\ \bibnamefont {Bianchi}}, \bibinfo {author} {\bibfnamefont {K.~A.}\ \bibnamefont {Ross}},\ and\ \bibinfo {author} {\bibfnamefont {B.~D.}\ \bibnamefont {Gaulin}},\ }\href {https://doi.org/10.1103/PhysRevX.12.021015} {\bibfield  {journal} {\bibinfo  {journal} {Phys. Rev. X}\ }\textbf {\bibinfo {volume} {12}},\ \bibinfo {pages} {021015} (\bibinfo {year} {2022})}\BibitemShut {NoStop}%
\bibitem [{\citenamefont {Bhardwaj}\ \emph {et~al.}(2022)\citenamefont {Bhardwaj}, \citenamefont {Zhang}, \citenamefont {Yan}, \citenamefont {Moessner}, \citenamefont {Nevidomskyy},\ and\ \citenamefont {Changlani}}]{Bhardwaj_CZO_2022}%
  \BibitemOpen
  \bibfield  {author} {\bibinfo {author} {\bibfnamefont {A.}~\bibnamefont {Bhardwaj}}, \bibinfo {author} {\bibfnamefont {S.}~\bibnamefont {Zhang}}, \bibinfo {author} {\bibfnamefont {H.}~\bibnamefont {Yan}}, \bibinfo {author} {\bibfnamefont {R.}~\bibnamefont {Moessner}}, \bibinfo {author} {\bibfnamefont {A.~H.}\ \bibnamefont {Nevidomskyy}},\ and\ \bibinfo {author} {\bibfnamefont {H.~J.}\ \bibnamefont {Changlani}},\ }\href {https://doi.org/10.1038/s41535-022-00458-2} {\bibfield  {journal} {\bibinfo  {journal} {npj Quantum Materials}\ }\textbf {\bibinfo {volume} {7}},\ \bibinfo {pages} {51} (\bibinfo {year} {2022})}\BibitemShut {NoStop}%
\bibitem [{\citenamefont {Hosoi}\ \emph {et~al.}(2022)\citenamefont {Hosoi}, \citenamefont {Zhang}, \citenamefont {Patri},\ and\ \citenamefont {Kim}}]{Hosoi_PRL_2022}%
  \BibitemOpen
  \bibfield  {author} {\bibinfo {author} {\bibfnamefont {M.}~\bibnamefont {Hosoi}}, \bibinfo {author} {\bibfnamefont {E.~Z.}\ \bibnamefont {Zhang}}, \bibinfo {author} {\bibfnamefont {A.~S.}\ \bibnamefont {Patri}},\ and\ \bibinfo {author} {\bibfnamefont {Y.~B.}\ \bibnamefont {Kim}},\ }\href {https://doi.org/10.1103/PhysRevLett.129.097202} {\bibfield  {journal} {\bibinfo  {journal} {Phys. Rev. Lett.}\ }\textbf {\bibinfo {volume} {129}},\ \bibinfo {pages} {097202} (\bibinfo {year} {2022})}\BibitemShut {NoStop}%
\bibitem [{\citenamefont {Por{\'e}e}\ \emph {et~al.}(2025)\citenamefont {Por{\'e}e}, \citenamefont {Yan}, \citenamefont {Desrochers}, \citenamefont {Petit}, \citenamefont {Lhotel}, \citenamefont {Appel}, \citenamefont {Ollivier}, \citenamefont {Kim}, \citenamefont {Nevidomskyy},\ and\ \citenamefont {Sibille}}]{Poree_2025}%
  \BibitemOpen
  \bibfield  {author} {\bibinfo {author} {\bibfnamefont {V.}~\bibnamefont {Por{\'e}e}}, \bibinfo {author} {\bibfnamefont {H.}~\bibnamefont {Yan}}, \bibinfo {author} {\bibfnamefont {F.}~\bibnamefont {Desrochers}}, \bibinfo {author} {\bibfnamefont {S.}~\bibnamefont {Petit}}, \bibinfo {author} {\bibfnamefont {E.}~\bibnamefont {Lhotel}}, \bibinfo {author} {\bibfnamefont {M.}~\bibnamefont {Appel}}, \bibinfo {author} {\bibfnamefont {J.}~\bibnamefont {Ollivier}}, \bibinfo {author} {\bibfnamefont {Y.~B.}\ \bibnamefont {Kim}}, \bibinfo {author} {\bibfnamefont {A.~H.}\ \bibnamefont {Nevidomskyy}},\ and\ \bibinfo {author} {\bibfnamefont {R.}~\bibnamefont {Sibille}},\ }\href {https://doi.org/10.1038/s41567-024-02711-w} {\bibfield  {journal} {\bibinfo  {journal} {Nature Physics}\ }\textbf {\bibinfo {volume} {21}},\ \bibinfo {pages} {83} (\bibinfo {year} {2025})}\BibitemShut {NoStop}%
\bibitem [{\citenamefont {Por\'ee}\ \emph {et~al.}(2025)\citenamefont {Por\'ee}, \citenamefont {Bhardwaj}, \citenamefont {Lhotel}, \citenamefont {Petit}, \citenamefont {Gauthier}, \citenamefont {Yan}, \citenamefont {Pomjakushin}, \citenamefont {Ollivier}, \citenamefont {Quilliam}, \citenamefont {Nevidomskyy}, \citenamefont {Changlani},\ and\ \citenamefont {Sibille}}]{Poree_CHO_2023}%
  \BibitemOpen
  \bibfield  {author} {\bibinfo {author} {\bibfnamefont {V.}~\bibnamefont {Por\'ee}}, \bibinfo {author} {\bibfnamefont {A.}~\bibnamefont {Bhardwaj}}, \bibinfo {author} {\bibfnamefont {E.}~\bibnamefont {Lhotel}}, \bibinfo {author} {\bibfnamefont {S.}~\bibnamefont {Petit}}, \bibinfo {author} {\bibfnamefont {N.}~\bibnamefont {Gauthier}}, \bibinfo {author} {\bibfnamefont {H.}~\bibnamefont {Yan}}, \bibinfo {author} {\bibfnamefont {V.}~\bibnamefont {Pomjakushin}}, \bibinfo {author} {\bibfnamefont {J.}~\bibnamefont {Ollivier}}, \bibinfo {author} {\bibfnamefont {J.~A.}\ \bibnamefont {Quilliam}}, \bibinfo {author} {\bibfnamefont {A.~H.}\ \bibnamefont {Nevidomskyy}}, \bibinfo {author} {\bibfnamefont {H.~J.}\ \bibnamefont {Changlani}},\ and\ \bibinfo {author} {\bibfnamefont {R.}~\bibnamefont {Sibille}},\ }\href {https://doi.org/10.1103/j451-ztvr} {\bibfield  {journal} {\bibinfo  {journal} {Phys. Rev. B}\ }\textbf {\bibinfo {volume} {112}},\ \bibinfo {pages} {L180404} (\bibinfo {year} {2025})}\BibitemShut {NoStop}%
\bibitem [{\citenamefont {Bhardwaj}\ \emph {et~al.}(2025)\citenamefont {Bhardwaj}, \citenamefont {Por\'ee}, \citenamefont {Yan}, \citenamefont {Gauthier}, \citenamefont {Lhotel}, \citenamefont {Petit}, \citenamefont {Quilliam}, \citenamefont {Nevidomskyy}, \citenamefont {Sibille},\ and\ \citenamefont {Changlani}}]{Bhardwaj_CHO_2025}%
  \BibitemOpen
  \bibfield  {author} {\bibinfo {author} {\bibfnamefont {A.}~\bibnamefont {Bhardwaj}}, \bibinfo {author} {\bibfnamefont {V.}~\bibnamefont {Por\'ee}}, \bibinfo {author} {\bibfnamefont {H.}~\bibnamefont {Yan}}, \bibinfo {author} {\bibfnamefont {N.}~\bibnamefont {Gauthier}}, \bibinfo {author} {\bibfnamefont {E.}~\bibnamefont {Lhotel}}, \bibinfo {author} {\bibfnamefont {S.}~\bibnamefont {Petit}}, \bibinfo {author} {\bibfnamefont {J.~A.}\ \bibnamefont {Quilliam}}, \bibinfo {author} {\bibfnamefont {A.~H.}\ \bibnamefont {Nevidomskyy}}, \bibinfo {author} {\bibfnamefont {R.}~\bibnamefont {Sibille}},\ and\ \bibinfo {author} {\bibfnamefont {H.~J.}\ \bibnamefont {Changlani}},\ }\href {https://doi.org/10.1103/PhysRevB.111.155137} {\bibfield  {journal} {\bibinfo  {journal} {Phys. Rev. B}\ }\textbf {\bibinfo {volume} {111}},\ \bibinfo {pages} {155137} (\bibinfo {year} {2025})}\BibitemShut {NoStop}%
\bibitem [{\citenamefont {B{\"a}rtschi}\ \emph {et~al.}(2024)\citenamefont {B{\"a}rtschi}, \citenamefont {Caravelli}, \citenamefont {Coffrin}, \citenamefont {Colina}, \citenamefont {Eidenbenz}, \citenamefont {Jayakumar}, \citenamefont {Lawrence}, \citenamefont {Lee}, \citenamefont {Lokhov}, \citenamefont {Mishra} \emph {et~al.}}]{bartschi2024potential}%
  \BibitemOpen
  \bibfield  {author} {\bibinfo {author} {\bibfnamefont {A.}~\bibnamefont {B{\"a}rtschi}}, \bibinfo {author} {\bibfnamefont {F.}~\bibnamefont {Caravelli}}, \bibinfo {author} {\bibfnamefont {C.}~\bibnamefont {Coffrin}}, \bibinfo {author} {\bibfnamefont {J.}~\bibnamefont {Colina}}, \bibinfo {author} {\bibfnamefont {S.}~\bibnamefont {Eidenbenz}}, \bibinfo {author} {\bibfnamefont {A.}~\bibnamefont {Jayakumar}}, \bibinfo {author} {\bibfnamefont {S.}~\bibnamefont {Lawrence}}, \bibinfo {author} {\bibfnamefont {M.}~\bibnamefont {Lee}}, \bibinfo {author} {\bibfnamefont {A.~Y.}\ \bibnamefont {Lokhov}}, \bibinfo {author} {\bibfnamefont {A.}~\bibnamefont {Mishra}}, \emph {et~al.},\ }\href {https://doi.org/10.48550/arXiv.2406.06625} {\bibfield  {journal} {\bibinfo  {journal} {arXiv preprint arXiv:2406.06625}\ } (\bibinfo {year} {2024})}\BibitemShut {NoStop}%
\bibitem [{\citenamefont {Laurell}\ \emph {et~al.}(2025)\citenamefont {Laurell}, \citenamefont {Scheie}, \citenamefont {Dagotto},\ and\ \citenamefont {Tennant}}]{laurell2024witnessing}%
  \BibitemOpen
  \bibfield  {author} {\bibinfo {author} {\bibfnamefont {P.}~\bibnamefont {Laurell}}, \bibinfo {author} {\bibfnamefont {A.}~\bibnamefont {Scheie}}, \bibinfo {author} {\bibfnamefont {E.}~\bibnamefont {Dagotto}},\ and\ \bibinfo {author} {\bibfnamefont {D.~A.}\ \bibnamefont {Tennant}},\ }\href {https://doi.org/10.1002/qute.202400196} {\bibfield  {journal} {\bibinfo  {journal} {Advanced Quantum Technologies}\ }\textbf {\bibinfo {volume} {8}},\ \bibinfo {pages} {2400196} (\bibinfo {year} {2025})}\BibitemShut {NoStop}%
\bibitem [{\citenamefont {Scheie}\ \emph {et~al.}(2024)\citenamefont {Scheie}, \citenamefont {Laurell}, \citenamefont {Simeth}, \citenamefont {Dagotto},\ and\ \citenamefont {Tennant}}]{scheie2024tutorial}%
  \BibitemOpen
  \bibfield  {author} {\bibinfo {author} {\bibfnamefont {A.}~\bibnamefont {Scheie}}, \bibinfo {author} {\bibfnamefont {P.}~\bibnamefont {Laurell}}, \bibinfo {author} {\bibfnamefont {W.}~\bibnamefont {Simeth}}, \bibinfo {author} {\bibfnamefont {E.}~\bibnamefont {Dagotto}},\ and\ \bibinfo {author} {\bibfnamefont {D.~A.}\ \bibnamefont {Tennant}},\ }\href {https://doi.org/10.1016/j.mtquan.2024.100020} {\bibfield  {journal} {\bibinfo  {journal} {Materials Today Quantum}\ ,\ \bibinfo {pages} {100020}} (\bibinfo {year} {2024})}\BibitemShut {NoStop}%
\bibitem [{\citenamefont {Hauke}\ \emph {et~al.}(2016)\citenamefont {Hauke}, \citenamefont {Heyl}, \citenamefont {Tagliacozzo},\ and\ \citenamefont {Zoller}}]{Hauke2016}%
  \BibitemOpen
  \bibfield  {author} {\bibinfo {author} {\bibfnamefont {P.}~\bibnamefont {Hauke}}, \bibinfo {author} {\bibfnamefont {M.}~\bibnamefont {Heyl}}, \bibinfo {author} {\bibfnamefont {L.}~\bibnamefont {Tagliacozzo}},\ and\ \bibinfo {author} {\bibfnamefont {P.}~\bibnamefont {Zoller}},\ }\href {https://doi.org/10.1038/nphys3700} {\bibfield  {journal} {\bibinfo  {journal} {Nature Physics}\ }\textbf {\bibinfo {volume} {12}},\ \bibinfo {pages} {778} (\bibinfo {year} {2016})}\BibitemShut {NoStop}%
\bibitem [{\citenamefont {Scheie}\ \emph {et~al.}(2021)\citenamefont {Scheie}, \citenamefont {Laurell}, \citenamefont {Samarakoon}, \citenamefont {Lake}, \citenamefont {Nagler}, \citenamefont {Granroth}, \citenamefont {Okamoto}, \citenamefont {Alvarez},\ and\ \citenamefont {Tennant}}]{PhysRevB.103.224434}%
  \BibitemOpen
  \bibfield  {author} {\bibinfo {author} {\bibfnamefont {A.}~\bibnamefont {Scheie}}, \bibinfo {author} {\bibfnamefont {P.}~\bibnamefont {Laurell}}, \bibinfo {author} {\bibfnamefont {A.~M.}\ \bibnamefont {Samarakoon}}, \bibinfo {author} {\bibfnamefont {B.}~\bibnamefont {Lake}}, \bibinfo {author} {\bibfnamefont {S.~E.}\ \bibnamefont {Nagler}}, \bibinfo {author} {\bibfnamefont {G.~E.}\ \bibnamefont {Granroth}}, \bibinfo {author} {\bibfnamefont {S.}~\bibnamefont {Okamoto}}, \bibinfo {author} {\bibfnamefont {G.}~\bibnamefont {Alvarez}},\ and\ \bibinfo {author} {\bibfnamefont {D.~A.}\ \bibnamefont {Tennant}},\ }\href {https://doi.org/10.1103/PhysRevB.103.224434} {\bibfield  {journal} {\bibinfo  {journal} {Phys. Rev. B}\ }\textbf {\bibinfo {volume} {103}},\ \bibinfo {pages} {224434} (\bibinfo {year} {2021})}\BibitemShut {NoStop}%
\bibitem [{\citenamefont {Xu}\ \emph {et~al.}(2013)\citenamefont {Xu}, \citenamefont {Xu},\ and\ \citenamefont {Tranquada}}]{Xu_AbsUnits}%
  \BibitemOpen
  \bibfield  {author} {\bibinfo {author} {\bibfnamefont {G.}~\bibnamefont {Xu}}, \bibinfo {author} {\bibfnamefont {Z.}~\bibnamefont {Xu}},\ and\ \bibinfo {author} {\bibfnamefont {J.~M.}\ \bibnamefont {Tranquada}},\ }\href {https://doi.org/10.1063/1.4818323} {\bibfield  {journal} {\bibinfo  {journal} {Review of Scientific Instruments}\ }\textbf {\bibinfo {volume} {84}},\ \bibinfo {pages} {083906} (\bibinfo {year} {2013})}\BibitemShut {NoStop}%
\bibitem [{\citenamefont {Laurell}\ \emph {et~al.}(2021)\citenamefont {Laurell}, \citenamefont {Scheie}, \citenamefont {Mukherjee}, \citenamefont {Koza}, \citenamefont {Enderle}, \citenamefont {Tylczynski}, \citenamefont {Okamoto}, \citenamefont {Coldea}, \citenamefont {Tennant},\ and\ \citenamefont {Alvarez}}]{Laurell2021Quantifying}%
  \BibitemOpen
  \bibfield  {author} {\bibinfo {author} {\bibfnamefont {P.}~\bibnamefont {Laurell}}, \bibinfo {author} {\bibfnamefont {A.}~\bibnamefont {Scheie}}, \bibinfo {author} {\bibfnamefont {C.~J.}\ \bibnamefont {Mukherjee}}, \bibinfo {author} {\bibfnamefont {M.~M.}\ \bibnamefont {Koza}}, \bibinfo {author} {\bibfnamefont {M.}~\bibnamefont {Enderle}}, \bibinfo {author} {\bibfnamefont {Z.}~\bibnamefont {Tylczynski}}, \bibinfo {author} {\bibfnamefont {S.}~\bibnamefont {Okamoto}}, \bibinfo {author} {\bibfnamefont {R.}~\bibnamefont {Coldea}}, \bibinfo {author} {\bibfnamefont {D.~A.}\ \bibnamefont {Tennant}},\ and\ \bibinfo {author} {\bibfnamefont {G.}~\bibnamefont {Alvarez}},\ }\href {https://doi.org/10.1103/PhysRevLett.127.037201} {\bibfield  {journal} {\bibinfo  {journal} {Phys. Rev. Lett.}\ }\textbf {\bibinfo {volume} {127}},\ \bibinfo {pages} {037201} (\bibinfo {year} {2021})}\BibitemShut {NoStop}%
\bibitem [{\citenamefont {Ramirez}\ \emph {et~al.}(1999)\citenamefont {Ramirez}, \citenamefont {Hayashi}, \citenamefont {Cava}, \citenamefont {Siddharthan},\ and\ \citenamefont {Shastry}}]{ramirez1999zero}%
  \BibitemOpen
  \bibfield  {author} {\bibinfo {author} {\bibfnamefont {A.~P.}\ \bibnamefont {Ramirez}}, \bibinfo {author} {\bibfnamefont {A.}~\bibnamefont {Hayashi}}, \bibinfo {author} {\bibfnamefont {R.~J.}\ \bibnamefont {Cava}}, \bibinfo {author} {\bibfnamefont {R.}~\bibnamefont {Siddharthan}},\ and\ \bibinfo {author} {\bibfnamefont {B.}~\bibnamefont {Shastry}},\ }\href {https://doi.org/10.1038/20619} {\bibfield  {journal} {\bibinfo  {journal} {Nature}\ }\textbf {\bibinfo {volume} {399}},\ \bibinfo {pages} {333} (\bibinfo {year} {1999})}\BibitemShut {NoStop}%
\bibitem [{\citenamefont {Pe\c{c}anha-Antonio}\ \emph {et~al.}(2017)\citenamefont {Pe\c{c}anha-Antonio}, \citenamefont {Feng}, \citenamefont {Su}, \citenamefont {Pomjakushin}, \citenamefont {Demmel}, \citenamefont {Chang}, \citenamefont {Aldus}, \citenamefont {Xiao}, \citenamefont {Lees},\ and\ \citenamefont {Br\"uckel}}]{Antonio2017}%
  \BibitemOpen
  \bibfield  {author} {\bibinfo {author} {\bibfnamefont {V.}~\bibnamefont {Pe\c{c}anha-Antonio}}, \bibinfo {author} {\bibfnamefont {E.}~\bibnamefont {Feng}}, \bibinfo {author} {\bibfnamefont {Y.}~\bibnamefont {Su}}, \bibinfo {author} {\bibfnamefont {V.}~\bibnamefont {Pomjakushin}}, \bibinfo {author} {\bibfnamefont {F.}~\bibnamefont {Demmel}}, \bibinfo {author} {\bibfnamefont {L.-J.}\ \bibnamefont {Chang}}, \bibinfo {author} {\bibfnamefont {R.~J.}\ \bibnamefont {Aldus}}, \bibinfo {author} {\bibfnamefont {Y.}~\bibnamefont {Xiao}}, \bibinfo {author} {\bibfnamefont {M.~R.}\ \bibnamefont {Lees}},\ and\ \bibinfo {author} {\bibfnamefont {T.}~\bibnamefont {Br\"uckel}},\ }\href {https://doi.org/10.1103/PhysRevB.96.214415} {\bibfield  {journal} {\bibinfo  {journal} {Phys. Rev. B}\ }\textbf {\bibinfo {volume} {96}},\ \bibinfo {pages} {214415} (\bibinfo {year} {2017})}\BibitemShut {NoStop}%
\bibitem [{\citenamefont {Lozano-G\'omez}\ \emph {et~al.}(2024{\natexlab{b}})\citenamefont {Lozano-G\'omez}, \citenamefont {Benton}, \citenamefont {Gingras},\ and\ \citenamefont {Yan}}]{lozano2024atlas}%
  \BibitemOpen
  \bibfield  {author} {\bibinfo {author} {\bibfnamefont {D.}~\bibnamefont {Lozano-G\'omez}}, \bibinfo {author} {\bibfnamefont {O.}~\bibnamefont {Benton}}, \bibinfo {author} {\bibfnamefont {M.~J.}\ \bibnamefont {Gingras}},\ and\ \bibinfo {author} {\bibfnamefont {H.}~\bibnamefont {Yan}},\ }\bibfield  {journal} {\bibinfo  {journal} {arXiv preprint arXiv:2411.03547}\ }\href {https://doi.org/10.48550/arXiv.2411.03547} {10.48550/arXiv.2411.03547} (\bibinfo {year} {2024}{\natexlab{b}})\BibitemShut {NoStop}%
\end{thebibliography}

%

\pagebreak 

\quad 

\newpage

\section*{Supplemental Information for Intrinsic quantum disorder in \yto{} and the quantum $S=1/2$ pyrochlore phase diagram} 

\quad 

\renewcommand{\thesection}{\Roman{section}}
\setcounter{section}{0}
\renewcommand\thefigure{S.\arabic{figure}}
\setcounter{figure}{0}
\renewcommand{\theequation}{S\arabic{equation}}
\setcounter{equation}{0}
\renewcommand{\thetable}{S\Roman{table}}
\setcounter{table}{0}

\section*{I. \> Pyrochlore geometry and effective spin-1/2 dipolar Hamiltonian}
\label{sec:pyrochlore_geom}
We summarize here some key details of the pyrochlore geometry and low-energy effective spin-1/2 dipolar Hamiltonian. Some of these details overlap with what has been reported in previous work (for example, see Refs.~\cite{Ross_Hamiltonian,Yan2017, Scheie2017, Changlani2017quantum, Scheie2020}) -- we restate them here for completeness.

\begin{figure}[h]
	\centering
	\includegraphics[width=0.98\linewidth]{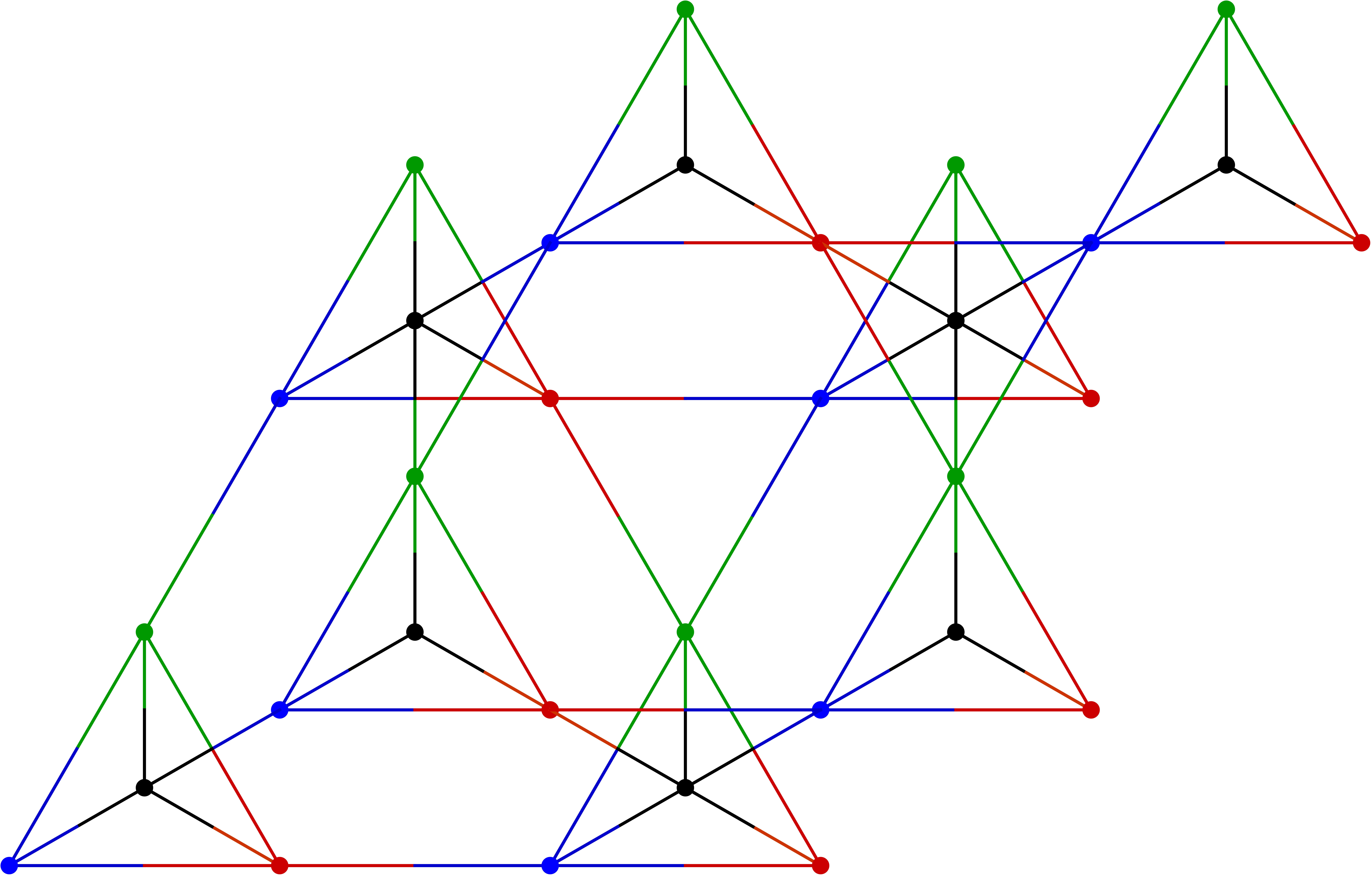}
	\caption{32 site cluster used in the ED calculations.}
	\label{fig:32sites}
\end{figure}

The pyrochlore lattice has four sublattices, which we label as $0,1,2,3$, and we take the relative locations of the sites on a single tetrahedron to be,
(in units of lattice constant $a$) ${\mathbf{r}_0}=(1/8,1/8,1/8)$, ${\mathbf{r}_1}=(1/8,-1/8,-1/8)$,
${\mathbf{r}_2}=(-1/8,1/8,-1/8)$ and ${\mathbf{r}_3}=(-1/8,-1/8,1/8)$. The centers of each such tetrahedron lie on the vertices of a FCC lattice. Thus a finite size cluster would have $4L^3$ sites where $L$ corresponds to the number of translations along each of the primitive lattice vectors. Our 32 site system corresponds to $L=2$, see Fig.~\ref{fig:32sites}.
Alternatively, the sites of a pyrochlore lattice can be thought of as those derived from a simple cubic lattice with a 16 site unit cell. A simple cubic lattice with edge length $L$ has $16 L^3$ sites, and thus our 16 site system corresponds to the case of $L=1$. Periodic boundary conditions along the directions of the primitive lattice vectors of the underlying Bravais lattice are considered in both cases.

As mentioned in the main text, the spin $1/2$ low-energy effective Hamiltonian on the pyrochlore lattice, with nearest neighbor interactions and Zeeman coupling to an external field ($h=(h_x,h_y,h_z)$) is given by,
\begin{equation}
H = \frac{1}{2} \sum_{ij} J^{\mu\nu}_{ij} S^{\mu}_{i} S^{\nu}_{j} - \mu_{B} h^{\mu} \sum_{i} g^{\mu \nu}_{i} S^{\nu}_{i}
\label{eq:Ham_supp}
\end{equation}
where $i,j$ are nearest neighbors and $\mu,\nu$ refer to $x,y,z$, $S^{\mu}_i$ refer to the spin-1/2 components at site $i$, and
 $\mathbf{J}_{ij}$ and $\mathbf{g}_i$ are bond and site dependent interactions and coupling matrices respectively. Symmetry considerations dictate that $\mathbf{J}_{ij}$ and $\mathbf{g}_{i}$ are
completely described by four ($J_1,J_2,J_3,J_4$) and two ($g_{xy}$,$g_z$) scalars respectively.
$\mathbf{J}_{ij}$ depends only on the sublattices that $i,j$ belong to, similarly $\mathbf{g}_i$ depends only on the sublattice of site $i$, and thus we use the notation
in terms of $i,j=0,1,2,3$. Also, since $\mathbf{J}_{ij}=\mathbf{J}_{ji}^T$, we write out only the $i<j$ matrices. The $\mathbf{J}_{ij}$ matrices are,
\begin{align*}
\mathbf{J}_{01} &\equiv
\left(\begin{array}{ccc}
 J_2 & J_4 & J_4 \\
-J_4 & J_1 & J_3 \\
-J_4 & J_3 & J_1 \end{array} \right), &
\mathbf{J}_{02} &\equiv
\left(\begin{array}{ccc}
 J_1 & -J_4 & J_3 \\
 J_4 & J_2 & J_4 \\
 J_3 & -J_4 & J_1 \end{array} \right), \\
\mathbf{J}_{03} &\equiv
\left(\begin{array}{ccc}
 J_1 & J_3 & -J_4 \\
 J_3 & J_1 & -J_4 \\
 J_4 & J_4 & J_2 \end{array} \right), &
\mathbf{J}_{12} &\equiv
\left(\begin{array}{ccc}
 J_1 & -J_3 & J_4 \\
-J_3 & J_1 & -J_4 \\
-J_4 & J_4 & J_2 \end{array} \right), \\
\mathbf{J}_{13} &\equiv
\left(\begin{array}{ccc}
 J_1 & J_4 & -J_3 \\
-J_4 & J_2 & J_4 \\
-J_3 & -J_4 & J_1 \end{array} \right), &
\mathbf{J}_{23} &\equiv
\left(\begin{array}{ccc}
 J_2 & -J_4 & J_4 \\
 J_4 & J_1 & -J_3 \\
-J_4 & -J_3 & J_1 \end{array} \right).  
\end{align*}

Defining $g_{+}=\frac{1}{3}(2g_{xy}+g_{z})$ and $g_{-}=\frac{1}{3}(g_{xy}-g_z)$, the
$\mathbf{g}_i$ matrices read as,
\begin{align*}
\mathbf{g}_{0} &\equiv
\left(\begin{array}{ccc}
 g_+ & -g_- & -g_- \\
 -g_- & g_+ & -g_- \\
 -g_- & -g_- & g_+ \end{array} \right), &
\mathbf{g}_{1} &\equiv
\left(\begin{array}{ccc}
 g_+ & g_- & g_- \\
 g_- & g_+ & -g_- \\
 g_- & -g_- & g_+ \end{array} \right), \\
\mathbf{g}_{2} &\equiv
\left(\begin{array}{ccc}
 g_+ & g_- & -g_- \\
 g_- & g_+ & g_- \\
-g_- & g_- & g_+ \end{array} \right), &
\mathbf{g}_{3} &\equiv
\left(\begin{array}{ccc}
 g_+ & -g_- & g_- \\
-g_- & g_+ & g_- \\
 g_- & g_- & g_+ \end{array} \right). 
\end{align*}

\section*{II. \>Additional neutron scattering data}

\begin{figure*}
 	\centering\includegraphics[width=0.96\textwidth]{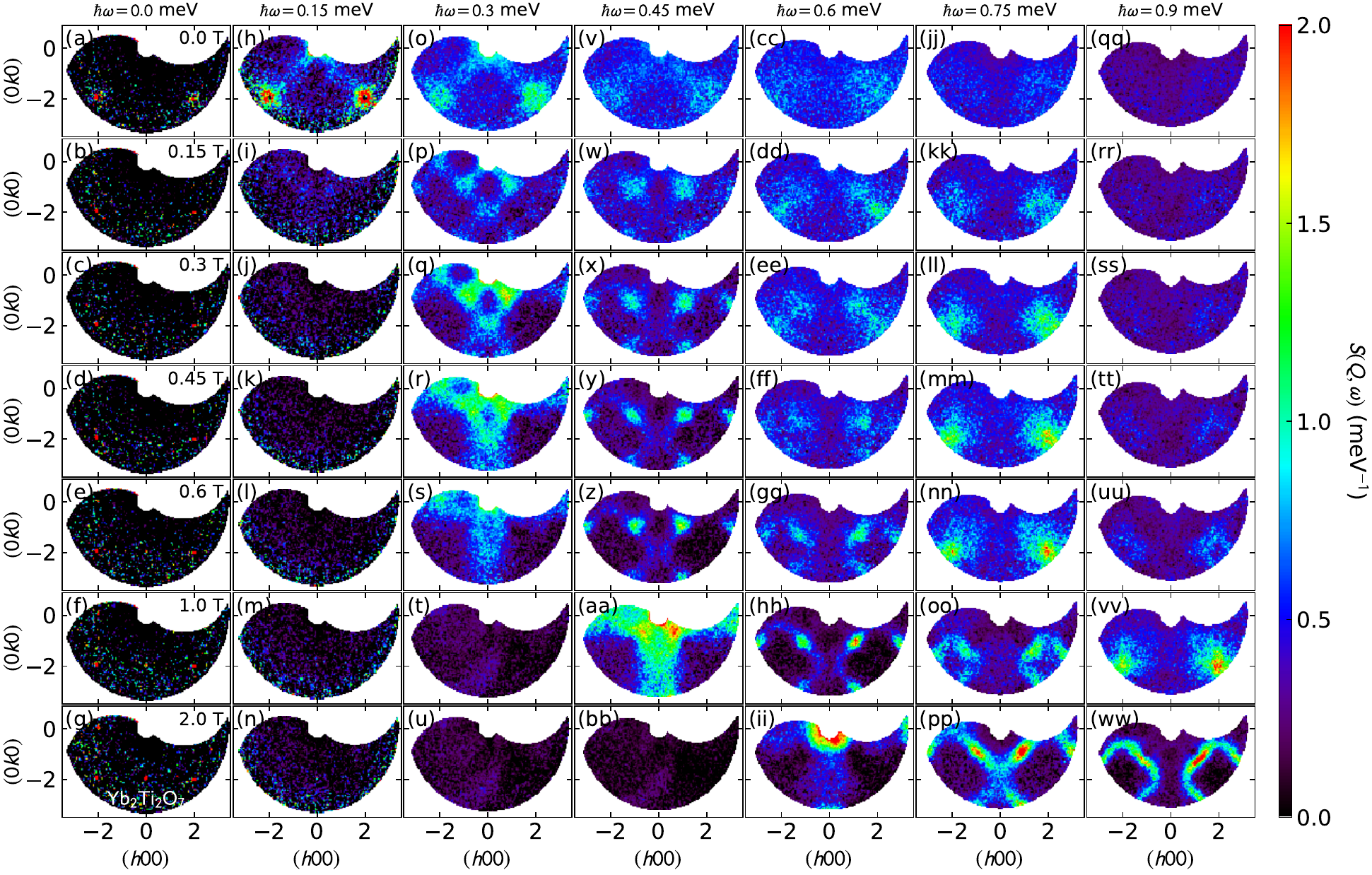}
 	
 	\caption{Constant energy slices of $(hk0)$ scattering in \yto{}. The rows show the seven measured fields from 0~T to 2~T, and the columns indicate different constant energy ranges, with a bin width $\hbar \omega \pm 0.05$~meV. A 12~K background was subtracted, as described in main text Appendix A.}
 	
 	\label{fig:constE}
 	
 \end{figure*}

\begin{figure*}
	\centering\includegraphics[width=0.72\textwidth]{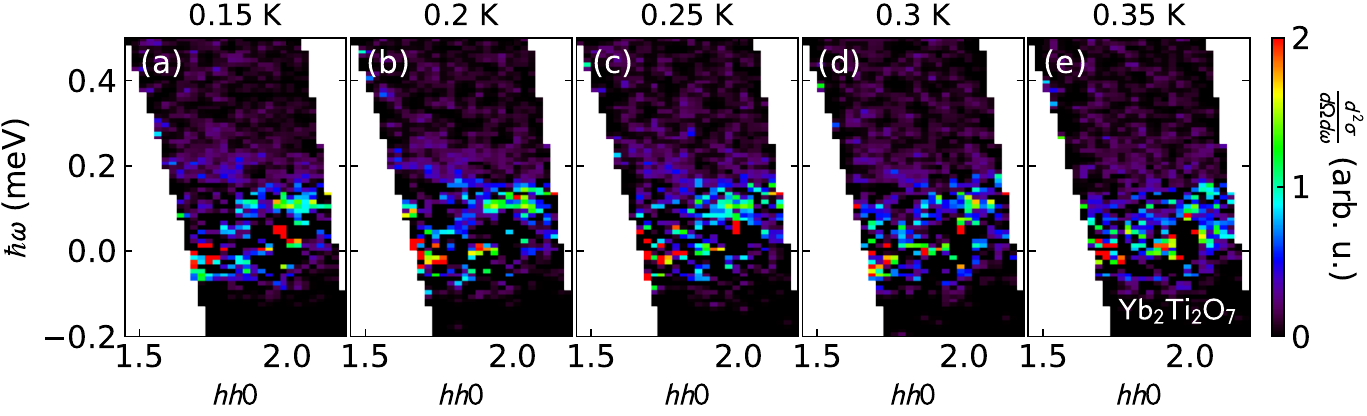}
	
	\caption{\yto{} scattering measured with $E_i = 2.49$~meV neutrons at temperatures between 0.15~K and 0.35~K, with 2~T scattering subtracted as a background. Note the resonance at 0.11 meV below $T_c=0.27$~K which disappears above the phase transition. Line cuts of this feature are shown in main text Fig. 8. }
	
	\label{fig:LowEslices}
	
\end{figure*}

The scattering data shown in the main text are cuts along high-symmetry directions in reciprocal space. 
In Fig. \ref{fig:constE} we plot the full $[hk0]$ scattering plane of data. 
The higher resolution data with  $E_i = 2.49$~meV  along the $hh0$ direction are shown in Fig. \ref{fig:LowEslices}. Even in the colormap plots, the resonance at 0.11~meV is distinctly visible at low temperatures. 

\begin{figure*}
	\centering
	\includegraphics[width=0.82\textwidth]{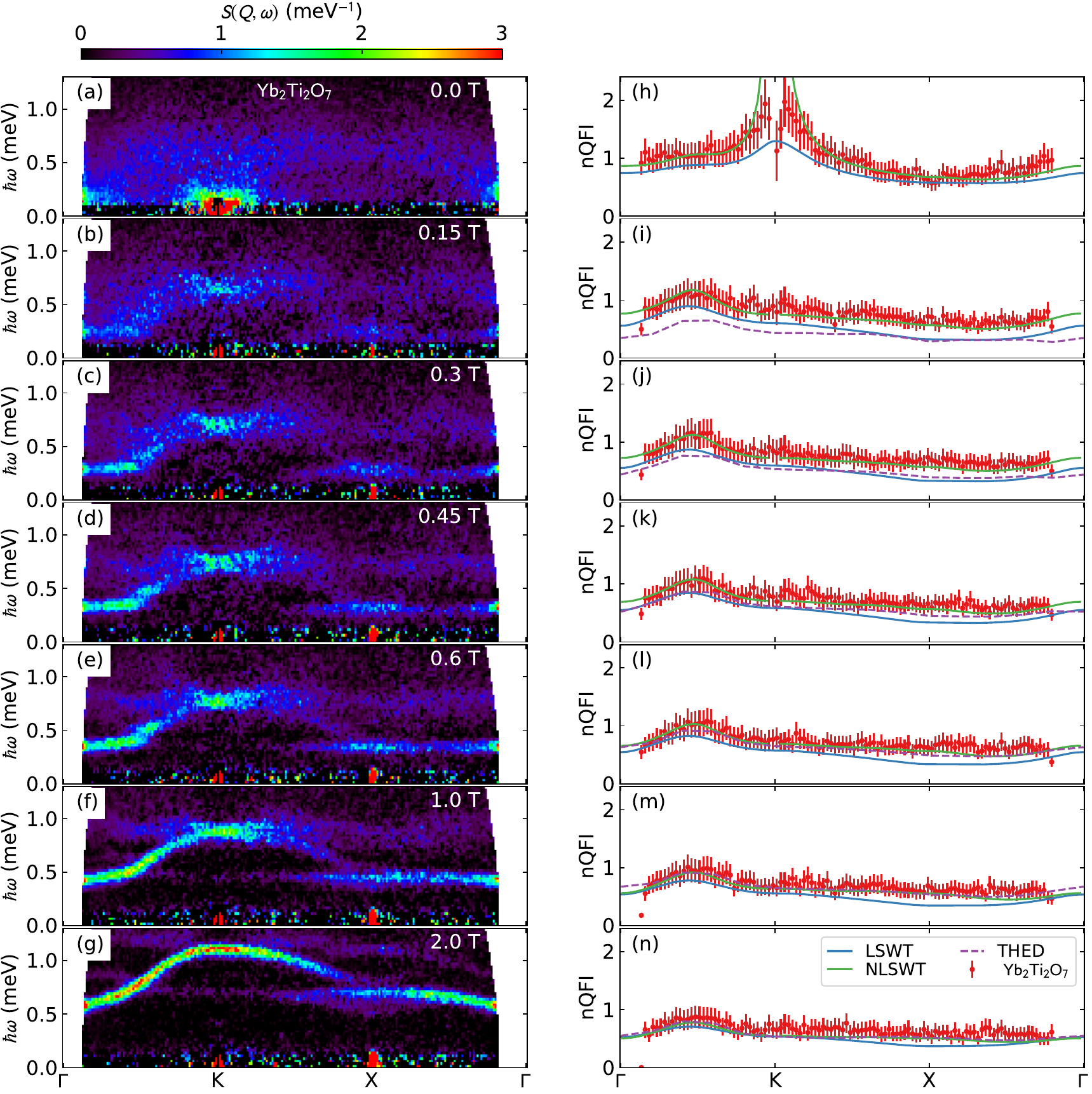}
	\caption{Quantum Fisher Information versus wavevector along high symmetry cuts. Panels (a)-(g) show the inelastic intensity normalized to $S=1/2$, and panels (h)-(n) show the calculated QFI of  \yto{} compared to different theoretical models.}
	\label{fig:QFIvsQ}
\end{figure*}

In Appendix B we compute the Quantum Fisher Information (QFI) of Yb$_2$Ti$_2$O$_7$ as a function of magnetic field. 
QFI is a wavevector-dependent quantity \cite{scheie2024tutorial}, and in Figure \ref{fig:QFIvsQ} we plot the  nQFI [normalized QFI, main text Eq. (B1)] along the high symmetry directions. Interestingly, the peak QFI is different between zero field and the finite field data. 
In Appendix B, we plot the maximal QFI across the Brillouin zone to give the most stringent bound on multipartite entanglement. 

An important part of the absolute unit conversion to $S=1/2$ (and thus being able to compute nQFI) is normalizing the $g$-tensor, as discussed in Appendix B. Figure \ref{fig:gtensorcomparison} shows the data compared to LSWT calculations with and without the anisotropic $g$-tensor $g_{xx}=g_{yy}=g_{xy} = 4.17$,  $g_{zz} = g_{z} = 2.14$ \cite{Thompson_2017} included in the intensity  calculation.  
To a rough approximation, the LSWT without a $g$-tensor is a simple intensity rescaling (though some details differ, particularly the visibility of the weak low-energy modes). 

\begin{figure*}
	\centering
	\includegraphics[width=0.96\textwidth]{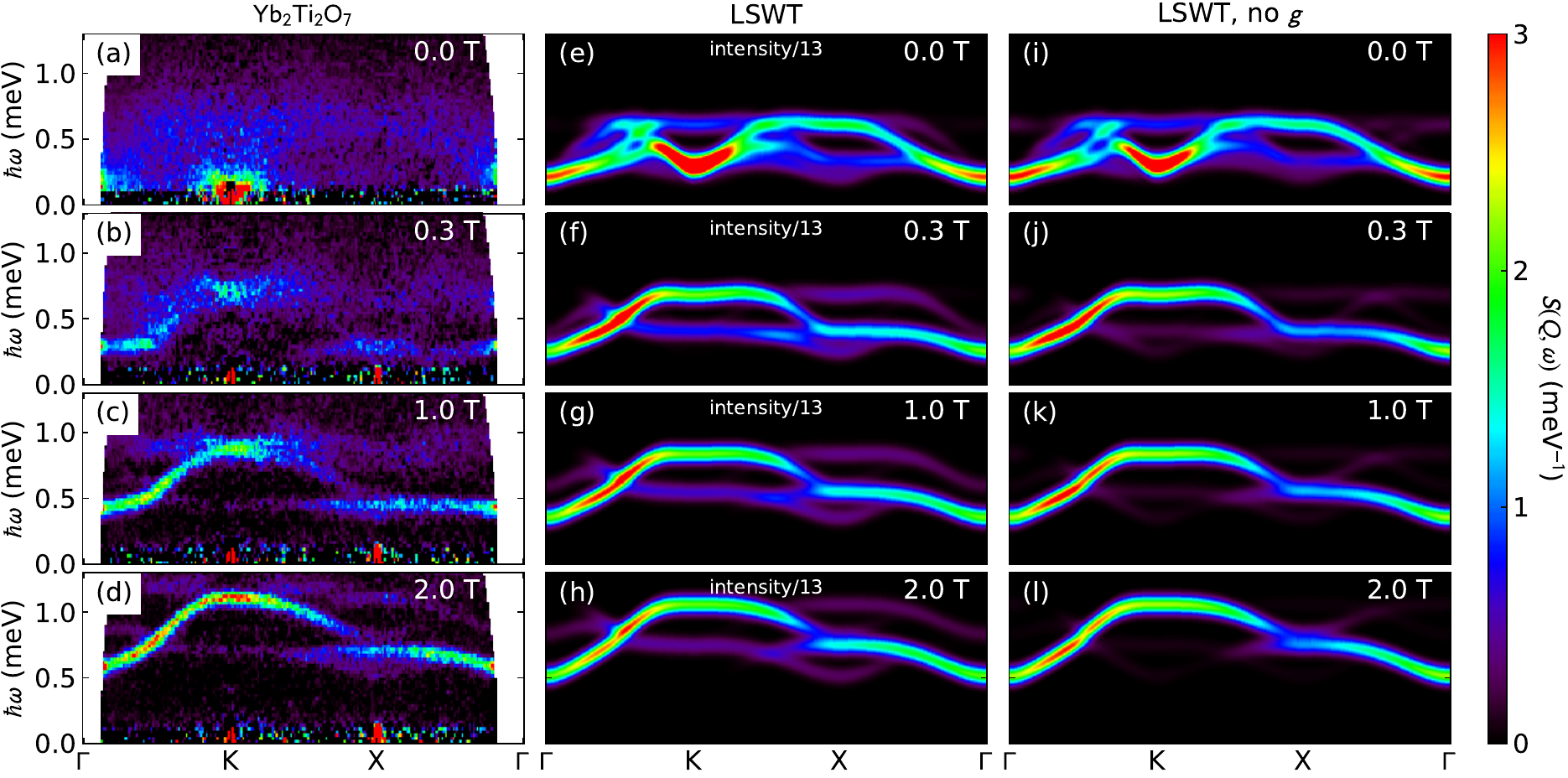}
	\caption{Comparison of the inelastic spectrum of \yto{} experiment (a)-(d), with the $g$-tensor (e)-(h), and without the $g$-tensor (i)-(l). 
		(The simulations with the $g$-tensor were plotted with intensity divided by a factor of 13 to keep all plots on the same color scale.) At all fields there are only minor differences between the LSWT calculations, and so we normalize the 2~T neutron scattering to the LSWT calculation in panel (l).}
	\label{fig:gtensorcomparison}
\end{figure*}

\section*{III. \>Details of nonlinear spin wave theory (NLSWT) calculations}

Our NLSWT calculations are carried out in the basis of the original Holstein-Primakoff bosons with both the normal and anomalous parts of the Green's functions kept at each stage of the calculation. For the dynamical structure factor the nonlinear spin-wave corrections both renormalize the transverse-transverse spin-spin correlation function that appears in linear spin-wave theory, but also provide contributions to the longitudinal-longitudinal and transverse-longitudinal spin-spin correlation functions~\cite{mourigal2013}. We calculated the field-dependent spectra for a wavevector grid corresponding to a finite system of size $N=4L^3$ with $L=32$, with the self-energy and other dynamical correlation functions discretized on a dense frequency grid. The relevant matrix Green's functions are computed directly with a very small artificial broadening and the final dynamical structure factor is then convolved with a Gaussian of width comparable to the experimental energy resolution.

\section*{IV. \>Details of the exact diagonalization calculations}

As stated in the main text, we have carried out Lanczos exact diagonalization (ED) calculations on two system sizes, $N=16$ and $N=32$ sites (see Sec.~\ref{sec:pyrochlore_geom}), with our main conclusions based on the latter. 
For $N=32 = 4 \times 2^3$, our ED calculations used translational symmetry, that block diagonalized the Hamiltonian into 8 sectors, each with a Hilbert space dimension of approximately 536 million. The sectors are labeled according to three dimensional momenta $(k_1,k_2,k_3)$, where $k_i = 0, \pi$. 

We work with the effective spin-1/2 nearest neighbor Hamiltonian (see main text and Sec.~\ref{sec:pyrochlore_geom}) and simulate approximately 450 parameter sets, sampling more points near phase boundaries. For all the parameter sets considered in this study, we find the ground state from ED to be in the $(0,0,0)$ sector. Expectation values of spin operators (static spin correlations) were computed in the ground state for each parameter set. 
These correlations were processed by utilizing the relationship between magnetic moments and spins (dipoles) given by,
\begin{equation}
M^{(i)}_{\mu} = \sum_{\rho} g^{(i)}_{\mu,\rho} S^{(i)}_{\rho}
\end{equation}
where $i$ is a site label and $\mu,\nu$ refer to $x,y,z$ (in the global frame). This yields,
\begin{equation}
\langle M^{(i)}_{\mu} M^{(j)}_{\nu} \rangle = \sum_{\rho,\gamma} g^{(i)}_{\mu,\rho} g^{(j)}_{\nu,\gamma} \langle S^{(i)}_{\rho} S^{(j)}_{\gamma}\rangle  
\end{equation}
where $\langle ... \rangle$ indicates expectation values computed in the ground state. 
Note that even though our ED based theoretical phase diagram was constructed for the case of zero external field, the $g$ matrices are required for the transformation between spins to magnetic moments.

These correlations were then Fourier transformed to compare our results with the measurements of energy-integrated neutron scattering.
\begin{equation}
S (\mathbf{Q}) = \frac{1}{N}\sum_{\mu, \nu} \Big( \delta_{\mu \nu} - \frac{Q_{\mu}Q_{\nu}}{Q^2}\Big) e^{i \mathbf{Q}\cdot ( \mathbf{r}_i - \mathbf{r}_j ) }\langle M^{(i)}_{\mu} M^{(j)}_{\nu} \rangle 
\end{equation}

\section*{V. \>Correlation analysis}

\begin{figure*}
	\centering
	\includegraphics[width=0.99\textwidth]{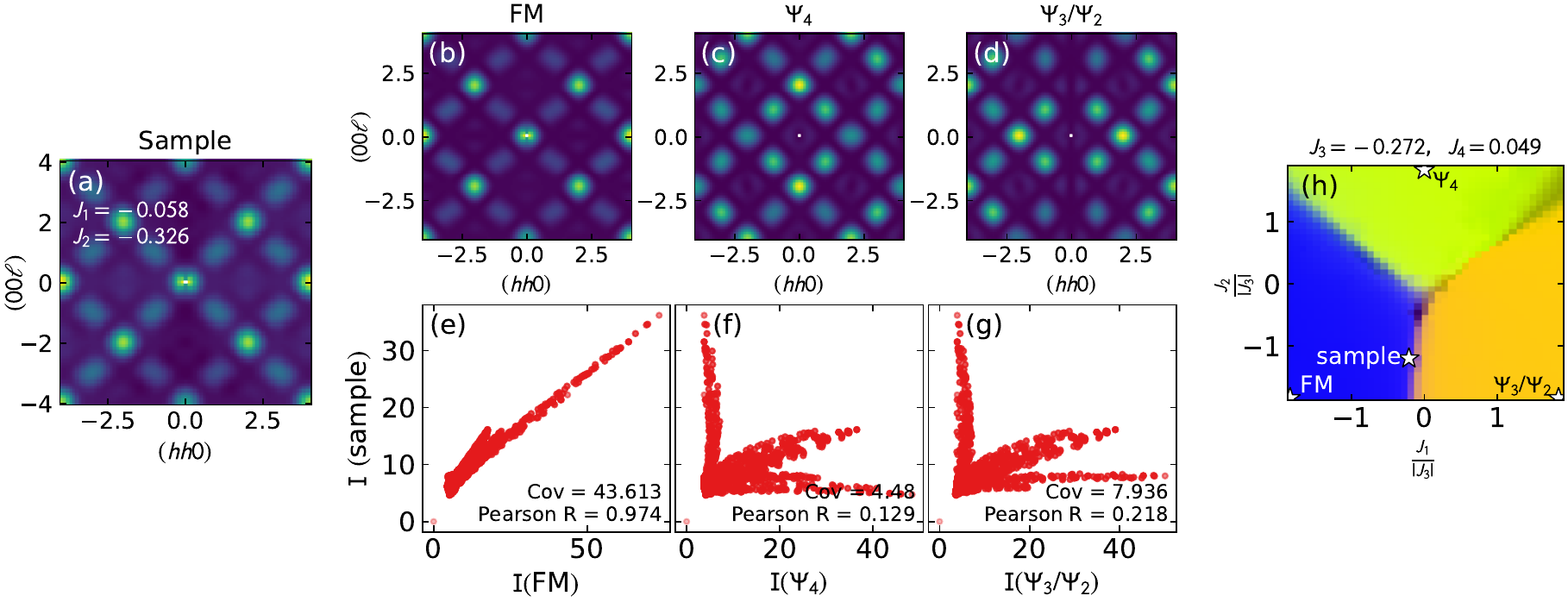}
	\caption{Correlation analysis to construct the phase diagram. Panel (a) shows a sample calculated $S(q)$ spectrum with $J = \{-0.1,\>-0.05,\>-0.272,\>0.049 \}$~meV. Panel (b) shows the characteristic FM spectrum at $J = \{-0.5,\>-0.5,\>-0.272,\>0.049 \}$~meV, panel (c) shows the characteristic $\Psi_4$ structure factor $J = \{0.0,\>0.5,\>-0.272,\>0.049 \}$~meV, and panel (d) shows the characteristic $\Psi_3/\Psi_2$ spectrum $J = \{0.5,\>-0.5,\>-0.272,\>0.049 \}$~meV. Panels (e)-(f) show the pixel intensities of the sample $S(\mathbf{Q})$ plotted against the FM, $\Psi_4$, and $\Psi_3/\Psi_2$ spectra. Each panel shows the covariance and Pearson R statistic, showing clearly that the spectra is dominantly FM. Panel (h) shows the phase diagram constructed from the correlation analysis.}
	\label{fig:CorrelationExplanation}
\end{figure*}

We construct the phase diagram from the calculated $S(\mathbf{Q})$ structure factors by computing correlation statistics with characteristic $S(\mathbf{Q})$ as shown in Fig. \ref{fig:CorrelationExplanation}.
The method is as follows: we select a ``characteristic structure factor'' from each phase deep inside the phase as defined by the classical phase boundaries found in Ref. \cite{Yan2017}. We then create a 1D vector of pixel intensities of each structure factor $S(\mathbf{Q})$, and compute the correlation with each of the characteristic spectra using the covariance and Pearson R statistic (as implemented in the \textit{Scipy} package \cite{virtanen2020scipy}).

From this, we build a phase diagram by assigning the Pearson R correlation with $\Psi_3/\Psi_2$, $\Psi_4$, and FM as red, green, and blue values respectively. By plotting the colors on a grid, the phase diagram naturally emerges. We then interpolate between the calculated points to fill out the phase diagram (the $\Psi_3/\Psi_2$ phase appears as orange because there is nonzero correlation with $\Psi_4$ antiferromagnetism even deep inside the phase). 

The only assumption we have made in this analysis is that there are three phases, but the boundaries between them are very clear. The weakness of this approach is also that we have assumed only three phases. The existence of a small region in the center of the phase diagram where no phase seems to be stabilized is an indication of a quantum disordered phase, as is the Heisenberg spin liquid region in the upper right hand side of the phase diagram. 

An alternative view of this data is plotting the covariance, which takes into account not only the correlation between different $S(\mathbf{Q})$, but the relative intensities of the correlated features. As such, this is an approximate way to estimate the strength of the ordered moment. From this, we see the dramatic moment reduction along the FM/AFM phase boundary discussed in the main text. 

Figure \ref{fig:PhaseDiagramYTO} shows the correlation analysis of the \yto{} phase diagram including a colormap of the energy landscape. Figure \ref{fig:PhaseDiagramJ4=0} shows the same phase diagram with  $J_3=-0.3$~meV, $J_4=0.0$ instead of the \yto{} values. This is very similar to the \yto{} phase diagram, although here the difference between the classical and quantum phase boundaries is slightly greater. 

\begin{figure*}
	\centering
	\includegraphics[width=\textwidth]{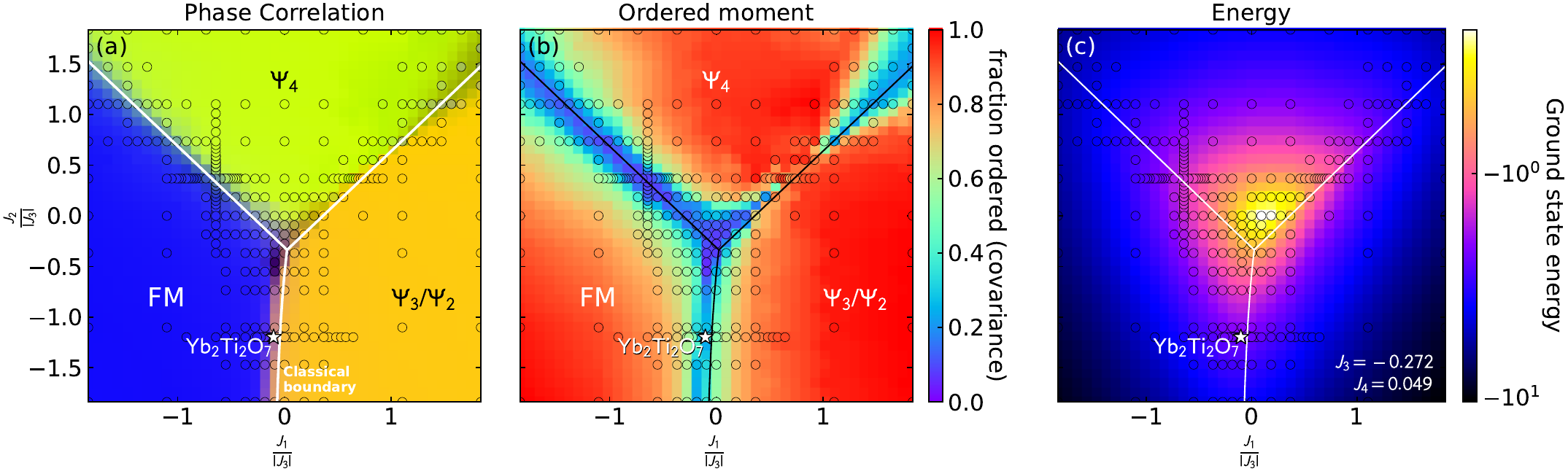}
	\caption{Calculated pyrochlore phase diagram assuming $J_3=-0.272$~meV, $J_4=0.049$~meV (the \yto{} parameters in Ref. \cite{Thompson_2017}). Panel (a) shows the phases determined by Pearson R correlations. Panel (b) shows the size of the ordered moment calculated via covariance with the reference $S(\mathbf{Q})$. Panel (c) shows the ground state energy; interestingly, the maximum energy is within the $\Psi_4$ phase rather than on a phase boundary. Black circles indicate parameters which were explicitly calculated, pixels in between were interpolated, solid lines are the classical phase boundary \cite{Yan2017}.}
	\label{fig:PhaseDiagramYTO}
\end{figure*}

\begin{figure*}
	\centering
	\includegraphics[width=\textwidth]{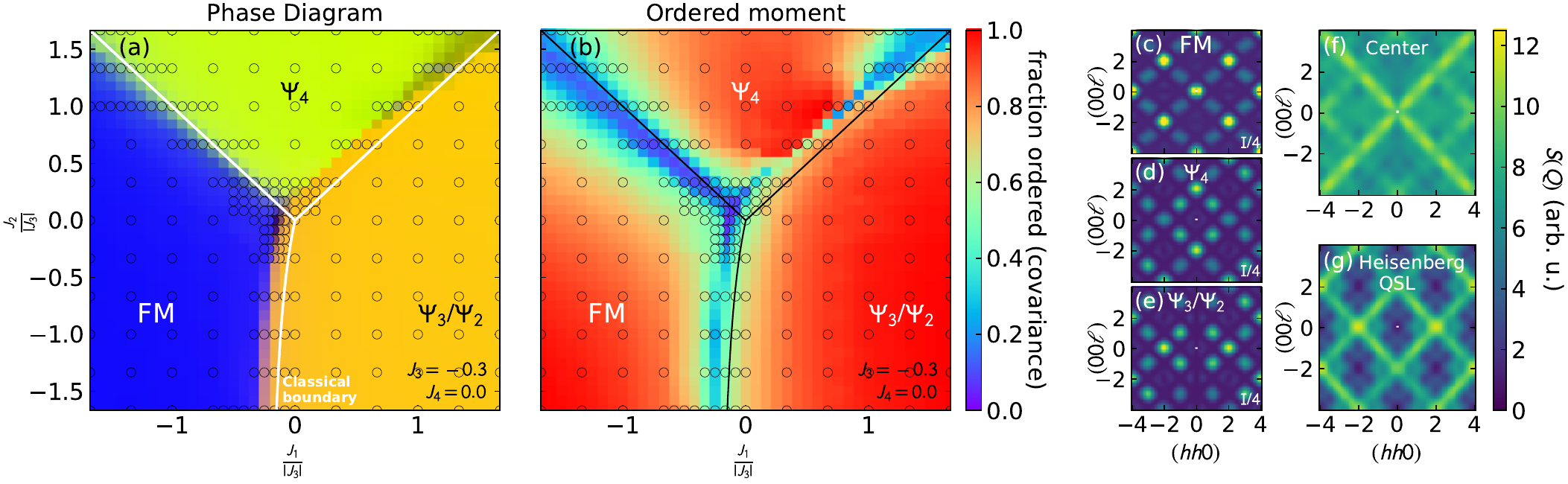}
	\caption{Calculated pyrochlore phase diagram assuming $J_3=-0.3$~meV, $J_4=0.0$. Panel (a) shows the phases determined by Pearson R correlations. Panel (b) shows the size of the ordered moment calculated via covariance with the reference spectra. 
    Panels (c)-(e) show the reference $S(\mathbf{Q})$ for FM, $\Psi_4$, and $\Psi_3/\Psi_2$. Panel (f) shows the spectrum for the center of the phase diagram, and panel (g) shows the spectrum for the Heisenberg quantum spin liquid region at the top right. 
    Qualitatively, these results are the same as in main text Fig. 2, but in this case the phase boundaries are shifted more from the classical values.}
	\label{fig:PhaseDiagramJ4=0}
\end{figure*}

\subsection*{Classical phase boundaries}

Overtop the phase diagram computed from ED, we also plot the phase boundaries derived in Ref. \cite{Yan2017}:
\begin{align*}
    J_2 &= J_1 - 2 J_4 \\
    J_2 &= \frac{-J_1 J_3 + 2 J_1 J_4 - 2 J_3 J_4 + 2 J_4^2}{J_3 - 2 J_4} \\
    J_2 &= \frac{4 J_1^2 - 5 J_1 J_3 - 2 J_1 J_4 + 2 J_3 J_4}{4 J_1 - J_3 - 2 J_4}
\end{align*}
where the degenerate point at the center of the phase diagram is at $J_1 = {J_4^2}/{(2J_4 - J_3)}$.

\section*{VI. \>Additional line scans for tracking the intermediate phase}

\begin{figure*}
	\centering
	\includegraphics[width=0.98\linewidth]{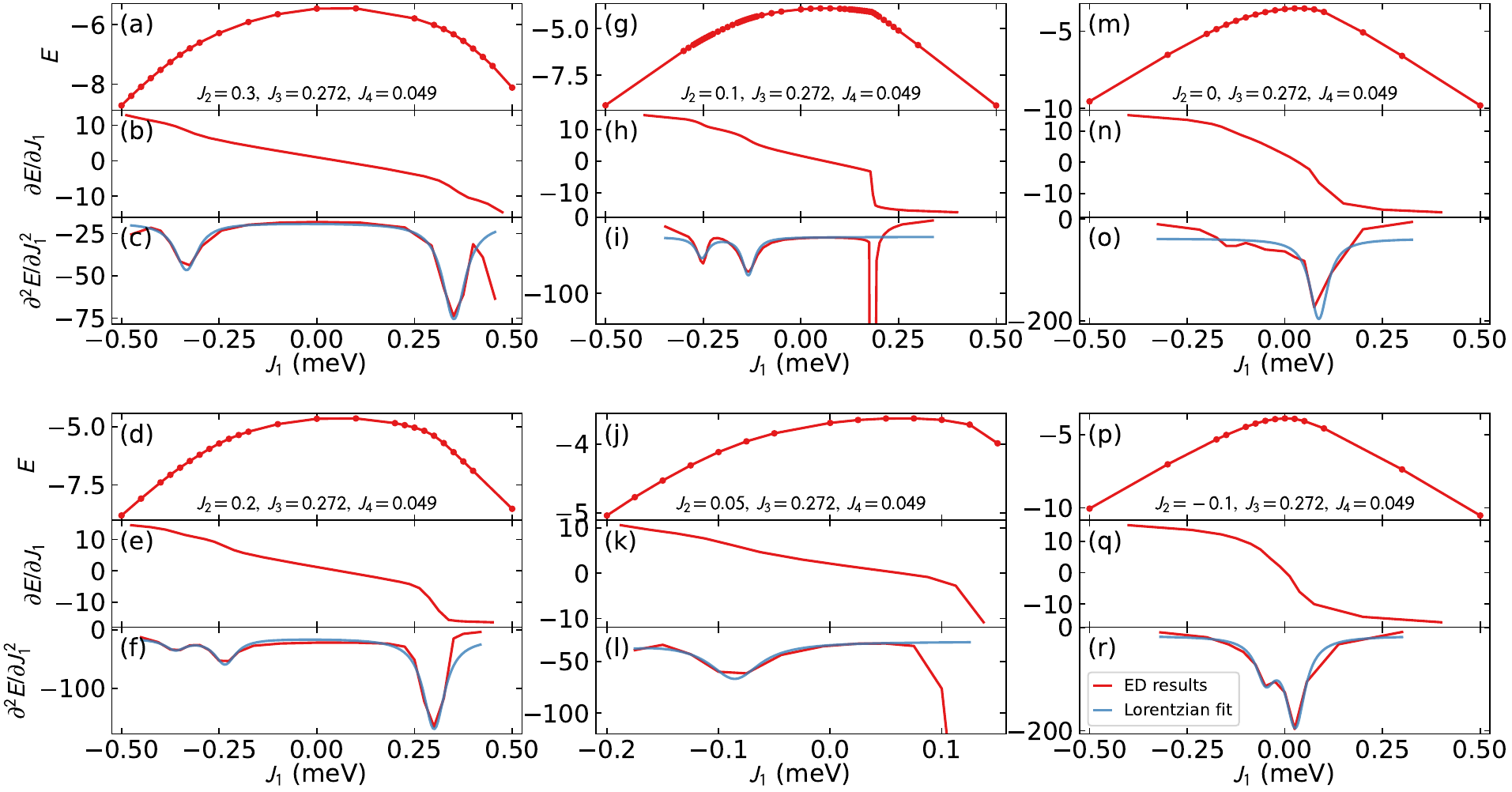}
	\caption{Energy scans used to identify phase boundaries in main text Fig. 1. Energy and its derivatives are plotted against $J_1$ for six different $J_2$ values. The boundary itself is identified by fitting the minimum in the second derivative with a Lorentzian lineshape. All energy units are in meV.}
	\label{fig:find-phase-boundaries}
\end{figure*}

Figure 1 of the main text shows our phase diagram with multiple boundaries; these were determined by analyses of the type shown in Fig.~3 and 4, and Fig.\ref{fig:find-phase-boundaries}.
Here we build on the observation that the second derivative of the energy with $J_1$, for fixed $J_2$=-0.326~meV, shows a spike near the FM/AFM phase boundary. [The magnitude of the spike is largely controlled by the size of the spacing of the parameter grid chosen.] Importantly, its presence indicates an additional intermediate phase between FM and AFM ($\Psi_3/\Psi_2$).

\begin{figure*}
\centering
	\includegraphics[width=0.95\textwidth]{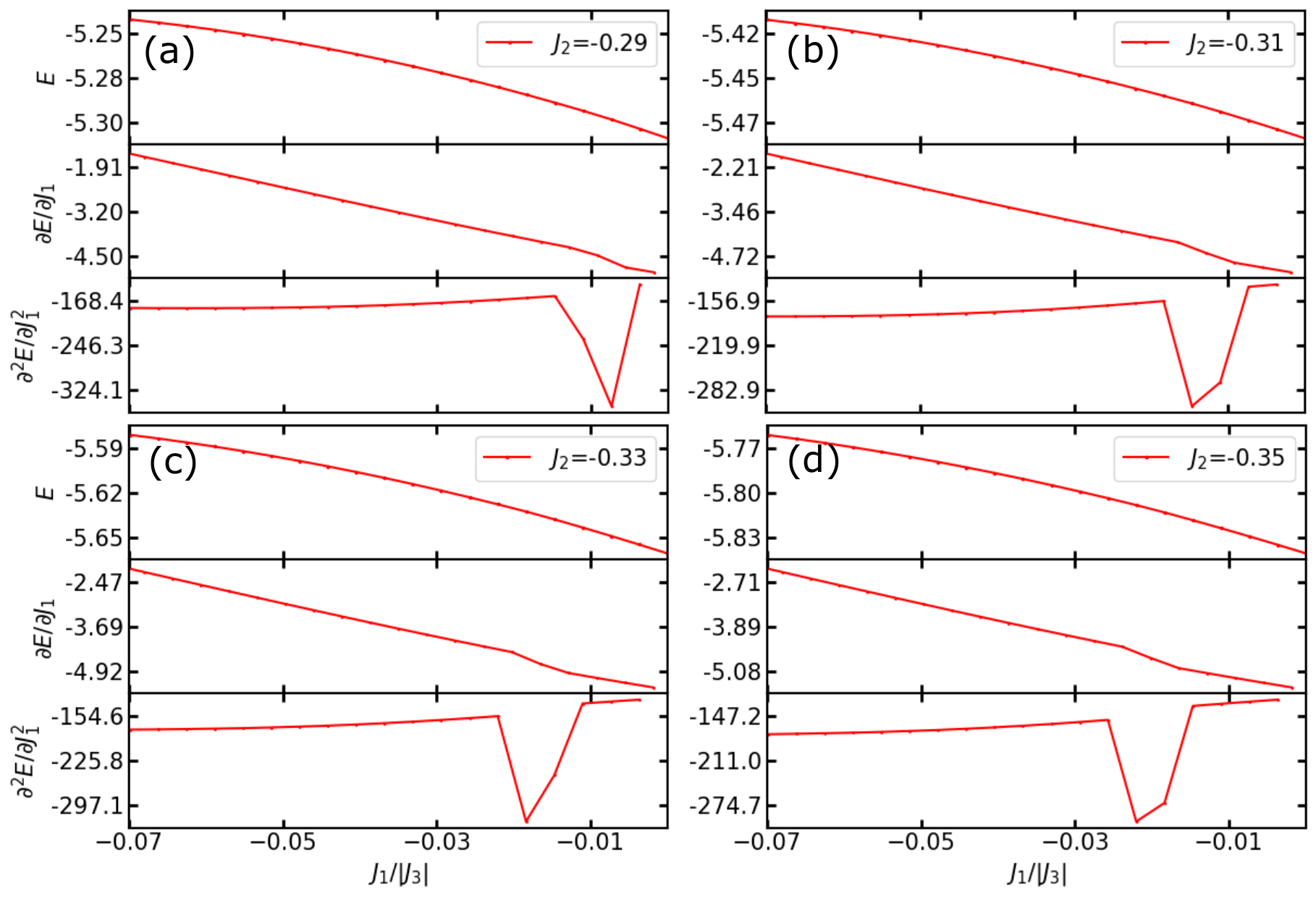}
	\caption{Ground state energy and its first and second derivative with $J_1$, as a function of $J_1/|J_3|$ for $J_2$ values fixed at (a) $-0.29$~meV,(b)$-0.31$~meV, (c) $-0.33$~meV and (d) $-0.35$ meV. $J_3=-0.272$~meV and $J_4=0.049$~meV, which correspond to the \yto{} Hamiltonian, were kept fixed. 
    All energy units are in meV.}
\label{fig:grid_scan}
\end{figure*}

To build confidence in the numerical results, we have carried out additional line scans with $J_1$ at other fixed values of $J_2$ (-0.29~meV,-0.31~meV,-0.33~meV and -0.35 meV). The $J_1$ values for each of these $J_2$ were varied from 0 to -0.02 meV in increments of 0.001 meV. Our results for the ground state energy, and its first and second derivatives with $J_1$ are shown in Fig.~\ref{fig:grid_scan} and clearly demonstrate that such spikes are also present at other values of $J_2$ near the FM/AFM phase boundary. We note that lowering $J_2$ from $-0.29$ to $-0.35$ meV leads to a movement of this feature to lower (more negative) values of $J_1$, consistent with the shape of the intermediate/AFM phase boundary shown 
in Fig. 1 of the main text.  

\section*{VII. \>Classical spin liquids}

\begin{figure*}
	\centering
	\includegraphics[width=0.98\textwidth]{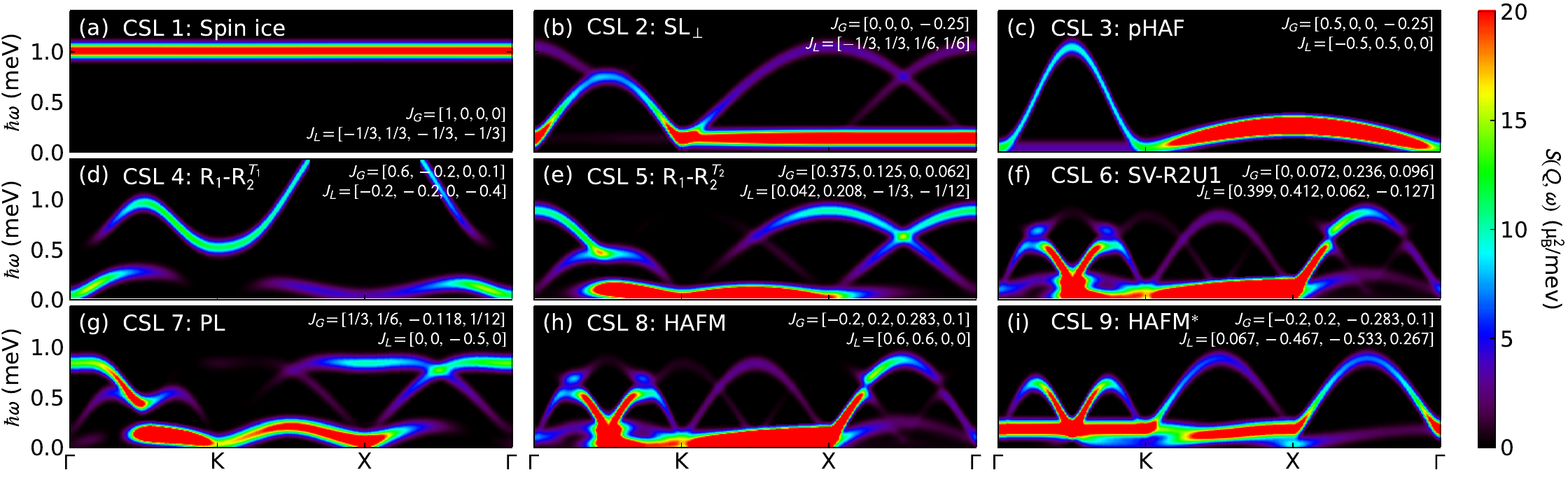}
	\caption{LSWT calculations of the nine classical spin liquids in Ref. \cite{lozano2024atlas} in a $\mu_0 H = 0.15$~T field along the $c$ axis. The Hamiltonians of each are listed in the upper right in the global basis $J_G = [J_{zz}, J_{\pm \pm}, J_{z \pm}, J_{\pm}]$ local basis $J_L = [J_1, J_2, J_3, J_4]$. None of these spectra resemble the experimental \yto{} spectra where an intense flat band runs from $X$ to $\Gamma$.}
	\label{fig:ClassicalComparison}
\end{figure*}

It is natural to ask whether the $X \rightarrow \Gamma $ flat band observed at 0.15~T in \yto{} can be reproduced by any of the highly degenerate points in the pyrochlore phase diagram \cite{lozano2024atlas,chung2024mapping,Francini_2025}. 
In Fig. \ref{fig:ClassicalComparison} we plot the LSWT calculated spectra of the nine classical spin liquids (CSLs) from Ref. \cite{lozano2024atlas} in a  $\mu_0 H = 0.15$~T field along [001]. Although there are many intense flat bands near zero energy which correpond to extensive degeneracies, none of them match precisely the flat modes observed in \yto{}. Therefore, despite the fact that the extended quantum phase is continuously connected to the ``pinch-line'' spin liquid (CSL 7 in Fig. \ref{fig:ClassicalComparison}) it is not possible to simplistically associate these features with this particular spin liquid. 

\begin{figure*}
	\centering
	\includegraphics[width=0.98\textwidth]{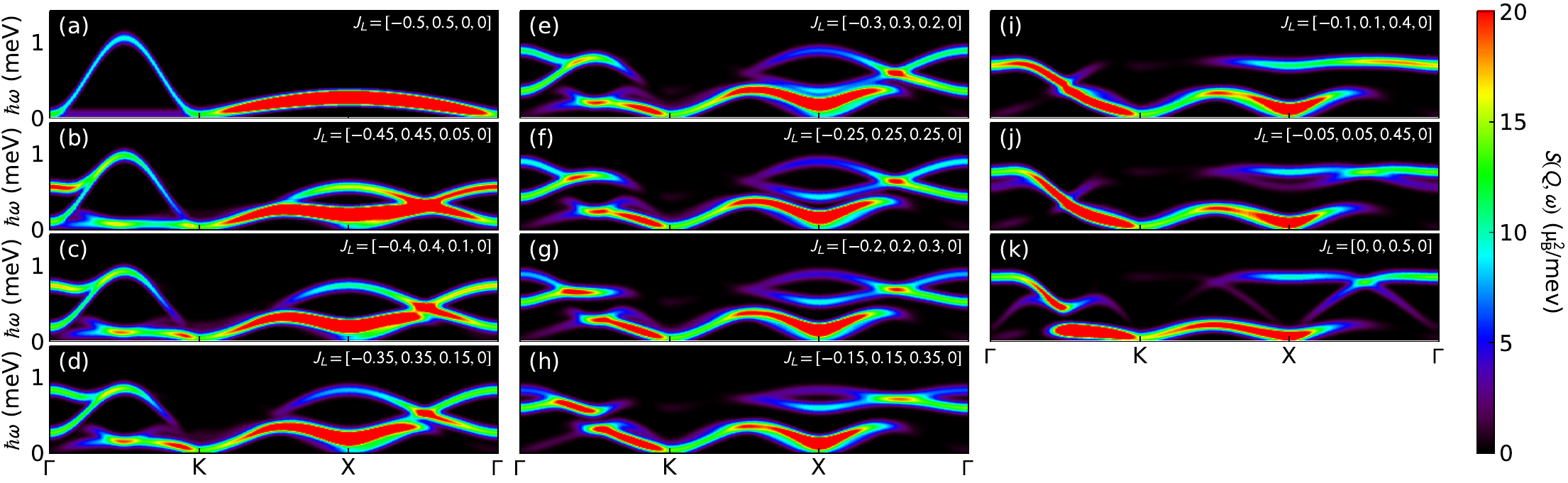}
	\caption{LSWT calculations in a $\mu_0 H = 0.15$~T field along $c$ tuning from Ref. \cite{lozano2024atlas} Classical Spin Liquid (CSL) 3 to CSL 7 (Fig. \ref{fig:ClassicalComparison}). None of the simulations reproduce the flat $\Gamma \rightarrow X$ mode noted in \yto{} and the interacting-magnon simulations.}
	\label{fig:ClassicalComparison2}
\end{figure*}

It is also worth asking whether one of the line-degeneracies (specifically the line where $J_2 = -J_1$) reproduces the flat $\Gamma \rightarrow X$ mode. In Fig. \ref{fig:ClassicalComparison2} we tune along this $J_2 = -J_1$ ``nematic phase'' boundary \cite{chung2024mapping,Francini_2025} connecting the CSL 3 to CSL 7 (pinch line) spin liquids \cite{lozano2024atlas}, but do not find a clear region where the flat $\Gamma \rightarrow X$ mode is reproduced. 

The failure of this exercise is not surprising as LSWT neglects interactions between magnons. Nevertheless, it would have been useful to know whether tuning to a known classical degeneracy would have reproduced the flat modes which seem to be driving the \yto{} to its exotic behaviors. 

\end{document}